\documentclass[aps,prev,onecolumn,preprintnumbers,floatfix,nofootinbib]{revtex4-1}
\pdfoutput=1

\usepackage[T1]{fontenc} 

\usepackage{geometry,amsmath,amsfonts}
\usepackage{slashed}
\usepackage{epsfig}
\usepackage{latexsym}
\usepackage{graphicx}
\usepackage{epstopdf}  % to use eps with Typeset -> Pdftex
\usepackage[justification=RaggedRight]{caption}
\usepackage{amssymb}
\usepackage{subfig}
\usepackage{color}
\usepackage{multirow}
\usepackage{color}
\usepackage{rotating}
\usepackage{ifthen}
\usepackage{epsfig}
\usepackage{graphics}
\usepackage{xcolor}
\usepackage{enumitem}

\begin{document}

\title{Gravitino Dark Matter and low-scale Baryogenesis}
\author{Giorgio Arcadi$^{a}$}
\email{giorgio.arcadi@th.u-psud.fr}
\author{Laura Covi$^{b}$}
\email{covi@theorie.physik.uni-goettingen.de}
\author{Marco Nardecchia $^{c}$}
\email{M.Nardecchia@dampts.ac.uk}

\vspace{0.1cm}
 \affiliation{
${}^a$ 
 Laboratoire de Physique Th\'eorique Universit\'e Paris-Sud, F-91405 Orsay, France
}

\affiliation{
${}^b$
Institute for Theoretical Physics, 
Georg-August University G\"ottingen, 
Friedrich-Hund-Platz~1, G\"ottingen, D-37077 Germany
 }

\affiliation{
${}^c$
DAMTP, CMS, University of Cambridge, Wilberforce Road, Cambridge CB3 0WA,
United Kingdom\\
Cavendish Laboratory, University of Cambridge, JJ Thomson Avenue, Cambridge
CB3 0HE, United Kingdom
 }

\begin{abstract} 

\noindent
A very simple way to obtain comparable baryon and DM densities in the early Universe is through their contemporary production from the out-of-equilibrium decay of 
a mother particle, if both populations are suppressed by comparably small numbers, i.e. the CP violation in the decay and the branching fraction respectively. 
We present a detailed study of this kind of scenario in the context of a R-parity violating realization of the MSSM in which the baryon asymmetry and the 
gravitino Dark Matter are produced by the decay of a Bino. 
The implementation of this simple picture in a realistic particle framework results, however, quite involving, due to the non trivial determination of the abundance 
of the decaying Bino, as well as due to the impact of wash-out processes and of additional sources both for the baryon asymmetry and the DM relic density. 
In order to achieve a quantitative determination of the baryon and Dark Matter abundances, we have implemented and solved a system of coupled Boltzmann 
equations for the particle species involved in their generation, including all the relevant processes. 
In the most simple, but still general, limit, in which the processes determining the abundance and the decay rate of the Bino are mediated by degenerate right-handed 
squarks, the correct values of the DM and baryon relic densities are achieved for a Bino mass between 50 and 100 TeV, Gluino NLSP mass in the range 15-60 TeV 
 and a gravitino  mass between 100 GeV and few TeV. These high masses are unfortunately beyond the kinematical reach of LHC. 
 On the contrary, an antiproton signal from the decays of the gravitino LSP might be within the sensibility of AMS-02 and gamma-ray telescopes.

\end{abstract}
\pacs{98.80.Cq}
\preprint{LPT-Orsay-15-43}

\maketitle

\section{Introduction}

\noindent
The origin of the Dark Matter (DM) component of the Universe and of the baryon asymmetry are two compelling puzzles of modern particle physics and cosmology. Conventionally, different and unrelated mechanisms are considered for the generation of these two quantities. Indeed, the generation of the correct baryon asymmetry is a rather difficult task to achieve and requires definite conditions~\cite{Sakharov:1988pm} to occur, mainly consisting in efficient $B$ and $CP$ violating processes occurring outside from thermal equilibrium. On the contrary the correct DM relic density can be generated by a very broad variety of mechanisms, also compatible with thermal relic DM particles, like the popular WIMP paradigm.

\noindent
A common generation mechanism for the DM and the baryon asymmetry is nonetheless a very intriguing possibility, also motivated by the similarity of the values of the two relic densities, being indeed $\Omega_{\Delta B}/\Omega_{\rm DM} \sim 0.2$.

\noindent
The most simple way to connect the baryon and the DM abundances is to assume also for the DM a generation through asymmetry. In the simplest realization of asymmetric DM models (see e.g.~\cite{Zurek:2013wia}) the ratio $\Omega_{\Delta B}/\Omega_{\rm DM}$ simply corresponds to the ratio between the mass of the proton and the mass of the DM.

\noindent
A viable alternative is however represented by the possibility of linking the generation of the baryon asymmetry to the popular WIMP mechanism. In this kind of scenarios the baryon asymmetry is produced after the chemical decoupling of a thermal relic, through $B$ and $CP$ violating annihilations~\cite{McDonald:2011sv,Baldes:2014gca,Baldes:2014rda,Baldes:2015lka} or decays~\cite{Cui:2012jh,Cheung:2013hza,Cui:2013bta,RompineveSorbello:2013xwa,Davoudiasl:2015jja}, or even of the DM itself~\cite{Cui:2011ab,Bernal:2012gv}.

\noindent
In a similar spirit, a rather simple mechanism allowing to achieve the contemporary production of the DM and of the baryon asymmetry~has been proposed in~\cite{Arcadi:2013jza}. Here these two quantities are contemporary generated by the out-of-equilibrium decay of a mother particle. The ratio between the DM and baryon density is expressed in terms of two analogously suppressed quantities, namely the CP-asymmetry and the branching ratio of decay of the mother particle into DM, and can be accommodated to be of the correct value, irrespectively of the initial abundance of the decaying particle, through a suitable choice of the parameters of the underlying particle theory.

\noindent
Although simple and elegant, this idea may be rather difficult to be implemented in concrete particle physics frameworks. For example, although the ratio between the baryon and DM abundances is independent from the one of the mother particle, this is not the case for the individual expectations of these two quantities. In order to match their rather precise experimental determinations~\cite{Planck:2015xua}, a similarly precise determination of the abundance of the mother particle is required. This is in general a not trivial task since it is determined by many different processes. In particular additional states of the underlying particle theory might play a relevant role through coannihilation effects. Furthermore, in presence of extra new particles, with respect to the mother particle and the DM, additional sources of baryon asymmetry and DM can be present, spoiling the simple picture discussed above. Analogously crucial is finally the determination of the impact of possible wash-out processes, namely the processes capable of depleting a possibly generated baryon asymmetry.

\noindent
In order to properly deal with these issues, a detailed numerical treatment, relying on suitable Boltzmann equations, is mandatory.

\noindent     
In this work we will investigate a definite case of study, being a R-parity violating realization of the MSSM with gravitino DM. SUSY models are a rather good playground for the scenario under consideration. Indeed, in absence of R-parity the SUSY superpotential features automatically sources of Baryon and Lepton number violation. At the same time, thanks to its Planck suppressed interactions, the gravitino DM remains stable on cosmological scales even in absence of symmetries forbidding its decay. In addition the decay branching fractions of supersymmetric particles into gravitino are as well Planck suppressed, thus not preventing an efficient generation of the baryon asymmetry from the out-of-equilibrium decay of a supersymmetric state.

\noindent
The mother particle is instead a Bino-like neutralino which generates the baryon asymmetry and the DM by late-time decays occurring after its chemical freeze-out. These two quantities are substantially determined, with mild assumptions on the cosmological history, by the underlying particle physics framework, in particular by the structure of the Supersymmetric spectrum. The requirement of the correct baryon and DM abundances can be translated into predictions on the particle content of the theory. These predictions can be tested by collider experiments if the particles involved in their generation are within their production reach.

\noindent
We have determined the baryon and DM relic densities, $\Omega_{\Delta B}$ and $\Omega_{\rm DM}$ through a system of coupled Boltzmann equations tracing the time evolutions of all the particle species involved in the generation of these two quantities. These are the Bino and the other two gauginos, the Gluino and the Wino, while scalar superpartners and the Higgsinos should be set, as clarified below, to very high scales such that they do not directly enter the system of equations as particle species, but only as mediators of the interactions of the gauginos. The system finally includes two additional equations, respectively for the baryon and the DM abundances. This kind of system has been solved as function of the relevant supersymmetric parameters. In order to provide a more clear understanding we will complement, where possible, our numerical treatment with analytical expressions for the relevant quantities. We will envisage, in particular, the dependence of the rates of the relevant processes, as well as the $CP$-asymmetry, on the flavor structure of the theory.

\noindent
The paper is organized as follows. After a brief review of the general idea of the contemporary generation of the DM and baryon densities from decay of a thermal relic, we will present in section III its MSSM realization. We will then present in section IV some analytical estimates of the relevant quantities. Section V will be instead dedicated to the numerical treatment and the quantitative determination of the parameters space compatible with the experimental expectation of the baryon asymmetry and of the DM relic density. Before stating our conclusions, we will finally briefly mention in section V the possible detection prospects relative to our setup.

\section{Dark Matter and Baryon production from out-of-equilibrium decay}

\noindent
A simple and elegant way to achieve the contemporary production of the baryon asymmetry and of the Dark Matter is, as proposed in~\cite{Arcadi:2013jza}, by out-of-equilibrium decay of a state $X$, featuring $B$ and $CP$ violating interactions, and thus capable, according the Sakharov conditions, of generating a baryon asymmetry. The resulting baryon density can be schematically expressed as:  
\begin{equation}
\label{eq:baryo_general}
\Omega_{\Delta B} = \xi_{\Delta B} \epsilon_{\rm CP} \frac{m_p}{m_{X}}\; BR(X \rightarrow b, \bar b)\;  \Omega_{X}
\end{equation}
where $m_p$ is the mass of the proton, $\epsilon_{\rm CP}$ is the CP asymmetry:
\begin{equation}
\epsilon_{\rm CP} = 
\frac{\Gamma (X \rightarrow b) - \Gamma (X \rightarrow \bar b)}{\Gamma (X \rightarrow b) +\Gamma (X \rightarrow \bar b)}\; .
\end{equation}
and $\Omega_{X}$ is the initial abundance of the state $X$. The factor $\xi_{\Delta B}$ encodes the effects of the sphaleron processes, as well possible wash-out and entropy dilution effects. 
The field $X$ features as well an additional decay channel, not necessarily B-violating, into DM such that its relic density is given by an analogous expression as above:
\begin{equation}
\label{eq:dm_general}
\Omega_{DM} = \xi_{DM} \frac{m_{DM}}{m_{X}} BR\left(X\rightarrow DM+\mbox{anything} \right) \Omega_{X}
\end{equation}
Expressions~(\ref{eq:baryo_general}) and~(\ref{eq:dm_general}) both feature suppression factors. For example the baryon density is suppressed by the ratio $m_p/m_X$. In addition, in most realistic particle frameworks, the CP asymmetry $\epsilon_{\rm CP}$ is a suppressed quantity. At the same time it is reasonable to expect that the branching ratio of decay of the $X$ particle into DM is as well suppressed in order to do not dangerously affect the baryon production. As a consequence in order to account for the experimental expectations of $\Omega_{\rm DM}$ and $\Omega_{\Delta B}$ a rather high value of the initial $\Omega_X$ is needed. In addition, under the assumption, performed in this work, that the initial abundance of $X$ is generated similarly to the WIMP mechanism, we need to require for it a sufficiently long lifetime such that it decays after chemical freeze-out.

\noindent
Interestingly, the ratio of the two densities is independent from the one of the $X$ state, being:
\begin{equation}
\label{eq:ratio_fit}
\frac{\Omega_{\Delta B}}{\Omega_{DM}}\!\! =\!\! \xi\,\epsilon_{\rm CP} \frac{m_p}{m_{DM}} 
\frac{BR(X \rightarrow b, \bar b)}{BR\left(X \rightarrow DM +\mbox{anything} \right)}\; ,\,\,\,\,\xi=\frac{\xi_{\Delta B}}{\xi_{\rm DM}}
\end{equation}
and its expected value $\sim 0.2$ is achieved by a suitable choice of the DM mass and of the parameters determining the $CP$ asymmetry and the two branching ratios. In this work we will embed this mechanism in a supersymmetric framework with gravitino DM while the decaying state $X$ is represented by a Bino-like neutralino.

\section{MSSM realization}

\noindent
In our investigation of the possibility of contemporary production of the baryon asymmetry and of gravitino DM we will focus on a MSSM realization with R-parity broken only by the operator $\lambda^{''} U^c D^c D^c$ which provides the breaking of the baryon number avoiding at the same time strong constraints from the stability of the proton~\footnote{The proton can actually decay into a gravitino, if kinematically possible, even in presence of only the $\lambda^{''}$ coupling~\cite{Barbier:2004ez}. As will be seen at the end of the paper the favored region of the parameter space will feature a gravitino much heavier than the proton, such that this kind of decay is forbidden.}, being lepton number violating operators absent. 
\noindent
The mother particle is chosen to be a Bino. As will be shown below, the rates of the processes governing its abundance and lifetime are set by the mass scales of the scalar superpartners and of the Higgsinos~\footnote{We can anticipate the most of the relevant processes occur before the Electroweak Phase Transition (EW) temperature. As a consequence the spectrum of the electroweakly interacting fermionic superpartners consists of two Majorana fermions, the Bino and the Wino, and two Dirac fermions, the Higgsinos.}. The requirement of overabundance and long lifetime of the Bino can be met for Supersymmetric spectra like the ones proposed in~\cite{Arvanitaki:2012ps,ArkaniHamed:2012gw} featuring a strong mass hierarchy between the three gauginos, namely the Bino, the Gluino and the Wino, and the scalar superpartners as well as the Higgsinos.
\noindent
The Bino is not, however, the Next-to-Light Supersymmetric particle (NLSP). The CP-asymmetry is created from the interference between tree-level and one loop processes involving a quark, a squark and another gaugino~\cite{Cui:2013bta}. According the Nanopolous-Weinberg theorem~\cite{Nanopoulos:1979gx,Kolb:1979qa,Claudson:1983js,RompineveSorbello:2013xwa}, this asymmetry is not null (at this order in perturbation theory) only if at least one between the squark and the gaugino running in the loop is lighter than the Bino. The case of a squark NLSP is however not feasible in a MSSM setup since a light-squark would enhance the annihilation and decay rate of the Bino making it not enough abundant and long-lived. A viable scenario can be obtained, for example, by extending the MSSM with an additional singlet playing the role of mother particle~\cite{Cui:2012jh,RompineveSorbello:2013xwa}. 

\noindent
In the scenario considered the lightest gaugino is the Gluino. It is possible to realize, alternatively, a leptogenesis scenario by considering the operator $\lambda^{'} QLD^c$, rather than the $B$-violating one, and considering the Wino as lightest particle, apart the gravitino DM.
\noindent
The Gluino and the Wino are instead not good candidates for the generation of the baryon asymmetry since they feature very efficient annihilation processes into gauge bosons, with rates depending on their same masses, which make their abundances too suppressed to generate sizable amounts of baryons and DM.

\noindent
The processes responsible for the generation of the baryon asymmetry can be described by the following effective lagrangian~\cite{Baltz:1997gd}, where, in agreement with the discussion above, the scalars and the Higgsinos have been integrated out (for simplicity we are omitting the mass terms and, from now on, indicate by $\lambda$, rather than, as conventional, $\lambda^{''}$, the RPV coupling.):
\begin{align}
\label{eq:asymmetry_lagrangian}
& \mathcal{L}=-2 \epsilon^{\alpha \beta \gamma}\left\{\lambda_{lij}\left[\overline{\tilde{B}}\left(G^{RL}_{u_{l,k}}P_L+G^{RR}_{u_{l,k}}P_R\right)u_{k\alpha}\overline{d}_{i\beta}^c P_R d_{j\gamma}+h.c.\right] \right. \nonumber \\
&\left. + \lambda_{klj}\left[  \overline{\tilde{B}}\left(G^{RL}_{d_{l,i}}P_L+G^{RR}_{d_{l,i}}P_R\right)d_{i\alpha}\overline{u}_{k\beta}^c P_R d_{j\gamma}+h.c.\right] \right. \nonumber\\
& \left. + \lambda_{kil}\left[ \overline{\tilde{B}}\left(G^{RL}_{d_{l,j}}P_L+G^{RR}_{d_{l,j}}P_R\right)d_{j\alpha}\overline{u}_{k\beta}^c P_R d_{i\gamma}+h.c. \right] \right \}\nonumber\\
& -2 \epsilon^{\alpha \beta \gamma}\left\{\lambda_{lij}\left[\overline{\tilde{G}}\left(G^{'\,RL}_{u_{l,k}}P_L+G^{'\,RR}_{u_{l,k}}P_R\right)u_{k\alpha}\overline{d}_{i\beta}^c P_R d_{j\gamma}+h.c.\right] \right. \nonumber \\
&\left. + \lambda_{klj}\left[  \overline{\tilde{G}}\left(G^{'\,RL}_{d_{l,i}}P_L+G^{'\,RR}_{d_{l,i}}P_R\right)d_{i\alpha}\overline{u}_{k\beta}^c P_R d_{j\gamma}+h.c.\right] \right. \nonumber\\
& \left. + \lambda_{kil}\left[ \overline{\tilde{G}}\left(G^{'\,RL}_{d_{l,j}}P_L+G^{'\,RR}_{d_{l,j}}P_R\right)d_{j\alpha}\overline{u}_{k\beta}^c P_R d_{i\gamma}+h.c. \right] \right \}\nonumber\\
& -\frac{1}{m_{\tilde{q}_{\alpha}}^2}\overline{\tilde{B}}\left(g^{LL}_{\tilde{B}}\Gamma^{U}_{L\alpha k}P_L+g^{RR}_{\tilde{B}}\Gamma^{U}_{R\alpha k}P_R\right)u_{k\alpha}\overline{u_{p\alpha}}\left(g^{LL\,*}_{\tilde{G}}\Gamma^{U\,*}_{L\alpha p}P_R+g^{RR\,*}_{\tilde{G}}\Gamma^{U\,*}_{R\alpha p}P_L\right)\tilde{G}+h.c.\nonumber\\
&  -\frac{1}{m_{\tilde{q}_{\alpha}}^2}\overline{\tilde{B}}\left(g^{LL}_{\tilde{B}}\Gamma^{D}_{L\alpha k}P_L+g^{RR}_{\tilde{B}}\Gamma^{D}_{R\alpha k}P_R\right)d_{i\alpha}\overline{d_{j\alpha}}\left(g^{LL\,*}_{\tilde{G}}\Gamma^{D\,*}_{L\alpha p}P_R+g^{RR\,*}_{\tilde{G}}\Gamma^{D\,*}_{R\alpha p}P_L\right)\tilde{G}+h.c.\nonumber\\
& - \frac{1}{2}\frac{g_1}{\mu}\cos\beta \sin\beta\overline{\tilde{B}}\tilde{B}HH^{*}
\end{align}
where:
\begin{align}
& G^{RL}_{f_{l,i}}=\Gamma^{F\,*}_{R\,l \alpha}\frac{1}{m_{\tilde{q}_\alpha}^2}\Gamma^{F}_{L\,\alpha i}g_{\tilde{B}}^{LL},\,\,\,\,\,\,G^{RR}_{f_{l,i}}=-\Gamma^{F\,*}_{R\,l \alpha}\frac{1}{m_{\tilde{q}_\alpha}^2}\Gamma^{F}_{R\,\alpha i}g_{\tilde{B}}^{RR} \nonumber \\
& G^{RL\,'}_{f_{l,i}}=\Gamma^{F\,*}_{R\,l \alpha}\frac{1}{m_{\tilde{q}_\alpha}^2}\Gamma^{F}_{L\,\alpha i}g_{\tilde{G}}^{LL},\,\,\,\,\,\,G^{RR}_{f_{l,i}}=-\Gamma^{F\,*}_{R\,l \alpha}\frac{1}{m_{\tilde{q}_\alpha}^2}\Gamma^{F}_{R\,\alpha i}g_{\tilde{G}}^{RR}
\end{align}
where $\Gamma^F_{R,L\,\alpha i}$ are $6 \times 3$ matrices defined by:
\begin{equation}
\tilde{q}_{\alpha}=\Gamma_{L \alpha i} \tilde{q}_{Li}+\Gamma_{R \alpha i} \tilde{q}_{Ri}
\end{equation}
where $\tilde{q}_{\alpha},\,\,\alpha=1,\cdots,6$ are the squark mass eigenstates while $\tilde{q}_{L,R\,i},\,\,i=1,\cdots,3$ are left-handed and right-handed squarks. The quantities $g_{\tilde{B},\tilde{G}}^{LL(RR)}$ are instead defined as:
\begin{align}
& g^{\rm LL}_{\tilde{B}}=-\sqrt{2} g_1 \left(Q_{f}-T_3\right)e^{i \phi_{\tilde{B}}}\,\,\,\,g^{\rm RR}_{\tilde{B}}=\sqrt{2} g_1Q_{f}e^{i \phi_{\tilde{B}}}\nonumber\\
& g^{\rm LL}_{\tilde{G}}=-\sqrt{2} g_3 e^{i \phi_{\tilde{G}}}\,\,\,\,g^{\rm RR}_{\tilde{G}}=\sqrt{2} g_3 e^{i \phi_{\tilde{G}}}
\end{align}
where $\phi_{\tilde{B}}$ and $\phi_{\tilde{G}}$ are $CP$-violating phases (see below for more details).

\noindent
The lagrangian (\ref{eq:asymmetry_lagrangian}) gives rise to CP violating decays of the gauginos into three SM fermions, e.g. $\tilde{B} \rightarrow udd$, as well as CP violating $2 \rightarrow 2$ scatterings. The baryon asymmetry arises in the decay (and annihilations) of the Bino, through the interference of tree-level diagrams, generated by the interactions proportional to the B-violating couplings $\lambda$, with loop level diagrams obtained by combining these operators with the effective $\tilde{B}-\tilde{G}$ interactions reported in the last lines of (\ref{eq:asymmetry_lagrangian}). Inverse decays and $2 \rightarrow 2$ scatterings involving all the gauginos, in particular the Gluino, represent the main wash-out processes which guarantee that no baryon asymmetry is created in thermal equilibrium. Moreover, B-violating single annihilation processes $\tilde{B}d(u) \rightarrow d(u)d$ play a prominent role in determining the abundance of the Bino, together with the Bino-Gluino coannihilations. The very last line of (\ref{eq:asymmetry_lagrangian}) is not associated to any process involved in the generation of the baryon asymmetry but triggers the pair annihilation $\tilde{B}\tilde{B} \rightarrow HH^{*}$ which is also relevant for determining the abundance of the Bino and, consequently, the one of DM and baryons. As evident the effective coupling depends on the combination $\sin\beta \cos\beta$. In the analytical expressions provided below we will implicitly assume the limit $\tan\beta \rightarrow 1$, in order to guarantee the correct EWSB and avoid tensions with the determination of the Higgs mass~\cite{Arvanitaki:2012ps,ArkaniHamed:2012gw,Bagnaschi:2014rsa}, given the high scalar mass scale~(See however the recent analysis~\cite{Vega:2015fna}).
   
\noindent
The expressions of the relevant interactions rates are in general very complicated, in particular because of a non-trivial interplay of the flavor structure, which is substantially free due to the very high scale of the scalar masses. However, as clarified in the next section, we can present our results, without loss of generality, in a simplified limit in which effects of flavor violation are neglected and the relevant interactions are mediated by only down-type right-handed squarks. This choice allows for simpler computations, since the number of possible processes is reduced; at the same time it guarantees the presence of all the possible topologies of diagrams responsible of the generation of the baryon asymmetry. For analogous reasons we have not reported the interactions of the sleptons which are assumed to be decoupled.
\noindent
In general one should consider analogous operators as the ones reported in eq. (\ref{eq:asymmetry_lagrangian}) also for the Wino. These however, as further discussed later on, do not significantly contribute to the generation of the baryon asymmetry and thus have been omitted for simplicity. 

\noindent
The DM candidate in this scenario is the gravitino. Although not exactly stable in an R-parity violating scenario, the Planck suppression of its interactions guarantees a lifetime largely exceeding the one of the Universe, even for $O(1)$ values of the RPV couplings~\cite{Buchmuller:2007ui,Bajc:2010qj}. The Bino (as well as the other superpartners) features a decay channel into the gravitino LSP with a Planck scale suppressed branching fraction. The generation of the baryon asymmetry is then insensitive to this decay channel which can nonetheless produce a sizable amount of Dark Matter since the suppressed branching ratio can be compensated by the overabundance of the decaying mother particle. The out-of-equilibrium decay of the other gauginos does not, instead, efficiently produce DM in view of their suppressed relic abundance. The Wino and the Gluino, as well as the Bino itself, can anyway copiously produce DM, by freeze-in, at early epochs while they are still in thermal equilibrium.

\noindent
In order to have a proper description of the generation of the baryon asymmetry, as well as the DM relic density one must rely on a solution of system of coupled Boltzmann equations. Before illustrating it, we will anyway provide some analytical approximations in order to provide a better understanding of the results, in particular the implications for the Supersymmetric spectrum.

\section{Analytical results}

\noindent
We will present in the following analytical expressions for both the baryon and the DM relic densities. These expressions are strictly valid only in definite regions of the parameter space while a systematic investigation requires the solution of a system of coupled Boltzmann equations, like the one presented in the next section. For greater clearness we will discuss separately, in the next two subsections, the generation of the baryon and of the DM abundances.

\subsection{Generation of baryon asymmetry}

\begin{figure}[htb]
\begin{center}
\subfloat[Diagram T1]{\includegraphics[width=4.5 cm]{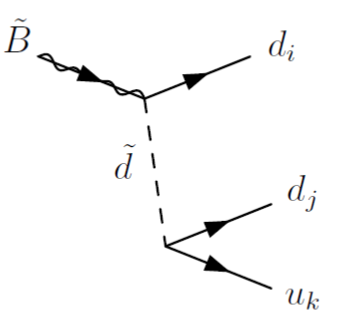}}
\hspace{2cm}
\subfloat[Diagram T2]{\includegraphics[width=4.5 cm]{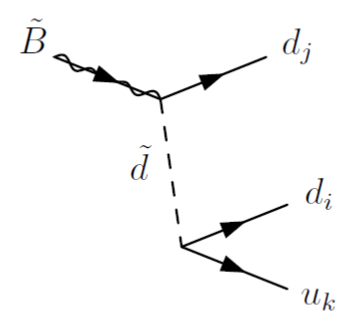}}
\caption{\footnotesize{Diagrams contributing, at the tree level, to the $B$ violating decay of the Bino in the case of mediation from only d-squarks.}}
\label{fig:treediagrams}
\end{center}
\end{figure}

\begin{figure}[htb]
\begin{center}
\subfloat[Diagram L1]{\includegraphics[width=4.5 cm]{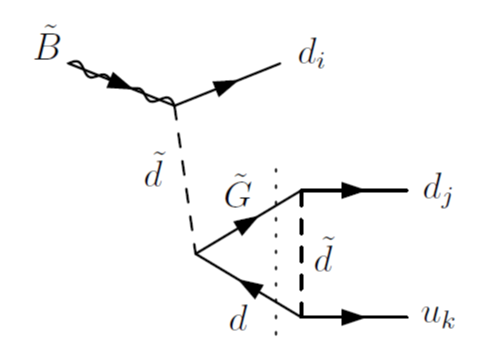}}
\hspace{2cm}
\subfloat[Diagram L2]{\includegraphics[width=4.5 cm]{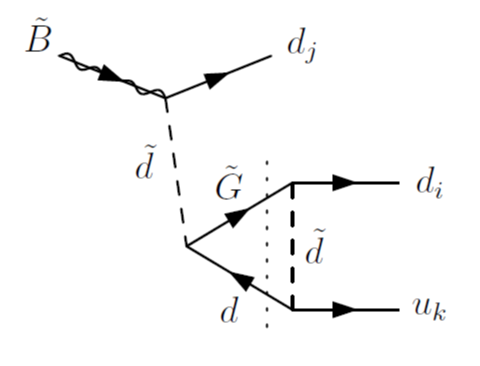}}\\
\subfloat[Diagram L3]{\includegraphics[width=4.5 cm]{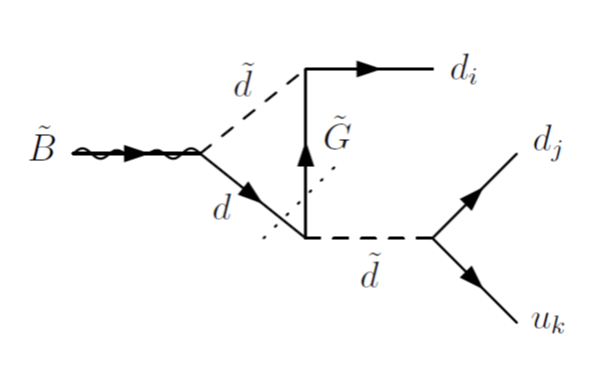}}
\hspace{2cm}
\subfloat[Diagram L4]{\includegraphics[width=4.5 cm]{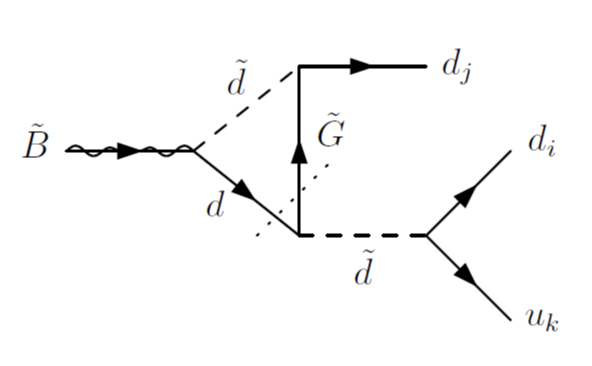}}
\caption{\footnotesize{Diagrams contributing, at the loop level, to the $B$ violating decay of the Bino in the case of mediation from only d-squarks. The CP-asymmetry is generated by the interference with tree-level diagrams reported in fig.(\ref{fig:treediagrams}).}}
\label{fig:loopdiagrams}
\end{center}
\end{figure}

\noindent
As already mentioned, there are actually two sources for the baryon asymmetry, namely the $B$ and $CP$ violating decays of the Bino as well as $2 \rightarrow 2$ scattering processes. In the first case the asymmetry is originated by the three-body decays, $\tilde{B} \rightarrow udd \left(\bar u \bar d \bar d \right)$. In the case that these processes are mediated by only right-handed d-squarks (see discussion below) the relevant tree-level and one-loop diagrams are shown, respectively, in fig.~(\ref{fig:treediagrams}) and~(\ref{fig:loopdiagrams}) (see also~\cite{Claudson:1983js,Cui:2013bta}).

\noindent
A CP asymmetry is generated as well by annihilation processes like, e.g., $\tilde{B}d \rightarrow \bar u \bar d+ \mbox{CP conjugate}$. The relevant diagrams are obtained, by crossing symmetry, from the ones shown in figs.~(\ref{fig:treediagrams})-(\ref{fig:loopdiagrams}).

\noindent
Assuming distinct timescales for the $B$-violating scatterings and decays, the total CP asymmetry $\epsilon_{CP}$ can be expressed as:
\begin{equation}
\epsilon_{CP}=\frac{\Delta \Gamma_{\rm dec}}{\Gamma_{\rm tot,\rm dec}}+\frac{\Delta \Gamma_{\rm ann}}{\Gamma_{\rm tot,\rm ann}}
\end{equation}
where $\Delta \Gamma_{\rm ann}$ and $\Delta \Gamma_{\rm dec}$ are, respectively, the differences between the rates $B$ violating scattering and decay processes and their $CP$ conjugates. The corresponding expressions are:
\begin{align}
\label{eq:decay_asymmetry}
& \Delta \Gamma_{\rm dec} =\sum_{\alpha \beta \gamma} \sum_{l,p,n} \frac{m_{\tilde{B}}^7}{m_{\tilde{q}_\alpha}^2 m_{\tilde{q}_\beta}^2 m_{\tilde{q}_\gamma}^2} \left[ \left(A_1 Im\left[g^{RR \,*}_{\tilde{B}}g^{RR}_{\tilde{B}}g^{RR \,*}_{\tilde{G}}g^{RR}_{\tilde{G}}\Gamma^{D\,*}_{R \alpha i}\Gamma^{D}_{R\alpha n}\Gamma^{D}_{R\gamma p}\Gamma^{D\,*}_{R\gamma j}\Gamma^{D}_{R\beta i}\Gamma^{D\,*}_{R \beta l} \lambda^{*}_{knj} \lambda_{kpl}\right] \right. \right. \nonumber\\
&\left. \left. + A_2 Im\left[g^{RR \,*}_{\tilde{B}}g^{RR}_{\tilde{B}}g^{RR \,*}_{\tilde{G}}g^{RR}_{\tilde{G}}\Gamma^{D\,*}_{R \alpha j}\Gamma^{D}_{R\alpha n}\Gamma^{D}_{R\gamma p}\Gamma^{D\,*}_{R\gamma j}\Gamma^{D}_{R\beta i}\Gamma^{D\,*}_{R \beta l} \lambda^{*}_{kni} \lambda_{kpl}\right]+ (i \leftrightarrow j)\right)f_1\left(\frac{m_{\tilde{G}}^2}{m_{\tilde{B}}^2}\right) \right. \nonumber\\
& \left. +\frac{m_{\tilde{G}}}{m_{\tilde{B}}}\left( B_1 Im\left[g^{RR\,*}_{\tilde{B}}g^{RR\,*}_{\tilde{B}}g^{RR}_{\tilde{G}}g^{RR}_{\tilde{G}}\Gamma^{D\,*}_{R \alpha i}\Gamma^{D}_{R\alpha n}\Gamma^{D\,*}_{R\gamma p}\Gamma^{D}_{R\gamma l}\Gamma^{D}_{R\beta i}\Gamma^{D\,*}_{R \beta l} \lambda^{*}_{knj} \lambda_{kpj}\right] \right. \right. \nonumber\\
& \left. \left. +\frac{m_{\tilde{G}}}{m_{\tilde{B}}} B_2 Im\left[g^{RR\,*}_{\tilde{B}}g^{RR\,*}_{\tilde{B}}g^{RR}_{\tilde{G}}g^{RR}_{\tilde{G}}\Gamma^{D\,*}_{R \alpha j}\Gamma^{D}_{R\alpha n}\Gamma^{D\,*}_{R\gamma p}\Gamma^{D}_{R\gamma l}\Gamma^{D}_{R\beta j}\Gamma^{D\,*}_{R \beta l} \lambda^{*}_{kni} \lambda_{kpj}\right]\right)f_2\left(\frac{m_{\tilde{G}}^2}{m_{\tilde{B}}^2}\right) \right],
\end{align}      
where:
\begin{equation}
f_1(x)=\left(1-x\right)^5,\,\,\,\,\,\,f_2(x)=1-8x+8x^3-x^4-12 x^2 \log(x)
\end{equation}

\begin{align}
\label{eq:decay_annihilation}
& \Delta \Gamma_{\rm ann}=\langle \sigma v \rangle \Delta n_{\tilde{B}},\,\,\,\,\,\,\Delta n_{\tilde{B}}=n_{\tilde{B}}-n_{\tilde{B},\rm eq}\\
& \langle \sigma v \rangle \Delta n_{\tilde{B}}=\sum_{\alpha \beta \gamma} \sum_{l,p,n} \frac{m_{\tilde{B}}^4}{m_{\tilde{q}_\alpha}^2 m_{\tilde{q}_\beta}^2 m_{\tilde{q}_\gamma}^2} \left[\left(C_1 Im\left[g^{RR \,*}_{\tilde{B}}g^{RR}_{\tilde{B}}g^{RR \,*}_{\tilde{G}}g^{RR}_{\tilde{G}}\Gamma^{D\,*}_{R \alpha n}\Gamma^{D}_{R\alpha i}\Gamma^{D}_{R\gamma p}\Gamma^{D\,*}_{R\gamma i}\Gamma^{D}_{R\beta l}\Gamma^{D\,*}_{R \beta j} \lambda^{*}_{knj} \lambda_{klp}\right]\right. \right. \nonumber\\
&\left. \left. + C_2 Im\left[g^{RR \,*}_{\tilde{B}}g^{RR}_{\tilde{B}}g^{RR \,*}_{\tilde{G}}g^{RR}_{\tilde{G}}\Gamma^{D\,*}_{R \alpha j}\Gamma^{D}_{R\alpha n}\Gamma^{D}_{R\gamma p}\Gamma^{D\,*}_{R\gamma i}\Gamma^{D}_{R\beta j}\Gamma^{D\,*}_{R \beta l} \lambda^{*}_{kni} \lambda_{klp}\right] \right) I\Delta \Sigma_1 \left(\frac{m_{\tilde{B}}}{T},\frac{m_{\tilde{G}}}{T}\right) \right. \nonumber\\
&+\left .\frac{m_{\tilde{G}}}{m_{\tilde{B}}}\left( D_1 Im\left[g^{RR\,*}_{\tilde{G}}g^{RR\,*}_{\tilde{G}}g^{RR}_{\tilde{B}}g^{RR}_{\tilde{B}}\Gamma^{D}_{R \alpha i}\Gamma^{D\,*}_{R\alpha l}\Gamma^{D}_{R\gamma n}\Gamma^{D\,*}_{R\gamma p}\Gamma^{D\,*}_{R\beta i}\Gamma^{D}_{R \beta p} \lambda^{*}_{knj} \lambda_{kpj}\right] \right. \right. \nonumber\\
& \left. \left. +\frac{m_{\tilde{G}}}{m_{\tilde{B}}} D_2 Im\left[g^{RR\,*}_{\tilde{B}}g^{RR\,*}_{\tilde{B}}g^{RR}_{\tilde{G}}g^{RR}_{\tilde{G}}\Gamma^{D\,*}_{R \alpha j}\Gamma^{D}_{R\alpha n}\Gamma^{D\,*}_{R\gamma p}\Gamma^{D}_{R\gamma l}\Gamma^{D}_{R\beta j}\Gamma^{D\,*}_{R \beta l} \lambda^{*}_{kni} \lambda_{klj}\right)I\Delta \Sigma_2 \left(\frac{m_{\tilde{B}}}{T},\frac{m_{\tilde{G}}}{T}\right)\right] \right]
\end{align}

\noindent
where:

\begin{align}
& I\Delta \Sigma_1 (x,y)=\frac{1}{x^4 K_2(x)}\int_x^{\infty} z^4\left(z^2-x^2\right)^2 {\left(1-\frac{y^2 x^2}{z^2}\right)}^2 K_1(z) \nonumber\\
& I\Delta \Sigma_2 (x,y)=\frac{1}{x^4 K_2(x)}\int_x^{\infty} z^2\left(z^2-x^2\right)^2 {\left(1-\frac{y^2 x^2}{z^2}\right)}^2 K_1(z)
\end{align}  
 
\noindent
while $A_{1,2},B_{1,2},C_{1,2},D_{1,2}$ are numerical coefficients which can be determined from the expressions provided in the appendices.

\noindent
A non-null $CP$ asymmetry originates from a non trivial combination of the phases coming from from the (Majorana) gaugino masses, encoded in the couplings $g_{\tilde{B}}$ and $g_{\tilde{G}}$, the squark mixing matrix and, possibly, the RPV couplings $\lambda$. We notice in particular that the terms proportional to the coefficients $A_{1,2}$ and $C_{1,2}$ are different from zero only in presence flavor violation since the combinations between the gauge couplings are automatically real and the phases in the RPV couplings would as well cancel in this limit. In absence of flavor violation the $CP$ violation arises from the differences of phases contained in the Maiorana masses of the Bino and the Gluino, which behave as effective $B$-violating terms~\cite{Claudson:1983js,Cui:2012jh}~\footnote{We are encoding the CP-phases in the vertices gaugino-quark-squark while the Majorana masses of the gauginos are assumed to be real. This configuration can be obtained through a suitable rotation of the superfields.}. In this case however the CP asymmetry is suppressed by the ratio $m_{\tilde{G}}/m_{\tilde{B}}$, as well as by the kinematical factor $f_2$. 
\noindent
As already mentioned we are considering a regime in which only right-handed d-type squarks contribute to the processes of interest. As clarified in the appendix additional contributions are originated by similar diagrams in which up-type right-handed squarks are exchanged. However the eventual increase of the CP asymmetry does not necessarily imply an increase of the baryon abundance. Indeed there is a tight relation between the processes governing the generation of CP asymmetry with the ones governing the abundance of the Bino as well as wash-out processes. In general an increase of the CP asymmetry is connected with an enhancement of the depletion rates of the Bino and of the baryon asymmetry itself and one has then to find a balance between the two effects. This provides a further indication that any analytical treatment should be complemented by the numerical solution of suitable Boltzmann equations. 

\noindent
The suppression $m_{\tilde{G}}/m_{\tilde{B}}$ can be also avoided in presence of mixing between left and right-handed squarks, which would make to arise analogous terms as the first in eq.~(\ref{eq:decay_asymmetry}-\ref{eq:decay_annihilation}). We remind however that the size of left-right mixing depends on the ratio $X_f/m_0^2$ where:
\begin{align}
& X_f=m_f \left(A_f - \mu q_\beta\right) \nonumber\\
& q_\beta=\left \{
\begin{array}{c}
\tan\beta \,\,\,\,\mbox{for d-type squarks}\\
\cot\beta \,\,\,\,\mbox{for u-type squarks}
\end{array}
\right.
\end{align}
where $m_f$ is the mass of the SM fermionic partner of the squark. This mixing is thus heavily suppressed as the ratio $m_f/m_{{\tilde{q}}_\alpha}$ with the only possible exception of the top squark, where the ratio $m_t/m_{{\tilde{q}}_\alpha}$ might be balanced by taking $\mu/m_{{\tilde{q}}_\alpha} \gg 1$ (The $A_f$ terms can differ at most by a $O(1)$ factor from $m_{\tilde{q}}$ in order to avoid breaking of the color.). As clarified below, an efficient production of the baryon asymmetry requires $m_{{\tilde{q}}_\alpha}> 10^{6}\,\mbox{GeV}$. For such values we can achieve values of $X_t/m_{{\tilde{q}}_\alpha}^2 \sim 10^{-(2 \div 1)}$ which do not induce sensitive variations of the total $CP$ asymmetry with respect to the simplified regime we are considering.   

\noindent
Even in presence of flavor violation the contribution from the coefficients $A_{1,2}$ and $C_{1,2}$ is limited since the combination between the flavor matrices is GIM suppressed and its imaginary part is zero in the limit of degenerate squarks. Even for non-degenerate squarks it is possible to achieve at most $O(1)$ variations of the CP-asymmetry with respect to the flavor universal scenario. The remark again that the impact of this variation is not trivial to identify at the analytical level because of the non trivial interplay with the wash-out processes and the ones responsible of the abundance of the Bino. We will thus postpone further discussion of this point to the section dedicated to the numerical analysis.

\noindent
We finally notice that the functions $f_1$ and $f_2$ in~(\ref{eq:decay_asymmetry}) as well as $I\Delta \Sigma_1$ and $I\Delta \Sigma_2$ in~(\ref{eq:decay_annihilation}) make the asymmetry zero, consistently with Nanopolous-Weinberg theorem, if $m_{\tilde{G}} \geq m_{\tilde{B}}$.  

\noindent
In agreement with the discussion above, without loss of generality, we will present our results in the limit of absence of flavor violation and degenerate squark masses $m_{\tilde{q}_{\alpha}}=m_0$. In this limit the expressions above simplify to:
\begin{align}
& \Delta \Gamma_{\rm dec}=\frac{\alpha_1 \alpha_s}{432 \pi^2}\sum_{kij} \lambda_{kij}^2 \frac{m_{\tilde{B}}^6 m_{\tilde{G}}}{m_0^6}f_2\left(\frac{m_{\tilde{G}}^2}{m_{\tilde{B}}^2}\right)Im\left[e^{2 i \phi}\right] \nonumber \\
& \Delta \Gamma_{\rm ann}= \frac{\sum \lambda_{ijk}^2 \alpha_1 \alpha_s}{384} \frac{m_{\tilde{B}}^4}{m_0^6} \frac{m_{\tilde{G}}}{m_{\tilde{B}}}  I\Delta \Sigma_2\left(x,\frac{m_{\tilde{G}}}{m_{\tilde{B}}}\right) \Delta n_{\tilde{B}}Im\left[e^{2 i \phi}\right],\,\,\,\,\,x=\frac{m_{\tilde{B}}}{T}
\end{align} 
where $\phi=\phi_{\tilde{G}}-\phi_{\tilde{B}}$. As further simplification we will assume that all the couplings $\lambda_{kij}$ (apart the ones set to zero by the asymmetry of the $ij$ indices) are equal to the same value $\lambda$.

\noindent
A sizable asymmetry from $2 \rightarrow 2$ scatterings can be created only for freeze-out temperatures of the Bino very close to its mass. For lower temperatures, indeed, it results drastically reduced by the Boltzmann suppression in $\Delta n_{\tilde{B}}$ \cite{Claudson:1983js}. On the other hand, at high temperature, wash-out processes are still active and tend again to reduce the contribution to the asymmetry. As shown in \cite{Baldes:2014rda} the correct amount of baryon asymmetry from $2 \rightarrow 2$ scatterings can arise only from a very restricted range of values of the relevant parameters. On the contrary out-of-equilibrium decay can lead to a very efficient baryon production since it occurs at later time and, as clarified in the following, can evade wash-out effects if the Bino is enough long-lived. As also confirmed by our numerical investigation the decay of the Bino accounts for substantially the total amount of baryon density in all the viable regions of the parameter space. The baryon density reduces to (\ref{eq:baryo_general}):
\begin{equation}
\label{eq:omegaBsimple}
\Omega_{\Delta B}= \xi_{\Delta B} \frac{m_p}{m_{\tilde{B}}} \epsilon_{\rm CP} \Omega_{\tilde{B}}^{\tau \rightarrow \infty}
\end{equation}
The parameter $\xi_{\Delta B}$ can be decomposed as the product $\xi_{\Delta B}=\xi_{\rm sp} \xi_{\rm w.o.} \xi_{\rm s}$. $\xi_{\rm sp}$ represents the effects of the sphaleron processes and can be set to $28/79$ or $1$ depending on whether the Bino decays before or after the temperature of electroweak phase transition, set to $T_{\rm EW}=140\,\mbox{GeV}$. $\xi_{\rm w.o.}$ and $\xi_{\rm s}$ represent instead the possible reduction of the baryon abundance due to wash-out effects while $\xi_{\rm s}$ is related to possible entropy dilution effects. An analytical estimate of the latter is provided at the end of this subsection. We have instead no analytical estimation for $\xi_{\rm w.o.}$. We can nonetheless identify, as explained below, two limit regimes, namely the case $\xi_{\rm w.o.} \ll 1$, corresponding to negligible baryon abundance, and $\xi_{\rm w.o}=1$, for which a viable phenomenology is instead achievable.

\noindent
In agreement with what stated above, the CP asymmetry $\epsilon_{\rm CP}$ is given by:
\begin{equation}
\epsilon_{\rm CP}= \frac{\Delta \Gamma_{\rm dec}}{\Gamma_{\rm tot}}
\end{equation}
with:
\begin{equation}
\Gamma_{\rm tot}=\Gamma \left(\tilde{B} \rightarrow udd + \bar u \bar d \bar d\right)+\left(\tilde{B} \rightarrow \tilde{G} d\bar d\right)+\Gamma\left(\tilde{B} \rightarrow \tilde{\psi}_{3/2}+X\right)
\end{equation}
where $X$ represents all the possible SM final states accompanying the gravitino and:
\begin{equation}
\Gamma\left(\tilde{B} \rightarrow udd+\overline{u}\overline{d}\overline{d}\right)= \frac{\lambda^2 \alpha_1}{16 \pi^2} \frac{m_{\tilde{B}}^5}{m_0^4}
\end{equation}
\begin{equation}
\Gamma\left(\tilde{B} \rightarrow \tilde{G} f \overline{f} \right)= \frac{\alpha_1 \alpha_3}{192 \pi^2} \frac{m_{\tilde{B}}^5}{m_0^4}f_2\left(\frac{m_{\tilde{G}}^2}{m_{\tilde{B}}^2}\right)
\end{equation}
\begin{equation}
\label{eq:gamma_sugra}
\Gamma \left(\tilde{B} \rightarrow \tilde{\psi}_{3/2}+ X\right)= \frac{1}{48 \pi} \frac{m_{\tilde{B}}^5}{m_{3/2}^2 M_{\rm Pl}^2}
\end{equation}
are, respectively, the tree-level B-violating decay rate of the Bino in three SM fermions and of the B-conserving channel into the gluino and a pair of d-quarks, and finally
the decay rate into any final state with gravitino, responsible of DM production. $M_{\rm Pl}$ is the reduced Planck mass $M_{\rm Pl}=2.43 \times 10^{18}$GeV. 
The last decay channel does not affect the baryogenesis mechanism in view of its very suppressed branching ratio: 
\begin{equation}
Br\left(\tilde{B} \rightarrow \tilde{\psi}_{3/2}+X\right) \approx 5.7 \times 10^{-10} {\left(1+\frac{6 \lambda^2}{ \pi  \alpha_s}\right)}^{-1} {\left(\frac{m_{3/2}}{1 \mbox{GeV}}\right)}^{-2}{\left(\frac{m_0}{10^6 \mbox{GeV}}\right)}^4
\end{equation}
The CP asymmetry is then given by:
\begin{equation}
\label{eq:epsilon_CP}
\epsilon_{\rm CP}=\frac{8}{3} Im\left[e^{2 i \phi}\right] \frac{m_{\tilde{B}} m_{\tilde{G}}}{m_0^2} \alpha_s {\left(1+\frac{ \pi  \alpha_s}{6 \lambda^2} \right)}^{-1} f_2\left(\frac{m_{\tilde{G}}^2}{m_{\tilde{B}}^2}\right)
\end{equation}
\noindent
From now on we will take the value of the phase giving maximal $ \epsilon_{\rm CP} $ and assume that $Im\left[e^{2 i \phi}\right]=1$.
We notice that for $\lambda > \sqrt{\frac{\alpha_s \pi}{6}}$ the $CP$ asymmetry is substantially independent from the amount of R-parity violation. For lower values it instead decreases as $\lambda^2$. The CP asymmetry is suppressed by the ratio $m_{\tilde{B}}m_{\tilde{G}}/m_0^2$ as well as by the kinematic function $f_2$. In order to achieve 
the correct baryon abundance this suppression should be compensated by a sufficiently high initial abundance of the Bino, which is set by its annihilation processes. 

\noindent
These are described by thermally averaged cross-sections which can schematically be expressed as~\footnote{The extrema of integration in principle exceeds the energy 
scales for which the effective description~(\ref{eq:asymmetry_lagrangian}) is valid. However, as will explained in the next subsection, in order to have a cosmologically viable
 scenario, we need to assume a low reheating temperature such $T_{\rm R} < m_0$. As a consequence, all the rates will be the computed at temperatures such that neglecting 
the momentum dependence of the propagators of the squarks and the Higgsinos, does not produce sensitive variations in the results.}:
\begin{align}
\label{eq:general_integral}
& \langle \sigma v \rangle \left(\chi_i \chi_j \rightarrow \chi_l \chi_k\right)=\frac{1}{8 T m_i^2 m_j^2 K_2 \left(\frac{m_i}{T}\right) K_2 \left(\frac{m_j}{T}\right)}\int_{{\left(m_i+m_j\right)}^2}^{\infty} ds\; p_{ij} W_{ij} K_1\left(\frac{\sqrt{s}}{T}\right) \nonumber\\
& W_{ij}=\frac{p_{kl}}{64\pi^2 \sqrt{s}} \int d\Omega\; |M|^2 \nonumber\\
& p_{ij}=\frac{\sqrt{s-{\left(m_i-m_j\right)}^2}\sqrt{s-{\left(m_i+m_j\right)}^2}}{2\sqrt{s}}
\end{align}

\noindent
The possible annihilation processes include, first of all, conventional pair annihilations; in our scenario the dominant ones are into two Higgses or into two SM fermion 
final states. The corresponding cross-sections are: 
\begin{align}
& \langle \sigma v \rangle \left(\tilde{B}\tilde{B} \rightarrow HH^{*}\right)=\frac{\alpha_1^2 \pi}{32 \mu^2} A\left(\frac{m_{\tilde{B}}}{T}\right)\nonumber\\
& A(x)=\frac{1}{x^4 K_2(x)^2} \int_{2 x}^{\infty} dz z {\left(z^2-4 x^2\right)}^{3/2} K_1(z)\nonumber\\
& \langle \sigma v \rangle \left(\tilde{B}\tilde{B} \rightarrow q \overline{q}\right) =\frac{16 \pi}{27} \alpha_1^2 \frac{m_{\tilde{B}}^2}{m_0^4}\left[{\left(\frac{K_3(x)}{K_2(x)}\right)}^2-{\left(\frac{K_1(x)}{K_2(x)}\right)}^2\right]
\end{align}
We have then coannihilation~\cite{Edsjo:1997bg} processes with the two other gauginos:
\begin{align}
&\langle \sigma v \rangle \left(\tilde{B}\tilde{G} \rightarrow u \overline{u}\right)+\langle \sigma v \rangle \left(\tilde{B}\tilde{G} \rightarrow d \overline{d}\right) =\frac{16 \pi \alpha_1 \alpha_s}{27} \frac{m_{\tilde{B}}^2}{m_0^4} \left(2 \frac{K_4(x)}{K_2(x)}+1\right) \nonumber\\
& \left \langle \sigma v \right \rangle \left(\tilde{B}\tilde{W} \rightarrow HH^{*}\right)=\frac{\alpha_1 \alpha_2 \pi}{32 \mu^2} B\left(\frac{m_{\tilde{B}}}{T},\frac{m_{\tilde{W}}}{m_{\tilde{B}}}\right)\nonumber\\
& B(x,y)=\frac{1}{x^4 y^2 K_2(x) K_2(y x)}\int dz\; {\left(z^2-4 x^2 (1+y)^2\right)}^{3/2}{\left(z^2-4 x^2 (1-y)^2\right)}^{1/2}K_1(z)
\end{align}
We remark that a sizable contribution from coannihilations with at least one gaugino is unavoidable in our scenario since, in order to have a non-zero baryon asymmetry, the presence of  a lighter gaugino with respect to the Bino is mandatory. 
\noindent
Contrary to conventional WIMPs, we have to take into account also single annihilation processes, both RPC and RPV, with a SM fermion as second initial state. The relevant cross-sections are:
\begin{align}
& \langle \sigma v \rangle \left(\tilde{B}u \rightarrow \tilde{G} \overline{u}\right)+\langle \sigma v \rangle \left(\tilde{B}d \rightarrow \tilde{G} \overline{d}\right) =\frac{4 \pi \alpha_1 \alpha_s}{27}\frac{m_{\tilde{B}}^2}{m_0^4}\left(8 \frac{K_4(x)}{K_2(x)}+1\right)\nonumber\\
& \langle \sigma v \rangle \left(\tilde{B}u_k\rightarrow \overline{d}_i \overline{d}_j\right)+\langle \sigma v \rangle \left(\tilde{B}d_i \rightarrow \overline{u}_k d_j\right)=\frac{\alpha_1 \lambda^2}{3} \frac{m_{\tilde{B}}^2}{m_0^4}\left(5 \frac{K_4(x)}{K_2(x)}+1\right)
\end{align}

\begin{figure}[htb]
\begin{center}
\subfloat{\includegraphics[width=6.8 cm]{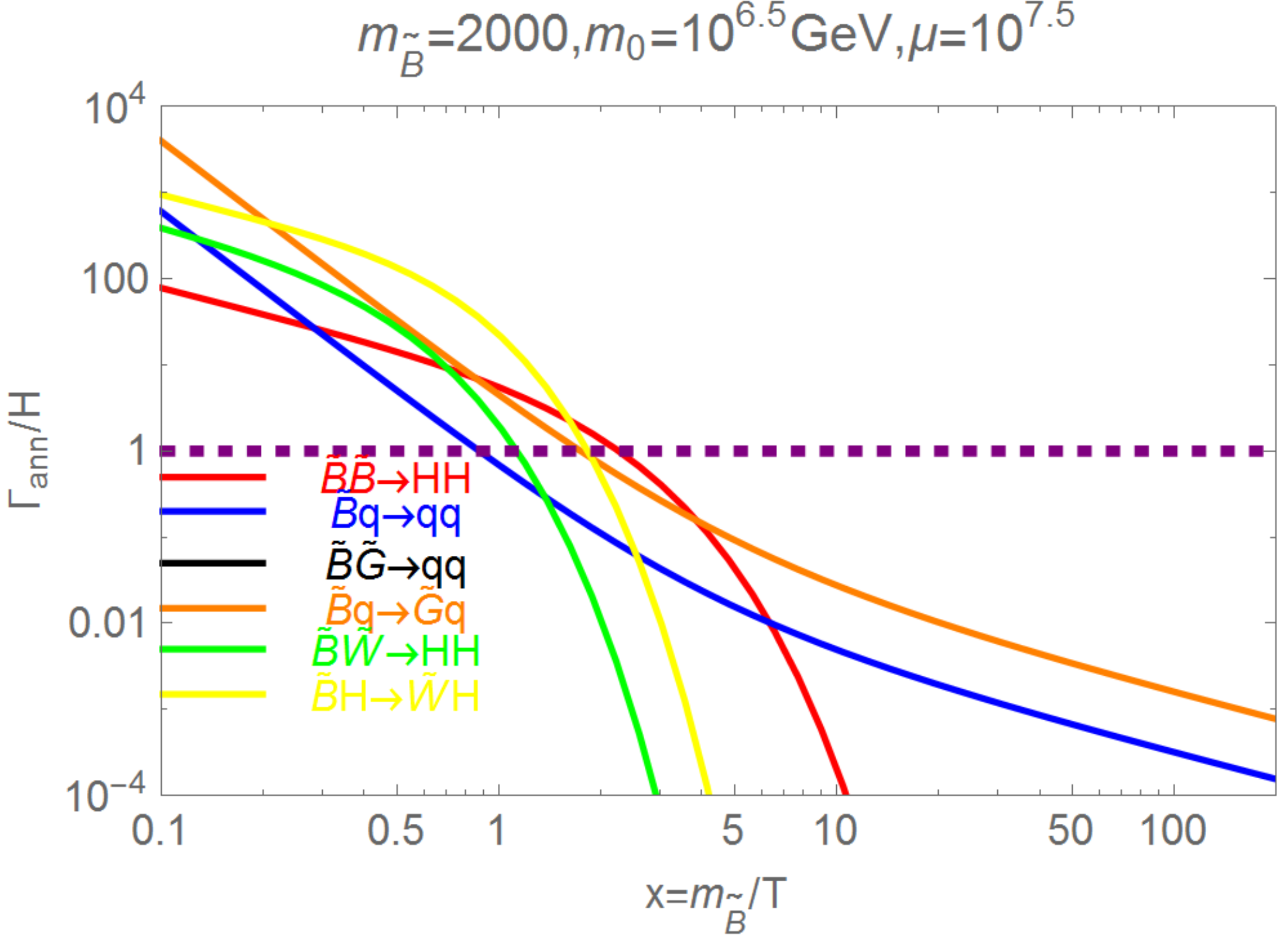}}
\subfloat{\includegraphics[width=6.8 cm]{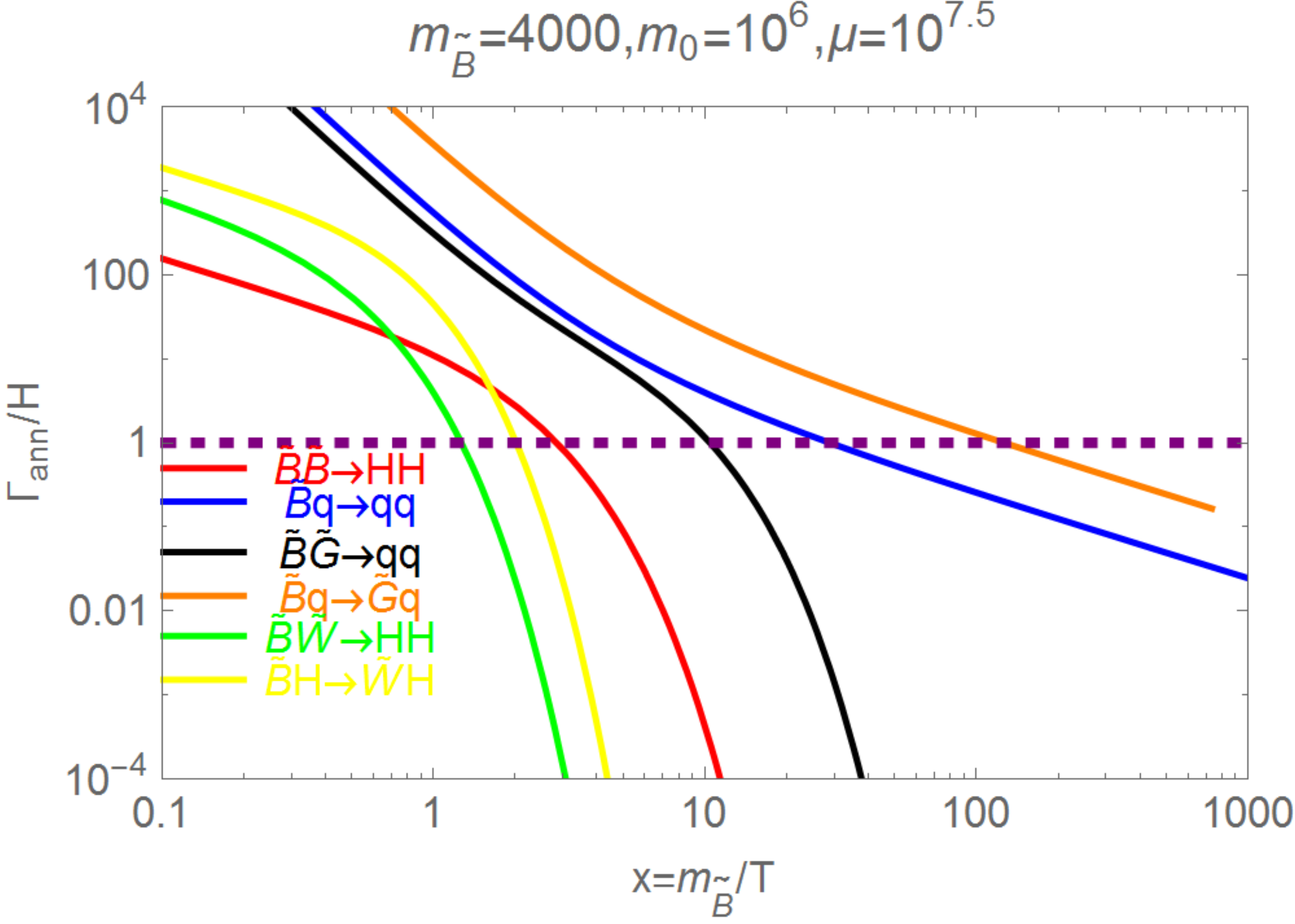}}\\
\subfloat{\includegraphics[width=6.8 cm]{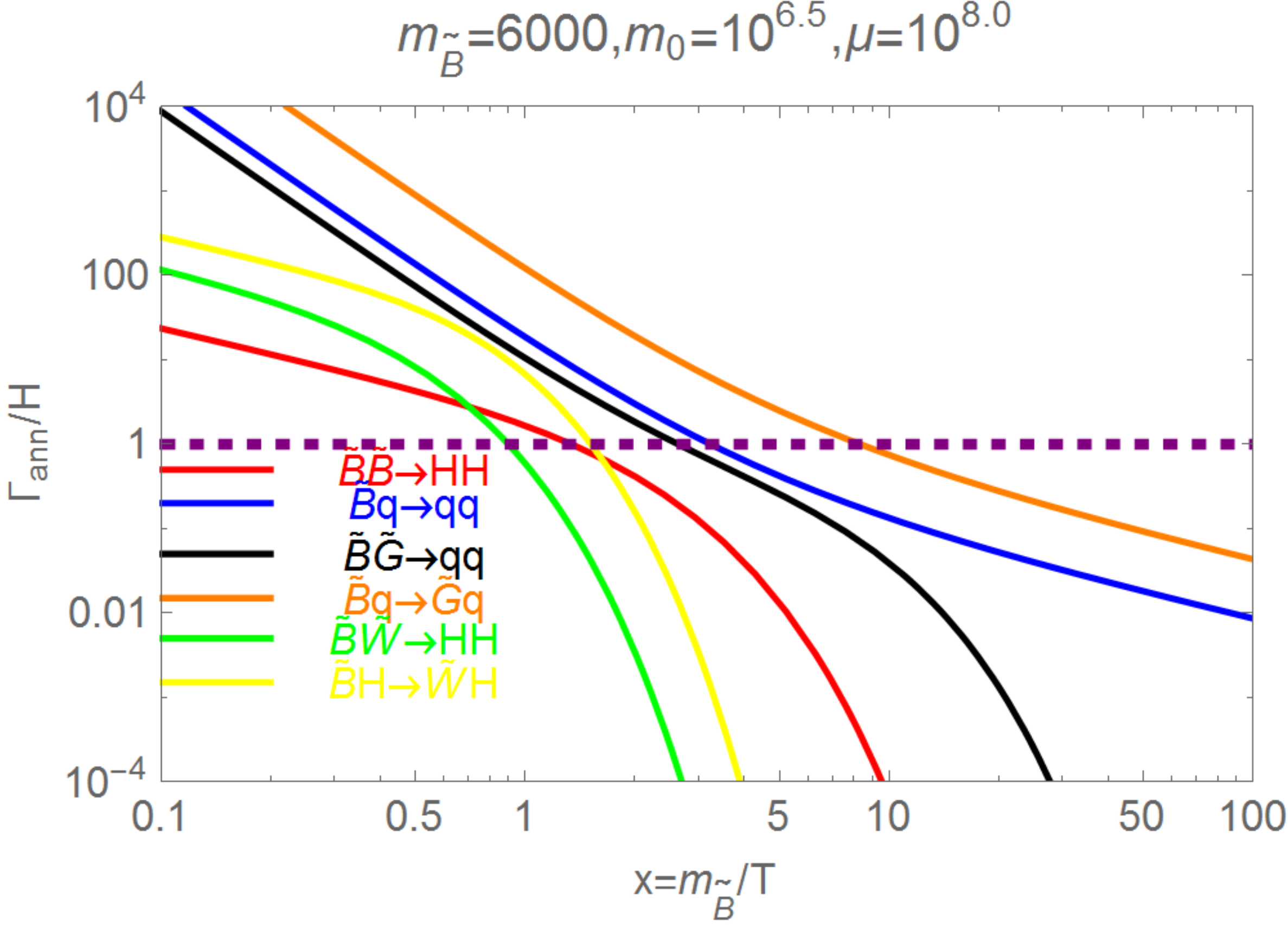}}
\subfloat{\includegraphics[width=6.8 cm]{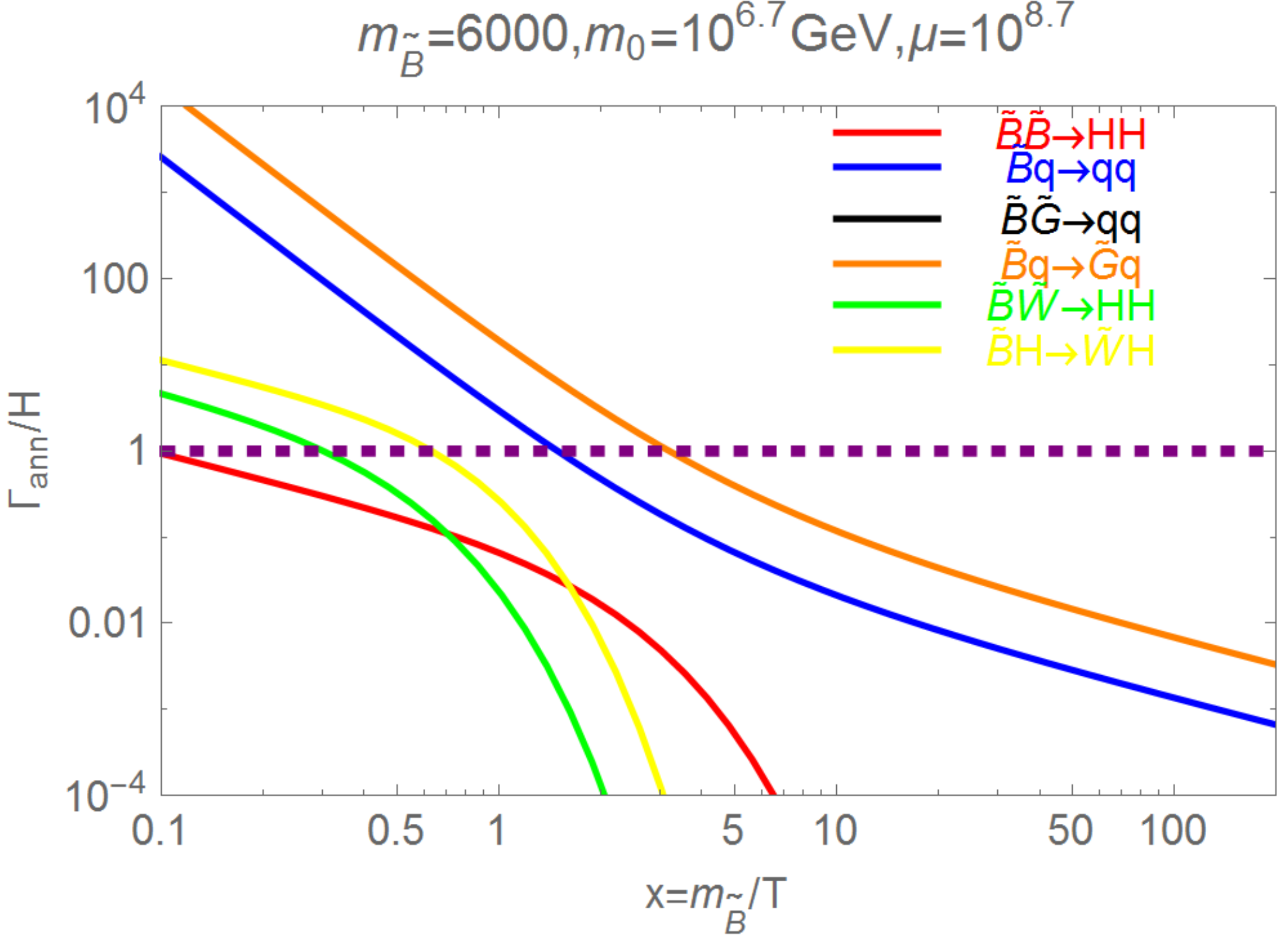}}
\caption{\footnotesize{Annihilation rates, normalized with the Hubble expansion factor $H$, for the channels reported in the plot, of the Bino, for four assignations of $(m_{\tilde{B}},m_0,\mu)$ reported in the panels and $m_{\tilde{G}}=0.35\; m_{\tilde{B}}, m_{\tilde{W}}=5\; m_{\tilde{B}}$, $\lambda=0.2$.}}
\label{fig:plotcross}
\end{center}
\end{figure}

\noindent
The relative contributions of the various annihilation channels, expressed in the form $\Gamma_{\rm ann}/H$ where $H$ is the Hubble expansion parameter and 
$\Gamma_{\rm ann}\equiv \langle \sigma v \rangle n_{X}^{\rm eq}$ where $X=\tilde{B}$ for pair annihilation processes, $X=\tilde{G},\tilde{W}$ for coannihilations, 
and $X=q$ for single annihilations, are shown in fig.~(\ref{fig:plotcross}). We have considered there four assignations of the set $(m_{\tilde{B}},m_0,\mu)$ while we have 
fixed the remaining parameters as $m_{\tilde{G}}=0.35 m_{\tilde{B}},\, m_{\tilde{W}}=5 m_{\tilde{B}},\, \lambda=0.2$. The pair annihilation cross-section, in particular the 
$HH^{*}$ channel, dominates for lower masses of the Bino and small hierarchy between $m_0$ and $\mu$ while, once increasing these quantities, single annihilation 
processes are the most important in determining the abundance and decoupling time of the Bino. Single annihilations are also in general dominant for low value of $m_0$. 
We notice from the second and third panel of fig.~(\ref{fig:plotcross}), with $m_0$ set, respectively, to $10^6$ and $10^{6.5}$ GeV, that single annihilation processes 
determine a very late (even more than conventional WIMP scenarios) chemical decoupling of the Bino, for which a very suppressed relic abundance is expected. 
The results shown in fig.~(\ref{fig:plotcross}) thus provide a first qualitative indication that very high values of the scale $m_0$ are required to generate a sizable baryon 
abundance.
\noindent
A quantitative determination of the abundance of the Bino necessarily relies on the numerical solution of Boltzmann equations, illustrated in the next section, in particular 
because single annihilation processes can induce deviations from the conventional WIMP scenarios. Indeed an analytical estimate of the baryon adundance is given by:
\begin{align}
\label{eq:bino_yield_analytical}
& Y_{\tilde{B}}(x_{\rm f})=M(x_{\rm f}){\left [\frac{M(x_{\rm i})}{Y_{\tilde{B}}(x_{\rm i})}+\frac{\langle \sigma v \rangle_{\rm p}}{\langle \sigma v \rangle_{\rm l}Y_{q,\rm eq}} \left(M(x_{\rm i})-M (x_f)\right)\right]}^{-1}\nonumber\\
& M(x)=\exp\left[\frac{a}{x} \langle \sigma v \rangle_{\rm l} Y_{q, \rm eq}\right]\nonumber\\
& a=\sqrt{\frac{\pi}{45}}m_{\tilde{B}}M_{\rm Pl}
\end{align}
where $\langle \sigma v \rangle_{\rm p}$ and $\langle \sigma v \rangle_{\rm l}$ represent, respectively, the sum of the thermally averaged pair (including coannihilations~\cite{Edsjo:1997bg}) and single annihilation cross-sections. $Y_{q,\rm eq} \equiv n_{q,\rm eq}/T$ represents the yield of the quarks (constant in the relativistic limit).~\footnote{In writing eq.~(\ref{eq:bino_yield_analytical}) we have neglected the time dependence of the annihilation cross-sections, in order to provide a simple expression. This is not fully motivated given the possibility, as shown below, of relativistic or semirelativistic decoupling of the Bino. A generalization of the expression is straightforwardly obtained by inserting the cross-sections in suitable integrals.} $x_{\rm i}$ and $x_{\rm f}$ represent, respectively, an initial time, which can determined through an analogous procedure 
as presented in~\cite{Gondolo:1990dk}, and a final time which can set to be the decay time scale of the Bino, as defined below.
\noindent
In the limit $a \langle \sigma v \rangle_{\rm l}/x \ll 1$ and for late enough decays (such that the first term in the parenthesis can be neglected) it is possible to recover the conventional WIMP behaviour, $Y\left(x_{\rm f}\right) \propto \frac{1}{\langle \sigma v \rangle_{\rm p}}$. The relic density, in this limit, is well approximated by the well known formula~\cite{Gondolo:1990dk}:
\begin{equation}
\label{eq:omega_db}
\Omega_{\tilde{B}}^{\tau \rightarrow \infty} \simeq \frac{3.9 \times 10^8 x_{\rm f.o.}\,{\mbox{GeV}}^{-1}}{g_{*}^{1/2} M_{\rm Pl} \left \langle \sigma v(x_{\rm f.o.}) \right \rangle_{\rm p}}
\end{equation}
where the effective thermally averaged pair annihilation cross-section is computed at $x_{\rm f.o} \equiv \frac{m_{\tilde{B}}}{T_{\rm f.o.}}$ with $T_{\rm f.o}$ being the freeze-out temperature. In the limit in which the dominant annihilation channel is the one into $HH^{*}$ the Bino abundance is given by the rather simple expression:
\begin{equation}
\label{eq:Omegab_only_pair}
\Omega_{\tilde{B}}^{\tau \rightarrow \infty} \approx 4.1 \times 10^{9} {\left(\frac{\mu}{10^8 \mbox{GeV}}\right)}^2 \frac{x_{\rm f.o.}}{A(x_{\rm f.o.})}
\end{equation}

\noindent
We can also expect that for high enough values of the scales $m_0$ and $\mu$, the consequent suppression of the annihilation cross-section leads to a relativistic decoupling of the Bino. In such a case its relic abundance would be even larger~\cite{Drees:2009bi}:
\begin{equation}
\Omega_{\tilde{B},\rm rel}^{\tau \rightarrow \infty}=7.8 \times 10^{10} \frac{g_{\tilde{B}}}{g_{*}(x_{\rm f.o.})} \left(\frac{m_{\tilde{B}}}{1\mbox{TeV}}\right)
\end{equation}
\noindent
By using eq.~(\ref{eq:Omegab_only_pair}) and~(\ref{eq:epsilon_CP}), and setting $\xi_{\rm sp}=\xi_{\rm w.o.}=\xi_{\rm s}=1$, we can write the baryon abundance as:
\begin{equation}
\label{eq:SWIMP_baryo}
\Omega_{\Delta B}h^2 \approx 3.3 \times 10^{-2}  \frac{x_{\rm f.o.}}{A(x_{\rm f.o.})} \left(\frac{m_{\tilde{B}}}{1 \mbox{TeV}}\right)\left(\frac{m_{\tilde{G}}}{m_{\tilde{B}}}\right)f_2\left(\frac{m_{\tilde{G}}^2}{m_{\tilde{B}}^2}\right) {\left(\frac{\mu}{10^{3/2} m_0}\right)}^2 \left(\frac{6\,\lambda^2}{\pi \alpha_s}\right) {\left(1+\frac{6\,\lambda^2}{\pi \alpha_s}\right)}^{-1}
\end{equation}
\noindent
In the limit considered the baryon density is not influenced by the absolute scale of $m_0$ but only by the ratio $\mu/m_0$ with the mass of the Bino $m_{\tilde{B}}$ being the only relevant scale. In particular, in order to achieve correct value of $\Omega_{\Delta B} h^2 \sim 0.02$~\cite{Planck:2015xua} a value $\mu/m_0 \gg 1$ appears favoured. We also notice that the factor $\left(\frac{m_{\tilde{G}}}{m_{\tilde{B}}}\right)f_2\left(\frac{m_{\tilde{G}}^2}{m_{\tilde{B}}^2}\right)$ suggests a suppression of the baryon abundance both for $m_{\tilde{G}} \ll m_{\tilde{B}}$ and $m_{\tilde{G}} \simeq m_{\tilde{B}}$. We have for it a maximal value $\sim 0.16$ for $m_{\tilde{G}}/m_{\tilde{B}} \sim 0.3$. 
We remind however, that eq.~(\ref{eq:SWIMP_baryo}), relies on assumptions valid only in a limited range of the parameter space. We will thus postpone a quantitative determination of $\Omega_{\Delta B}$, as function of the MSSM parameters, to the next section, once the detailed numerical treatment is considered.
%\noindent
Note that the expressions above are valid only in the limit in which it is possible to neglect the impact of wash-out processes and entropy dilution.

\begin{figure}[htb]
\begin{center}
\subfloat{\includegraphics[width=6.8 cm]{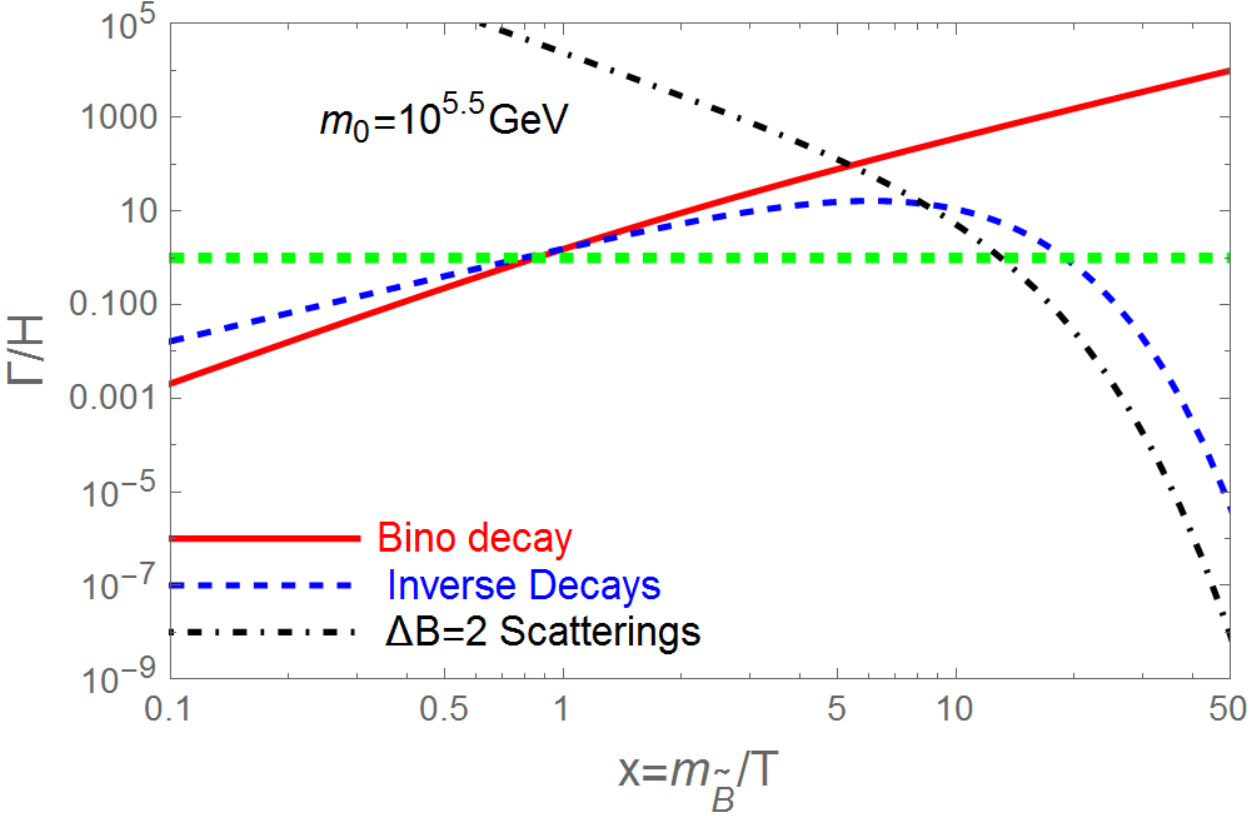}}
\subfloat{\includegraphics[width=6.8 cm]{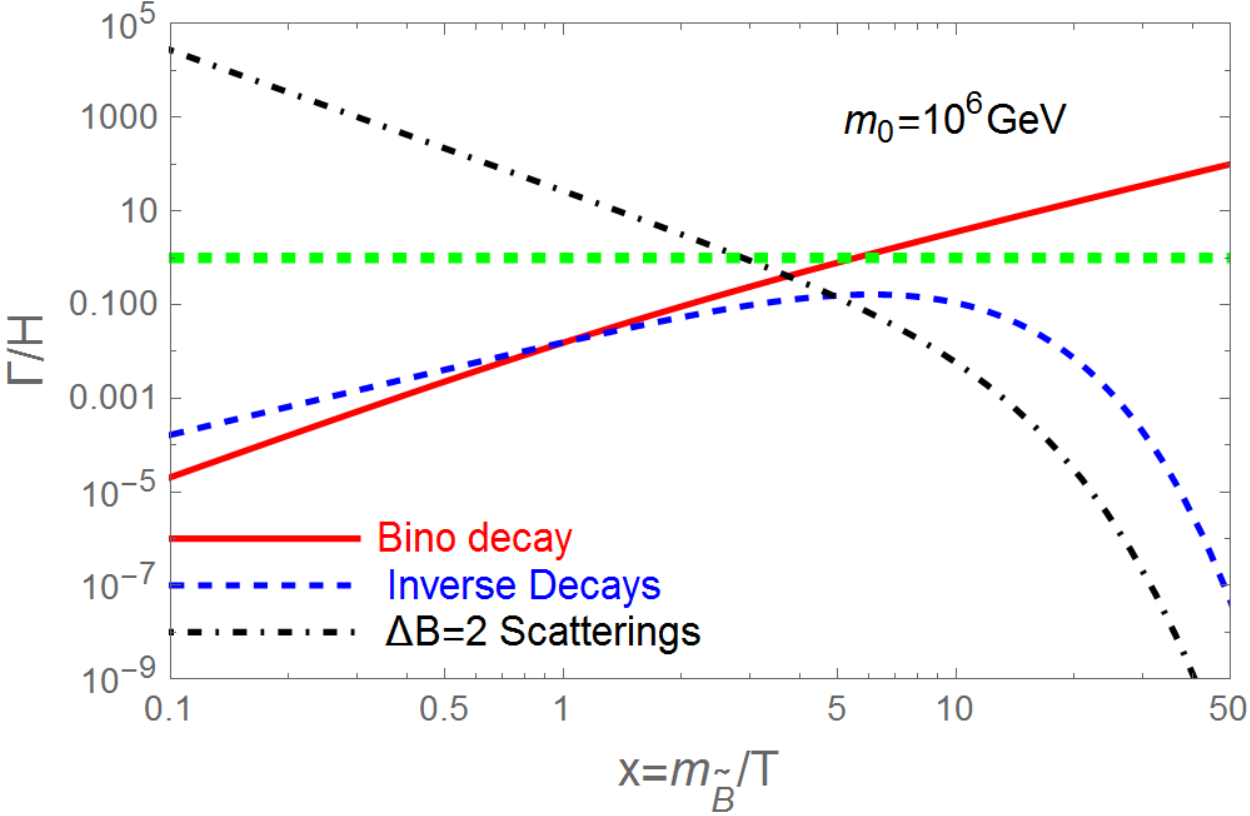}}
\caption{\footnotesize{Ratio of the total decay rate of the Bino (red solid curves) and of the two wash-out processes, namely inverse decays (blu dashed curves) and $2 \rightarrow 2$ processes (black dot-dashed curves), related to the gluino, over the Hubble expansion rate for the two values of $m_0$ reported in the plot and for $m_{\tilde{B}}=2\, \mbox{TeV}$. In the left plot the decay timescale of the Bino, namely $\Gamma \sim H$, is much lower than the one at which washout-processes become inactive. As a consequence the baryon asymmetry is expected to be at least partially erased. For the higher value of $m_0$ the rates of the wash-out processes are instead below $H$ and the generation of the baryon asymmetry is maximally efficient.}} 
\label{fig:washout}
\end{center}
\end{figure}

\noindent
Wash-out processes guarantee that no baryon asymmetry is created in thermal equilibrium and, if they are efficient up to rather late times, they can deplete partially, or even completely, the asymmetry created by the decay and the annihilations of the Bino. The main wash-out processes are inverse decays of three quarks into a Bino or a Gluino, as well as $2 \rightarrow 2$ scatterings of the type $u\tilde{B}(\tilde{G}) \leftrightarrow \bar d_i \bar d_j$, $d_i\tilde{B}(\tilde{G}) \leftrightarrow \bar u \bar d_j$ (and their CP conjugates). In addition one should also consider $3 \rightarrow 3$ scatterings of the type $udd \rightarrow \overline{u}\overline{d}\overline{d}$, mediated by two scalars and a off-shell gaugino, and, similarly, $2 \rightarrow 4$ scatterings~\cite{Cui:2012jh}. However these last two kinds of processes have very suppressed rates, as $m_0^{-8}$, and thus have been neglected in our analysis. A quantitative computation of the abundance of baryons including the effects of wash-out processes requires the solution of Boltzmann equations and will be discussed in detail in the next section. We can nonetheless distinguish two simple limit cases. As mentioned before the baryon asymmetry is mostly generated by the out-of-equilibrium decay of the Bino with a typical time scale determined by:
\begin{equation}
\Gamma_{\tilde{B},\rm tot}(x_{\rm d}) \approx H(x_{\rm d})
\end{equation}
By an analogous rule of thumb we can define the scale $x_{\rm w.o.}$ at which wash-out processes become inefficient. If $x_{\rm d} \ll x_{\rm w.o.}$, the baryon asymmetry is 
produced when the wash-out processes are very efficient and, as consequence, it is partially or totally depleted. In the opposite case the baryon production occurs, instead, 
when wash-out processes are not important anymore and, hence, the Bino abundance, weighted by the branching ratio of the $B$-violating processes, is totally converted in the 
baryon abundance.
\noindent
For kinematical reasons, as well as the presence of the strong coupling, the most important wash-out processes are the ones related to the gluino, with corresponding rates:
\begin{equation}
\label{eq:washoutID}
\Gamma_{\rm ID}=\frac{\lambda^2 \alpha_s}{\pi^2} z^7 \frac{m_{\tilde{B}}^5}{m_0^4} x^2 K_2(zx)
\end{equation}
where $z=\frac{m_{\tilde{G}}}{m_{\tilde{B}}}$,
\begin{equation}
\label{eq:washoutS}
\Gamma_{\rm S}= \frac{16 \alpha_s}{9 \pi^2} |\lambda|^2 z^4 \frac{m_{\tilde{B}}^5}{m_0^4} \frac{1}{x}\left[5 \frac{K_4(zx)}{K_2(zx)}+1\right]K_2(zx)
\end{equation}
describing, respectively, inverse decays $udd(\bar u \bar d \bar d) \rightarrow \tilde{G}$ and $2 \rightarrow 2$ scatterings, like e.g. $u d \rightarrow \bar d \tilde{G}$.
\noindent
These two rates, normalized with $H$, have been compared with the decay rate of the Bino in fig.~(\ref{fig:washout}). Here we have considered the following assignments of the parameters: 
$\lambda=0.1$, $m_{\tilde{B}}=2\,\mbox{TeV}$, $z=0.5$ and $m_0=10^{5.5}\,\mbox{GeV}$ (left plot) and $m_0=10^{6}\,\mbox{GeV}$ (right plot). In both cases $\mu$ has been kept fixed at $10^8\,\mbox{GeV}$.

\noindent
For the lowest value of $m_0$ the decay of the Bino occurs before the wash-out processes become ineffective and we thus expect that at least part of the generated baryon asymmetry is erased. As $m_0$ increases the rates of wash-out processes become more suppressed; the decay rate of the Bino is analogously suppressed such that its decay occurs at later times. As shown by the last panel of fig.~\ref{fig:washout} wash-out processes become negligible for $m_0=10^{6.0}\,\mbox{GeV}$. 

\noindent
We have therefore a further indication that the efficient production of the baryon abundance requires high values of $m_0$, at least $\gtrsim 10^6 \,\mbox{GeV}$. 
On the other hand we remind that the CP asymmetry $\epsilon$ is suppressed by $m_0^{-2}$ and as a consequence a too high $m_0$ would lead again to an 
insufficient amount of baryon asymmetry. We thus expect that the correct amount of the baryon asymmetry is achieved for a rather definite range of values of $m_0$.

\noindent
In addition to the wash-out processes, the produced baryon asymmetry can be as well reduced by entropy injection effects. Indeed, as already noticed 
in~\cite{Cui:2013bta,Baldes:2014rda}, an high enough abundance of the Bino can dominate the energy density of the Universe such that its decay is accompanied 
by a sizable entropy injection. We can thus define a dilution factor~\cite{Baldes:2014rda}:
\begin{equation}
\xi_{\rm s}=MAX\left[1,1.8 g_{*,s}^{1/4} \frac{Y_{\tilde{B}}(x_{\rm f.o.}) m_{\tilde{B}}}{\sqrt{\Gamma_{\tilde{B},\rm tot} M_{\rm pl}}}\right]
\end{equation}

\noindent
From the discussion above it is evident that the correct generation of the baryon asymmetry depends from two absolute scales, being the mass of the Bino $m_{\tilde{B}}$ 
and the scalar mass scale $m_0$. The other two scales, namely $\mu$ (entering only into the pair annihilation processes into two Higgs) and $m_{\tilde{G}}$ can be 
instead determined, as function of, respectively, $m_0$ and $m_{\tilde{B}}$, by requiring the generation of the maximal amount of asymmetry.

\begin{figure}[htb]
\begin{center}
\subfloat{\includegraphics[width=6.0 cm]{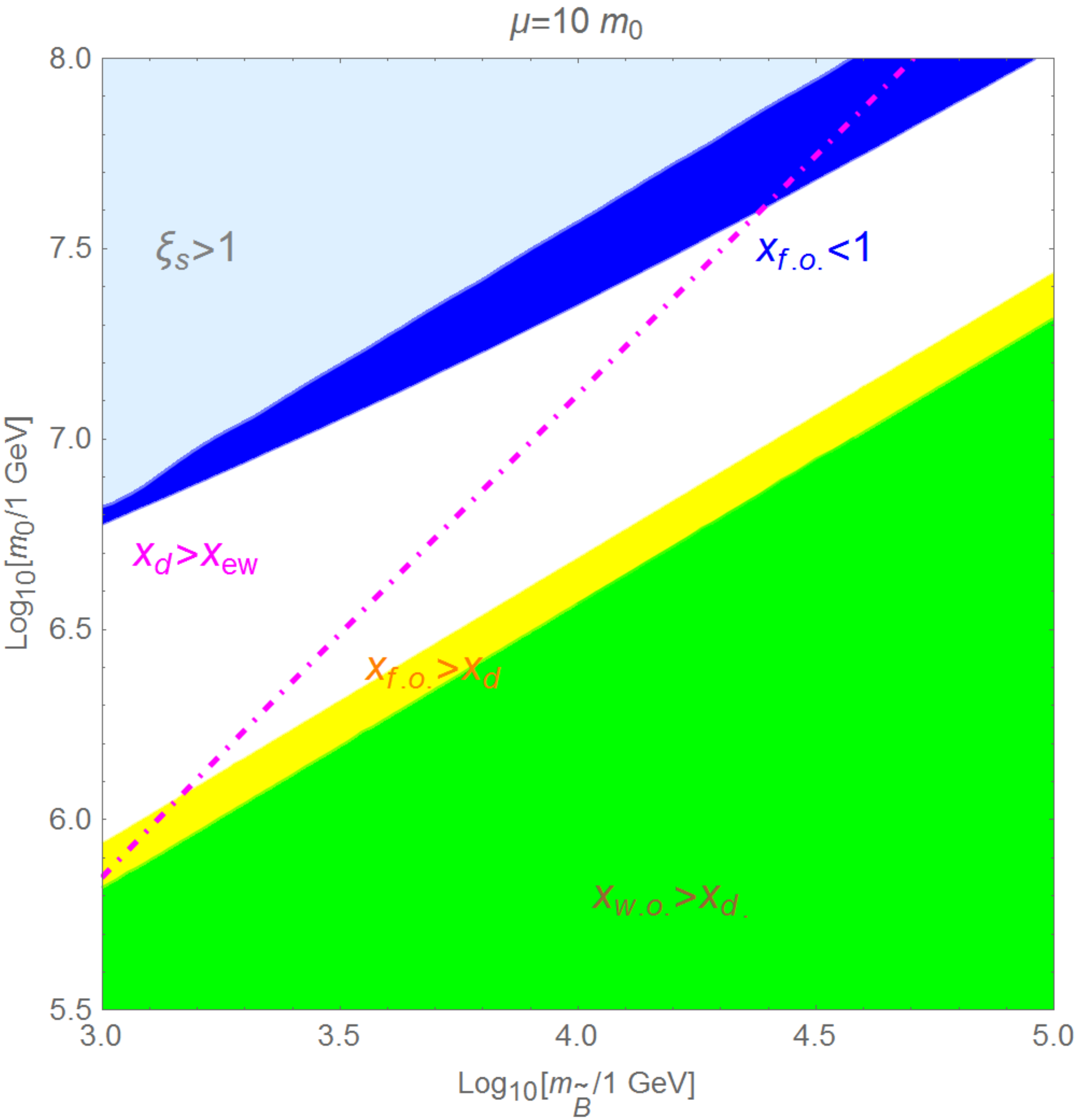}}
\subfloat{\includegraphics[width=6.0 cm]{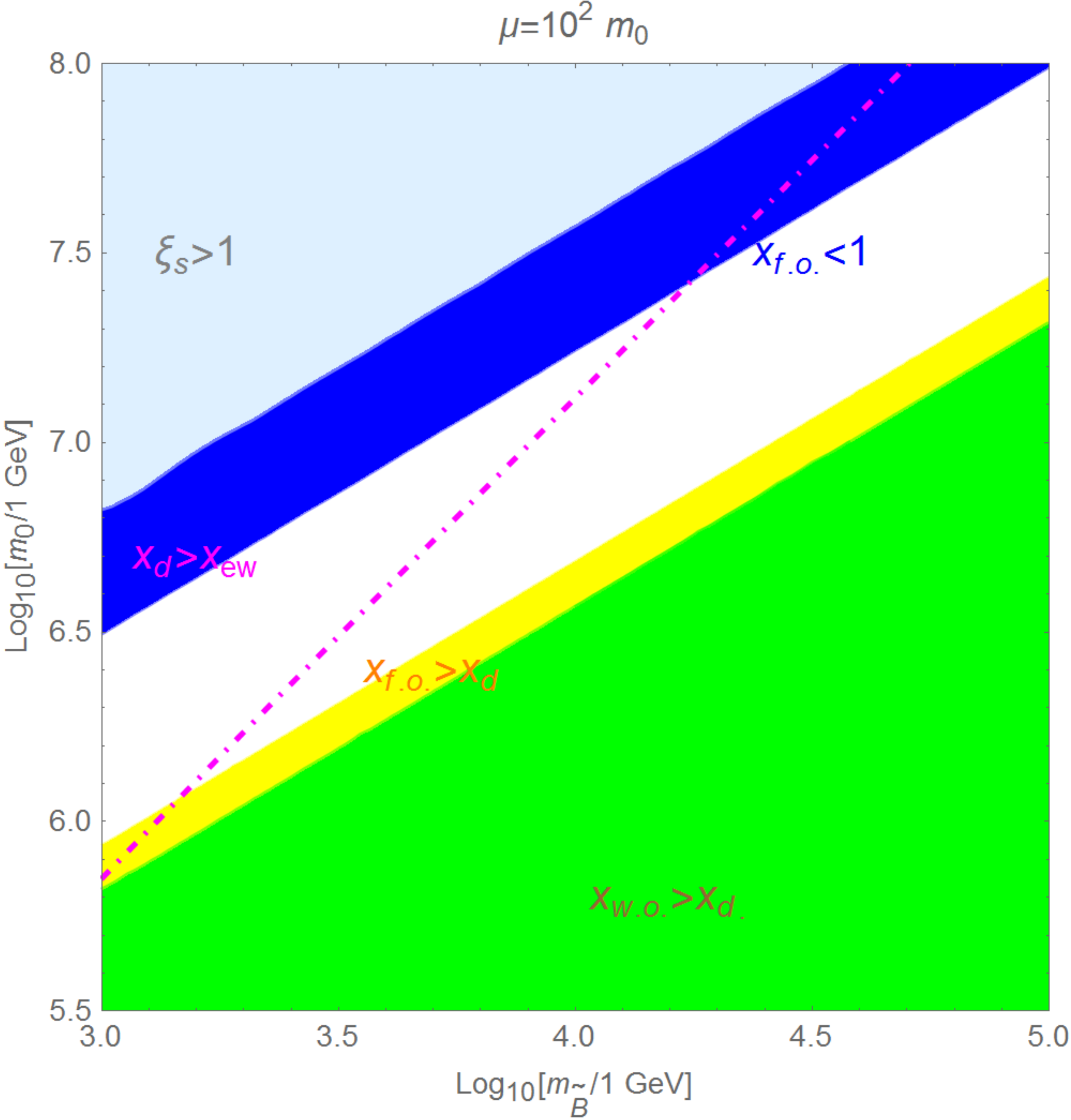}}
\caption{\footnotesize{Summary plots describing the regions of validity of the analytical estimates in the plane $(m_{\tilde{B}},m_0)$, for two values of the ratio $\mu/m_0$, 
namely 10 (left panel) and 100 (right panel). The mass of the Gluino has been set to $m_{\tilde{G}}=0.35\; m_{\tilde{B}}$ while $\lambda=0.3$. The Wino, finally, has been assumed to be very heavy and decoupled from the theory. The green region corresponds to typical decay times of the Bino smaller than the ones at which wash-out 
processes become ineffective. This region substantially overlaps with the yellow region corresponding to decay time, $x_d$, lower than the freeze-out time. In the blue 
region the Bino decouples while relativistic while the light blue region indicates sizable amounts of entropy injected at its decay. The production the baryon asymmetry 
is mostly efficient in the white strip outside the regions described above. The dot-dashed magenta line curves correspond to the case $x_d=x_{\rm ew}$ (see text for details). 
In the region below this curves the baryon asymmetry is produced before the EW phase transition and it thus depleted by a factor $\frac{28}{79}$.}}
\label{fig:ranges}
\end{center}
\end{figure}

\noindent
The impact on the parameters space of the effects, namely wash-out and entropy dilution, described above, as well as the range of validity of the analytical expressions are, 
qualitatively, described in fig.~(\ref{fig:ranges}). Here we show the  bidimensional plane $(m_{\tilde{B}},m_0)$ for two assignments of the parameter $\mu$, 
namely $\mu=10,100 m_0$, with $m_{\tilde{G}}=0.35 m_{\tilde{B}}$ and $\lambda=0.3$, while the mass of the Wino has been set to a much higher scale with respect 
to the other gauginos in order to decouple possible effects. As already argued wash-out processes are active at the lower values of $m_0$. In order to avoid these effects 
we need to require $m_0$ to be at least 2-3 orders of magnitude above the scale of the gauginos involved in the generation of the baryon asymmetry. The region of impact 
of wash-out processes (green region) substantially corresponds to the scenario in which the typical decay time of the Bino, set, by rule of thumb, by the condition 
$\Gamma_{\rm tot}=H$, is close to the one of freeze-out (yellow region). The production of the baryon asymmetry is instead very efficient for much later decay times. 
Entropy dilution effects (light-blue region) occur instead for very high values of $m_0$ (and, in turn, $\mu$) for which the decoupling of the Bino is relativistic (blue region) 
while result negligible for a production of the baryon asymmetry from out-of-equilibrium decay in the non-relativistic regime. We have finally inserted, for reference, the 
isolines $x_d=x_{\rm EW}\equiv\frac{m_{\tilde{B}}}{T_{\rm EW}}$. The regions at the right of the curves correspond to a production of the baryon asymmetry before the 
EW phase transition, with its consequent reduction due to sphaleron processes.

\noindent
From the discussion above it is thus evident that an optimal production of the baryon asymmetry corresponds to a rather definite range of values of $m_0$, 
$m_0 \sim 10^{6 \div 7}\,\mbox{GeV}$.

\subsection{Production mechanisms for the gravitino DM}

\noindent
In a supersymmetric scenario there are in general three production mechanisms for the gravitino.
There is first of all the contribution from thermal scatterings occurring at high temperatures in the Early Universe giving a contribution to the relic density sensitive to the 
gravitino and gaugino masses as well as to the reheating temperature after the inflationary phase~\cite{Bolz:2000fu,Pradler:2006qh,Rychkov:2007uq}. 
The contribution to the DM relic density is given by \cite{Pradler:2006qh,Olechowski:2009bd}:
\begin{equation}
\label{eq:omega_thermal}
\Omega_{\rm DM}^{\rm Th}h^2 =\left(\frac{m_{3/2}}{1\mbox{GeV}}\right)\left(\frac{T_{\rm R}}{10^{10} \mbox{GeV}}\right) \sum_{r=1,3} y^{'}_r g_r^2(T_{\rm R})\left(1+\delta_r\right)\left(1+\frac{M_r^2(T_{\rm R})}{3 m_{3/2}^2}\right) \log\left(\frac{k_r}{g_r(T_{\rm R})}\right)  
\end{equation}
where $y^{'}_r$, $k_r$ and $\delta_r$ are numerical coefficients defined in~\cite{Pradler:2006qh}.
In addition we have a contribution from the freeze-in mechanism originated by the decays of the superpartners while they are still in thermal equilibrium~\cite{Cheung:2011nn}. The expression of the relic density can be written as:
\begin{equation}
\Omega^{\rm FIMP}_{DM} h^2 =\frac{1.09 \times 10^{27}}{g_{*}^{3/2}}m_{3/2} \sum_{i} g_i \frac{\Gamma_i}{m_i^2}
\end{equation}
where $\Gamma_i$ is the decay rate of the i-th superpartner, which can be a gaugino or a scalar, and it is given by eq. (\ref{eq:gamma_sugra}) by substituting the suitable mass, while $g_i$ represent the internal degrees of freedom to the i-th state. Since the decay rate depends on the fifth power of the mass of the decaying particle the DM relic density is mainly determined by the decays of the heaviest particles. It can be easily seen that it largely exceeds the experimental value because of the high scale of the scalars in this setup. The only way out is to impose the condition $T_{\rm R} < m_0$, in such a way we have no equilibrium population of the heaviest states in the early Universe. From now on we will thus assume the condition $m_{\tilde{B}} < T_{\rm R} < m_0$, i.e. a reheating temperature below the mass of the scalars, in order to avoid the existence of a thermal population of these particles, but still sensitively above the mass scale of the Bino, in order of not affect the generation of its abundance. This requirement is not problematic in our scenario since, as shown in the previous subsection, an efficient generation of the baryon asymmetry requires a 2-3 orders of magnitude separation between the scales $m_{\tilde{B}}$ and $m_0$. The freeze-in relic density thus reduces just to the contribution of the three gauginos which can be written as:
\begin{equation}
\label{eq:FIMP}
\Omega_{\rm DM}^{\rm FI} h^2 \approx 0.7 \times 10^{-3} {\left(\frac{m_{\tilde{B}}}{10\mbox{TeV}}\right)}^{3} {\left(\frac{m_{3/2}}{1\mbox{TeV}}\right)}^{-1}\left[1+3 {\left(\frac{m_{\tilde{W}}}{m_{\tilde{B}}}\right)}^{3}+8 {\left(\frac{m_{\tilde{G}}}{m_{\tilde{B}}}\right)}^{3}\right]
\end{equation}
\noindent 
We also remark that our requirement on the reheating temperature implies, as by-product, a suppression of the contribution from thermal scatterings, eq.~(\ref{eq:omega_thermal}). The DM relic density is then totally accounted by the decays of the gauginos.

\noindent
We have finally, for the contribution from the out-of-equilibrium decay of the Bino, the SuperWIMP contribution:
\begin{equation}
\label{eq:gravitino_swimp}
\Omega_{\rm DM}^{\rm SW} = \xi_s \frac{m_{3/2}}{m_{\tilde{B}}} Br(\tilde{B} \rightarrow \tilde{\psi}_{3/2}+X) \Omega_{\tilde{B}}^{\tau \rightarrow \infty}
\end{equation}
\noindent
Contrary to the baryon density the only suppression term present is $\xi_{\rm s}$, which accounts for possible entropy dilution effects.
\noindent
A similar contribution to eq.(\ref{eq:gravitino_swimp}) originates also from the decays of the Wino and the Gluino after they undergone chemical freeze-out. However these two particles have a much lower relic density, compared to the Bino, in virtue of their efficient annihilation processes and thus the corresponding contribution is negligible.

\noindent
Assuming for $\Omega_{\tilde{B}}^{\tau \rightarrow \infty}$ the expression given in eq.~(\ref{eq:Omegab_only_pair}), we can write:
\begin{equation}
\label{eq:SWIMP}
\Omega_{DM}^{\rm SW} \approx 2.34 \times 10^{-3} {\left(\frac{\mu}{10^{3/2}m_0}\right)}^2 {\left(\frac{m_0}{10^6 \mbox{GeV}}\right)}^6 {\left(\frac{m_{\tilde{B}}}{1 \mbox{TeV}}\right)}^{-1}{\left(\frac{m_{3/2}}{1 \mbox{GeV}}\right)}^{-1}
\frac{x_{\rm f.o.}}{A(x_{\rm f.o.})}{\left(1+\frac{6 \lambda^2}{\pi \alpha_s}\right)}^{-1}
\end{equation}

\noindent
By comparing eq.~(\ref{eq:SWIMP}) and~(\ref{eq:FIMP}) we notice that the SuperWIMP contribution tends to dominate at higher values of $m_0$ and of the ratio 
$\mu/m_0$ while the freeze-in one becomes more important once increasing the masses of the gauginos. In particular the DM relic density can result dominated 
by a heavy Wino, as required, as will be clarified in the next section, by avoiding coannihilation effects reducing the baryon abundance.

\noindent
If the out-of-equilibrium decay of the Bino is the main source of the DM abundance, the ratio $\Omega_{\Delta B}/\Omega_{\rm DM}$ assumes the simple form, 
as function of the supersymmetric parameters:
\begin{align}
\label{eq:ratio_fit}
\frac{\Omega_{\Delta B}}{\Omega_{DM}} &= \frac{m_p}{m_{3/2}} \frac{\epsilon_{\rm CP}}{Br\left(\tilde{B} \rightarrow \tilde{\psi}_{3/2}+X\right)} 
\nonumber\\
& \approx 3.3\, {\left(\frac{\lambda}{0.1}\right)}^2 \left(\frac{m_{3/2}}{m_p}\right)\left(\frac{m_{\tilde{G}}}{m_{\tilde{B}}}\right) f_2\left(\frac{m_{\tilde{G}}^2}{m_{\tilde{B}}^2}\right) {\left(\frac{m_{\tilde{B}}}{1 \mbox{TeV}}\right)}^2 {\left(\frac{m_0}{10^6 \mbox{GeV}}\right)}^{-6}\nonumber\\
& \approx 0.6 {\left(\frac{\lambda}{0.1}\right)}^2 \left(\frac{m_{3/2}}{m_p}\right) {\left(\frac{m_{\tilde{B}}}{1 \mbox{TeV}}\right)}^2 {\left(\frac{m_0}{10^6 \mbox{GeV}}\right)}^{-6}
\end{align}
where, in the last line, we have taken the maximal value for $\left(\frac{m_{\tilde{G}}}{m_{\tilde{B}}}\right) f_2\left(m_{\tilde{G}}^2/m_{\tilde{B}}^2\right)\sim 0.16$.
Note that this ratio is independent of the abundance of the decaying Bino and that. interestingly, the correct ratio between the two relic densities is achieved, for a 
Bino at the TeV scale, when the gravitino mass is of the same order as the mass of the proton. 
Unfortunately, as clarified by the numerical treatment in the next section, the requirement of the correct abundance of the Bino, mandatory for the matching of the 
individual quantities with their observed values, will point towards sensitively higher masses for the Bino and the gravitino.
%\pagebreak
\section{Numerical Analysis}
\subsection{Boltzmann equations}
\label{sec:boltzmann}

\noindent
The generation of the baryon asymmetry and of the DM, including additional effects like wash-out and coannihilations, in the scenario under study, 
can be traced through a system of five coupled Boltzmann equations. The first three describe the evolution of the yields, namely $Y =n/s$, of the three gauginos: 
\begin{align}
\label{eq:system_Wino}
\frac{dY_{\tilde{W}}}{dx} &=-\frac{1}{H x} \Gamma_{\tilde{W},\Delta B \neq 0} \left(Y_{\tilde{W}}-Y^{\rm eq}_{\tilde{W}}\right)-\frac{s}{H x}\langle \sigma v \rangle_{\tilde{W},\Delta B \neq 0} Y_q^{\rm eq} \left(Y_{\tilde{W}}-Y_{\tilde{W}}^{\rm eq}\right) \nonumber\\
& -\frac{s}{Hx}\langle \sigma v \rangle \left(\tilde{W}\tilde{G} \rightarrow \bar f f\right) \left(Y_{\tilde{W}}Y_{\tilde{G}}-Y_{\tilde{W}}^{\rm eq} Y_{\tilde{G}}^{\rm eq}\right)
\nonumber\\
& -\frac{s}{H x}\left(\langle \sigma v \rangle \left(\tilde{B}\tilde{W} \rightarrow \bar f f\right)+\langle \sigma v \rangle \left(\tilde{B}\tilde{W} \rightarrow H H^{*}\right)\right) \left(Y_{\tilde{B}}Y_{\tilde{W}}-Y_{\tilde{B}}^{\rm eq} Y_{\tilde{W}}^{\rm eq}\right)\nonumber\\
& - 2 \frac{s}{H x}\langle \sigma v \rangle_{\tilde{W}\tilde{W}} \left(Y_{\tilde{W}}^2-Y_{\tilde{W}}^{\rm eq \, 2}\right) \nonumber\\ 
& -\frac{1}{H x} \Gamma \left(\tilde{W} \rightarrow \tilde{G} \bar f f \right) \left(Y_{\tilde{W}}-Y^{\rm eq}_{\tilde{W}}\frac{Y_{\tilde{G}}}{Y_{\tilde{G}}^{\rm eq}}\right)-\frac{s}{H x}\langle \sigma v \rangle \left(\tilde{W}f \rightarrow \tilde{G}f\right) Y_q^{\rm eq} \left(Y_{\tilde{W}}-Y_{\tilde{W}}^{\rm eq}\right) \nonumber\\
& -\frac{1}{H x} \left(\Gamma\left(\tilde{W}\rightarrow \tilde{B}\bar f f\right)+\Gamma\left(\tilde{W}\rightarrow \tilde{B}H H^{*}\right)\right)\left(Y_{\tilde{W}}-\frac{Y_{\tilde{W}}^{\rm eq}}{Y_{\tilde{B}}^{\rm eq}}Y_{\tilde{B}}\right)\nonumber\\
&-\frac{s}{H x}\left(\langle \sigma v \rangle \left(\tilde{W}f \rightarrow \tilde{B} f\right)Y_q^{\rm eq}+\langle \sigma v \rangle \left(\tilde{W}H \rightarrow \tilde{B} H\right)Y_h^{\rm eq}\right)\left(Y_{\tilde{W}}-\frac{Y_{\tilde{W}}^{\rm eq}}{Y_{\tilde{B}}^{\rm eq}}Y_{\tilde{B}}\right)\nonumber\\
& -\frac{1}{H x} \Gamma \left(\tilde{W} \rightarrow \tilde{\psi}_{3/2}+X\right) Y_{\tilde{W}}
\end{align}
\begin{align}
\label{eq:system_Bino}
\frac{dY_{\tilde{B}}}{dx} &=-\frac{1}{H x} \Gamma_{\tilde{B},\Delta B \neq 0} \left(Y_{\tilde{B}}-Y^{\rm eq}_{\tilde{B}}\right)-\frac{s}{H x}\langle \sigma v \rangle_{\tilde{B},\Delta B \neq 0} Y_q^{\rm eq} \left(Y_{\tilde{B}}-Y_{\tilde{B}}^{\rm eq}\right) \nonumber\\
& -\frac{s}{H x}\langle \sigma v \rangle \left(\tilde{B}\tilde{G} \rightarrow \bar f f\right) \left(Y_{\tilde{B}}Y_{\tilde{G}}-Y_{\tilde{B}}^{\rm eq} Y_{\tilde{G}}^{\rm eq}\right)
\nonumber\\
&-\frac{s}{H x}\left(\langle \sigma v \rangle \left(\tilde{B}\tilde{W} \rightarrow \bar f f\right)+\langle \sigma v \rangle \left(\tilde{B}\tilde{W} \rightarrow H H^{*}\right)\right) \left(Y_{\tilde{B}}Y_{\tilde{W}}-Y_{\tilde{B}}^{\rm eq} Y_{\tilde{W}}^{\rm eq}\right)\nonumber\\
&- 2 \frac{s}{H x}\langle \sigma v \rangle_{\tilde{B}\tilde{B}} \left(Y_{\tilde{B}}^2-Y_{\tilde{B}}^{\rm eq \, 2}\right) \nonumber\\ 
& -\frac{1}{H x} \Gamma \left(\tilde{B} \rightarrow \tilde{G} \bar f f\right) \left(Y_{\tilde{B}}-Y^{\rm eq}_{\tilde{B}}\frac{Y_{\tilde{G}}}{Y_{\tilde{G}}^{\rm eq}}\right)-\frac{s}{H x}\langle \sigma v \rangle \left(\tilde{B}f \rightarrow \tilde{G}f\right) Y_q^{\rm eq} \left(Y_{\tilde{B}}-Y_{\tilde{B}}^{\rm eq}\right) \nonumber\\
& +\frac{1}{H x} \left(\Gamma\left(\tilde{W}\rightarrow \tilde{B}\bar f f\right)+\Gamma\left(\tilde{W}\rightarrow \tilde{B}H H^{*}\right)\right)\left(Y_{\tilde{W}}-\frac{Y_{\tilde{W}}^{\rm eq}}{Y_{\tilde{B}}^{\rm eq}}Y_{\tilde{B}}\right)\nonumber\\
&+\frac{s}{H x}\left(\langle \sigma v \rangle \left(\tilde{W}f \rightarrow \tilde{B} f\right)Y_q^{\rm eq}+\langle \sigma v \rangle \left(\tilde{W}H \rightarrow \tilde{B} H\right)Y_h^{\rm eq}\right)\left(Y_{\tilde{W}}-\frac{Y_{\tilde{W}}^{\rm eq}}{Y_{\tilde{B}}^{\rm eq}}Y_{\tilde{B}}\right)\nonumber\\
& -\frac{1}{H x} \Gamma \left(\tilde{B} \rightarrow \tilde{\psi}_{3/2}+X\right) Y_{\tilde{B}}
\end{align}
%\pagebreak
\begin{align}
\label{eq:system_gluino}
\frac{dY_{\tilde{G}}}{dx} &=-\frac{1}{H x} \Gamma_{\tilde{G},\Delta B \neq 0} \left(Y_{\tilde{G}}-Y^{\rm eq}_{\tilde{G}}\right)-\frac{s}{H x}\langle \sigma v \rangle_{\tilde{G},\Delta B \neq 0} Y_q^{\rm eq} \left(Y_{\tilde{G}}-Y_{\tilde{G}}^{\rm eq}\right) \nonumber\\
& -\frac{s}{H x}\langle \sigma v \rangle \left(\tilde{B}\tilde{G} \rightarrow \bar f f\right) \left(Y_{\tilde{B}}Y_{\tilde{G}}-Y_{\tilde{B}}^{\rm eq} Y_{\tilde{G}}^{\rm eq}\right)- 2 \frac{s}{H x}\langle \sigma v \rangle_{\tilde{G}\tilde{G}} \left(Y_{\tilde{G}}^2-Y_{\tilde{G}}^{\rm eq \, 2}\right) \nonumber\\ 
& -\frac{s}{H x}\langle \sigma v \rangle \left(\tilde{W}\tilde{G} \rightarrow \bar f f\right) \left(Y_{\tilde{W}}Y_{\tilde{G}}-Y_{\tilde{W}}^{\rm eq} Y_{\tilde{G}}^{\rm eq}\right)\nonumber\\
& +\frac{1}{H x} \Gamma\left(\tilde{B} \rightarrow \tilde{G} \bar f f\right) \left(Y_{\tilde{B}}-Y^{\rm eq}_{\tilde{B}}\frac{Y_{\tilde{G}}}{Y_{\tilde{G}}^{\rm eq}}\right)+\frac{s}{H x}\langle \sigma v \rangle \left(\tilde{B}f \rightarrow \tilde{G}f\right) Y_q^{\rm eq} \left(Y_{\tilde{B}}-Y_{\tilde{B}}^{\rm eq}\right)\nonumber\\
& +\frac{1}{H x} \Gamma\left(\tilde{W} \rightarrow \tilde{G} \bar f f\right) \left(Y_{\tilde{W}}-Y^{\rm eq}_{\tilde{W}}\frac{Y_{\tilde{G}}}{Y_{\tilde{G}}^{\rm eq}}\right)+\frac{s}{H x}\langle \sigma v \rangle \left(\tilde{W}f \rightarrow \tilde{G}f\right) Y_q^{\rm eq} \left(Y_{\tilde{W}}-Y_{\tilde{W}}^{\rm eq}\right)\nonumber\\
& -\frac{1}{H x} \Gamma \left(\tilde{G} \rightarrow \tilde{\psi}_{3/2}+X\right) Y_{\tilde{B}}
\end{align}

\noindent
In each equation the first row represents B-violating decay and single annihilation processes. The second to the fourth lines represent coannihilation and pair annihilation processes. The remaining lines, apart the last, give rise to transition processes, either decays or scatterings, between gauginos. 
The last line in each equation represents finally the production of the gravitino. These last decay terms are proportional only to the yields of the gauginos since 
we assume that the initial gravitino abundance is negligible and remains low enough to neglect inverse decay processes. Under this assumption the equation 
for the gravitino abundance assumes a rather simple form: 
\begin{equation}
\label{eq:system_gravitino}
\frac{dY_{3/2}}{dx}=\frac{1}{H x} \sum_{\tilde{X}} \Gamma \left(\tilde{X} \rightarrow \tilde{\psi}_{3/2}+X\right)Y_{\tilde{X}}\,\,\,\,\,\,\,\,\tilde{X}=\tilde{B},\tilde{W},\tilde{G}
\end{equation}       
\noindent
For simplicity we are neglecting the possibility that the Bino dominates the energy density of the Universe since, as already argued in the previous section and further confirmed by the results presented below, this occurs in a region of the parameter space of marginal relevance. As a consequence, the expression of the Hubble expansion parameter 
is the one typical of radiation domination, $H \approx 1.66 g_{*} \frac{m_{\tilde{B}}^2}{M_{\rm Pl}} x^{-2}$. In order to properly account entropy injection effects it should be modified similarly to what proposed e.g in~\cite{Gelmini:2006pw,Gelmini:2006pq,Arcadi:2011ev,Kane:2015qea}. 
\noindent
We have finally the equation for the baryon asymmetry which is casted as an equation for $Y_{\Delta B-L}$ in order to get rid of the effects of the sphalerons:
\begin{align}
\label{eq:system_BL}
& \frac{dY_{\Delta B-L}}{dx}=\frac{1}{H x} \Delta \Gamma_{\tilde{B},\Delta B \neq 0} \left(Y_{\tilde{B}}-Y_{\tilde{B}}^{\rm eq}\right)+\frac{s}{H x} \langle \Delta \sigma v \rangle_{\tilde{B}} \left(Y_{\tilde{B}}-\frac{Y_{\tilde{B}}^{\rm eq}}{Y_{\tilde{G}}^{\rm eq}}Y_{\tilde{G}}\right) \nonumber\\
&- \frac{3}{H x} \left(\langle \Gamma\left(\tilde{B} \rightarrow udd+\bar u \bar d \bar d\right) \rangle Y_{\tilde{B}}^{\rm eq}+\langle \Gamma\left(\tilde{G} \rightarrow udd+\bar u \bar d \bar d\right) \rangle Y_{\tilde{G}}^{\rm eq}\right.\nonumber\\
&\left.+\langle \Gamma\left(\tilde{W} \rightarrow udd+\bar u \bar d \bar d\right) \rangle Y_{\tilde{W}}^{\rm eq}\right) \frac{m_{\tilde{B}}}{x}\left[\mu_u+\mu_c+\mu_t+2 \left(\mu_d+\mu_s+\mu_b\right)\right] \nonumber\\
& -\frac{6 s}{H x}\{\langle \sigma v \left(u\tilde{B} \rightarrow \bar d \bar d\right) \rangle \left[\left(\mu_u+\mu_c+\mu_t\right) Y_{\tilde{B}}+2 \left(\mu_d+\mu_s+\mu_b\right)Y_{\tilde{B}}^{\rm eq}\right]\nonumber\\
& +\langle \sigma v \left(u\tilde{W} \rightarrow \bar d \bar d\right) \rangle \left[\left(\mu_u+\mu_c+\mu_t\right) Y_{\tilde{W}}+2 \left(\mu_d+\mu_s+\mu_b\right)Y_{\tilde{W}}^{\rm eq}\right]\nonumber\\
& +\langle \sigma v \left(u\tilde{G} \rightarrow \bar d \bar d\right) \rangle \left[\left(\mu_u+\mu_c+\mu_t\right) Y_{\tilde{G}}+2 \left(\mu_d+\mu_s+\mu_b\right)Y_{\tilde{G}}^{\rm eq}\right]\} Y_q^{\rm eq} \frac{m_{\tilde{B}}}{x}\nonumber\\
&-\frac{12 s}{H x}\{\langle \sigma v \left(d\tilde{B} \rightarrow \bar u \bar d\right) \rangle \left[\left(\mu_d+\mu_s+\mu_b\right) Y_{\tilde{B}}+2 \left(\mu_d+\mu_s+\mu_b+\frac{1}{2}\mu_u+\frac{1}{2}\mu_c+\frac{1}{2}\mu_t\right)Y_{\tilde{B}}^{\rm eq}\right]\nonumber\\
& +\langle \sigma v \left(d\tilde{W} \rightarrow \bar u \bar d\right) \rangle \left[\left(\mu_d+\mu_s+\mu_b\right) Y_{\tilde{W}}+2 \left(\mu_d+\mu_s+\mu_b+\frac{1}{2}\mu_u+\frac{1}{2}\mu_c+\frac{1}{2}\mu_t\right)Y_{\tilde{W}}^{\rm eq}\right]\nonumber\\
& +\langle \sigma v \left(d\tilde{G} \rightarrow \bar u \bar d\right) \rangle \left[\left(\mu_d+\mu_s+\mu_b\right) Y_{\tilde{G}}+2 \left(\mu_d+\mu_s+\mu_b+\frac{1}{2}\mu_u+\frac{1}{2}\mu_c+\frac{1}{2}\mu_t\right)Y_{\tilde{G}}^{\rm eq}\right]\} Y_q^{\rm eq} \frac{m_{\tilde{B}}}{x}
\end{align}

\noindent
The first row represents the source terms associated to the B-violating decays of the Bino and, as already mentioned, to the scatterings of both Binos and Gluinos. CPT invariance imposes a relation between the asymmetries generated by Binos and Gluinos \cite{Claudson:1983js,Baldes:2014gca,Baldes:2014rda}:
\begin{equation}
\langle \Delta\sigma v \rangle_{\tilde{B}} Y_{\tilde{B}}^{\rm eq}=-\langle \Delta \sigma v \rangle_{\tilde{G}} Y_{\tilde{G}}^{\rm eq}
\end{equation}
In general we could expect analogous source terms associated to decay and scattering processes with Wino initial state. As already discussed in the previous section and 
shown in an explicit example below, the Wino is always kept very close to thermal equilibrium by its efficient interactions and thus contribute to a negligible amount to the generation of the baryon asymmetry. 
\noindent
The last two rows describe instead the wash-out processes related to inverse decays and to the CP even component of the baryon number violating $2 \rightarrow 2$ 
scattering of both Binos and Gluinos.
\noindent
The equation for the baryon asymmetry depends as well on the chemical potentials $\mu_{f=u,d,s,c,b,t}$ of the right-handed quarks. These chemical potential can be expressed in terms to of the $B-L$ abundance~\cite{Buchmuller:2000as}. We have in reality different relations between the chemical potentials and $B-L$ according on whether the temperature lies above or below the one of the EW phase transition. Since, as will be discussed below, the production of the baryon asymmetry can occur, according the 
values of the relevant parameters, both above and below this critical temperature, we have employed, similarly to what done in~\cite{Baldes:2014rda}, a two step solution 
of the system eq.(\ref{eq:system_Wino})-(\ref{eq:system_BL}). We have first solved the system with initial conditions 
$Y_{\tilde{B}}(x \ll 1)=Y_{\tilde{B}}^{\rm eq}(x \ll 1)$, $Y_{\tilde{W}}(x \ll 1)=Y_{\tilde{W}}^{\rm eq}(x \ll 1)$, $Y_{\tilde{G}}(x \ll 1)=Y_{\tilde{G}}^{\rm eq}(x \ll 1)$ and 
$Y_{\Delta B-L}(x \ll 1)=0$ and:  
\begin{equation}
\mu_u=\mu_c=\mu_t=-\frac{10}{79}\frac{Y_{\Delta B-L} s}{T^2},\,\,\,\,\mu_d=\mu_s=\mu_b=\frac{38}{79}\frac{Y_{\Delta B-L} s}{T^2}
\end{equation}  
until $x=m_{\tilde{B}}/T_{\rm EW}$. Below the EW phase transition the sphalerons freeze-out and we can replace eq.(\ref{eq:system_BL}) with an equation for just 
$Y_{\Delta B}$ with initial condition, set at $T_{\rm EW}$, $Y_{\Delta B}=\frac{28}{79}Y_{B-L}$ and:
\begin{align}
& \mu_u=\left\{L+B\left[\frac{1}{3}+\frac{1}{2 N_d}+\frac{1}{2 N_e}\right]\right\}\times {\left[1+\frac{3 N_u}{N_e}+\frac{N_u}{N_d}+2N_u\right]}^{-1}\nonumber\\
& \mu_d=\frac{B-2 N_u \mu_u}{2 N_d}
\end{align}
where:
\begin{align}
& B=\frac{12 \pi^2 g_{*S}}{45}m_{\tilde B} x Y_{\Delta B} \nonumber\\
& L=\frac{12 \pi^2 g_{*S}}{45}m_{\tilde B} x Y_{L}(T_{\rm EW})
\end{align}

\noindent
The structure of the system makes evident the tight relation, already envisaged in the analytical treatment, between the generation of the CP asymmetry, the abundance of the Bino and the wash-out processes. The baryon asymmetry is originated by the decays (and annihilations) of the Bino into SM fermions. Inverse decay and scatterings, as well as analogous processes involving the Gluino (the processes related to the Gluino are generated by the same diagrams and then the rate differ only by the couplings and by the Gluino distribution) are responsible of the wash-out of the baryon asymmetry. Single RPV annihilations of the Bino can, finally, be the dominant contribution in determining its abundance. Any variation in the CP asymmetry, as originated, e.g. by flavor or left-right mixing effects, is reflected also in this last rates. The optimal production of the baryon asymmetry is thus achieved once a balance is found between a large enough CP-asymmetry and not excessive depletion of the Bino abundance, or excessive efficient wash-out.  

\noindent
The system has been solved for several assignments of the relevant parameters. Differently from the analytical treatment we have considered also the cases in which the 
relevant processes are mediated by left/right-handed top squarks, as well as generic effects of flavor violation in the right-handed down squark sector by assigning arbitrary 
entries and CP violating phases to the matrix $\Gamma^D_R$ and taking non-degenerate squark masses. In both these two cases we have found no sensitive variations with 
respect to the flavor universal scenario. In the case of left-right mixing in the top squark sector this is due to the fact that the (very moderate) enhancement of the CP asymmetry 
is actually compensated by the presence of the electric charge $Q_u$ of the up-quarks in the couplings of the Bino which translates into an overall increase by a factor 4 of 
the annihilation rates of the Bino (it can be easily seen from the analytical expressions which, on the contrary, the value of the CP-asymmetry is insensitive to this quantity.). 
In the flavor violating case instead the small variation in the total CP-asymmetry is due to the already mentioned GIM suppression. 
\noindent
For this reason we will discuss our results in the same flavor universal limit of the analytical treatment in order to profit of the more limited set of parameters, 
being~$(\lambda, m_{\tilde{B}},m_{\tilde{W}},m_{\tilde{G}},m_0,\mu)$. In all cases it has been found that the dominant contribution to the baryon asymmetry is originated 
by the out-of-equilibrium decays of the Bino. 
\noindent
Several examples of numerical solutions will be illustrated in the next subsections. We will first of all show quantitatively the effects of wash-out processes and the impact of 
the Wino in the generation of the baryon asymmetry. We will then determine the regions of the parameter space which provide the correct baryon abundance and the correct 
DM relic density.    

\subsection{Effects of coannihilations and wash-out}

\noindent
We show in the following some examples of numerical solutions of the system of Boltzmann equations highlighting in particular the impact of coannihilations and wash-out effects.

\begin{figure}[htb]
\begin{center}
\includegraphics[width=8 cm]{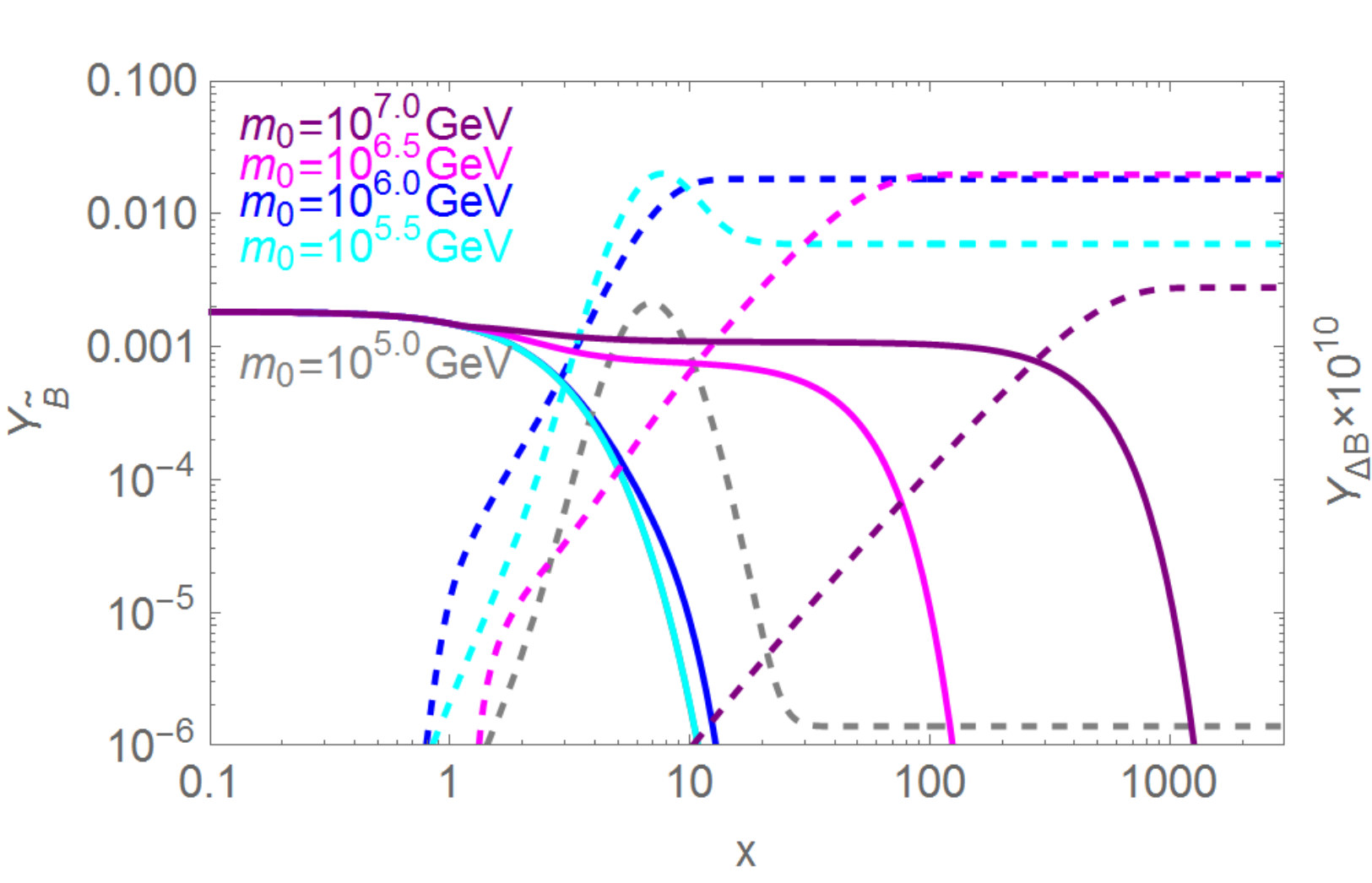}
\caption{\footnotesize{Evolution of the abundance of the yield $Y_{\tilde{B}}$ (solid lines) and of the baryon abundance (dashed lines) with $x=m_{\tilde{B}}/T$ for $m_{\tilde{B}}=2\,\mbox{TeV}$,$m_{\tilde{G}}=1\,\mbox{TeV}$, $\lambda=0.1$, $\mu=10^8\,\mbox{GeV}$ and four values of $m_0$ ranging from $10^{5.0}$ to $10^{7}\,\mbox{GeV}$ reported in the plot.}} 
\label{fig:boltzman_washout}
\end{center}
\end{figure}

\noindent
Fig.~(\ref{fig:boltzman_washout}) shows the Bino  (solid lines) and the baryon yield (dashed lines) for several values of $m_0$, ranging from 
$10^{5.5}$ to $10^{7}$ GeV, and with the following assignment for the remaining parameters:$m_{\tilde{B}}=2\,\mbox{TeV}$,$m_{\tilde{G}}=1\,\mbox{TeV}$, $\lambda=0.1$, 
$\mu=10^8\,\mbox{GeV}$. For the lowest values of $m_0$ we have a low baryon abundance as consequence of the suppressed abundance of the Bino, whose yield 
remains close to the equilibrium distribution until late times. Moreover the baryon abundance is almost completely depleted for $m_0=10^5\,\mbox{GeV}$ since for this 
values of the scalar mass scale the Bino decays before that wash-out processes become ineffective. The baryon abundance is maximal in the intermediate mass range, 
order of $10^{6.5}\,\mbox{GeV}$, where the Bino features a rather early decoupling and it is long-lived enough to evade the wash-out regime. 
The baryon density then decreases again at higher masses when the Bino gets close the relativistic decoupling. Indeed its relic abundance is poorly sensitive to the increase 
of $m_0$ while the CP asymmetry $\epsilon_{\rm CP}$ still features a sensitive suppression. This result justifies our choice to neglect eventual deviations from standard 
cosmology in the numerical system. Indeed entropy production occurs in the very high $m_0$ region which is not relevant for our analysis since we expect a suppressed 
asymmetry.       

\begin{figure}[htb]
\begin{center}
\subfloat{\includegraphics[width=6.8 cm]{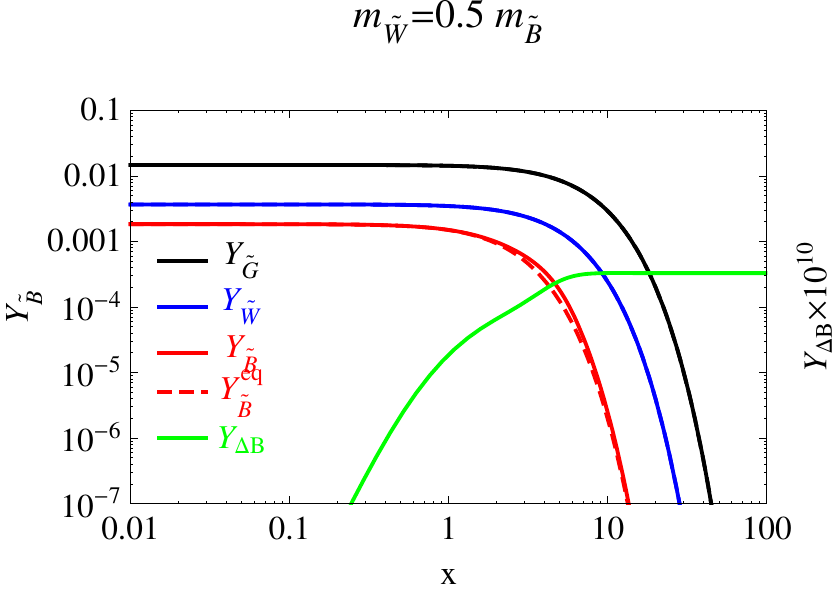}}
\subfloat{\includegraphics[width=6.8 cm]{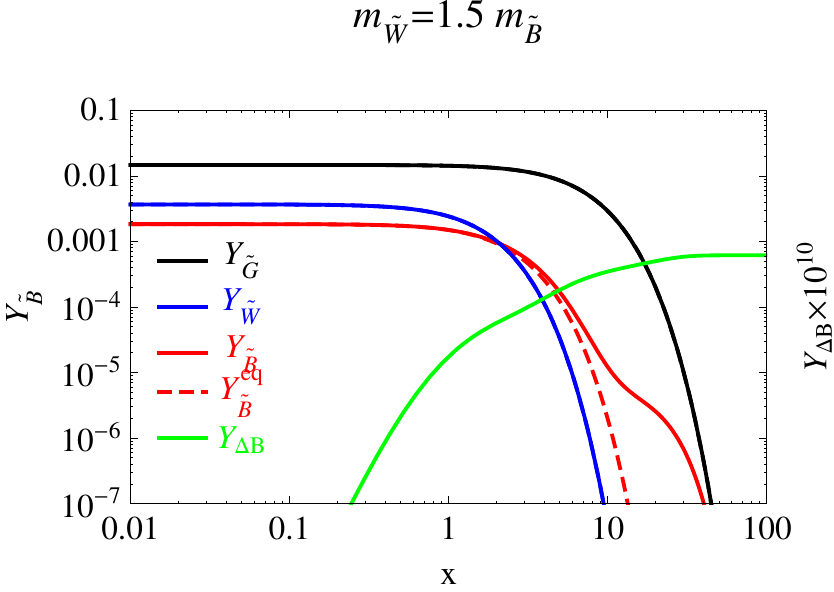}}\\
\subfloat{\includegraphics[width=6.8 cm]{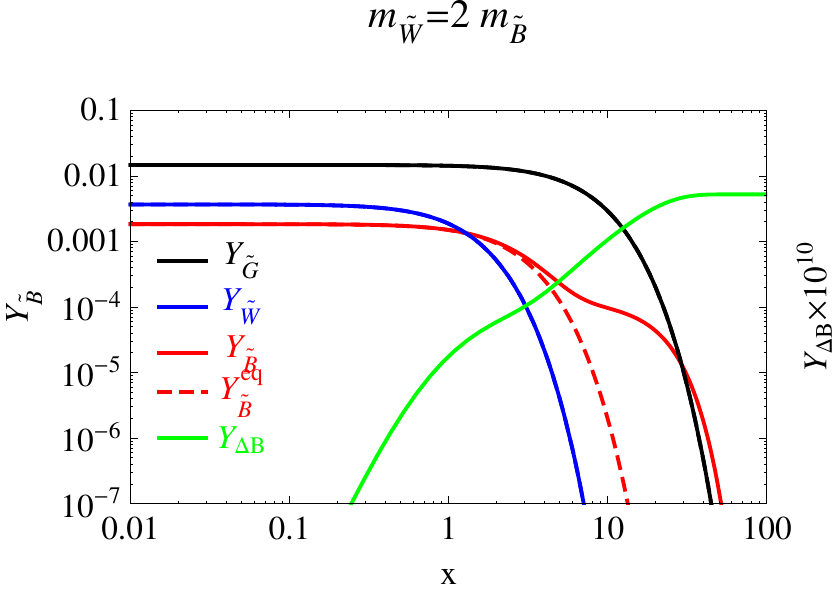}}
\subfloat{\includegraphics[width=6.8 cm]{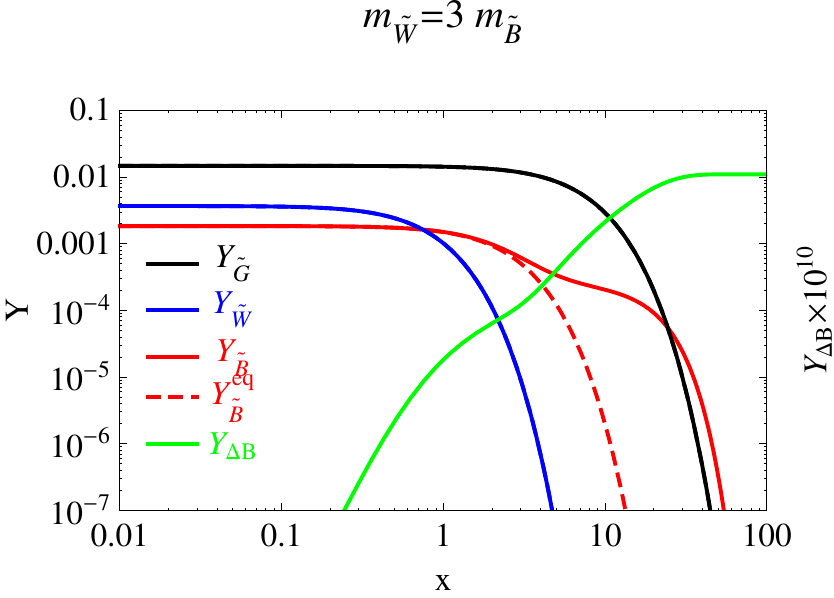}}
\end{center}
\caption{\footnotesize{Evolution of the Yields of the Bino (red solid line), the Gluino (black solid line), Wino (blue solid line) and of the baryons (green solid line), for three assignments of the mass of the Wino. For reference has been reported as well the equilibrium distribution of the Bino (red dashed line).}}
\label{fig:fullcoannihilation}
\end{figure}

\begin{figure}[htb]
\begin{center}
\includegraphics[width=8 cm]{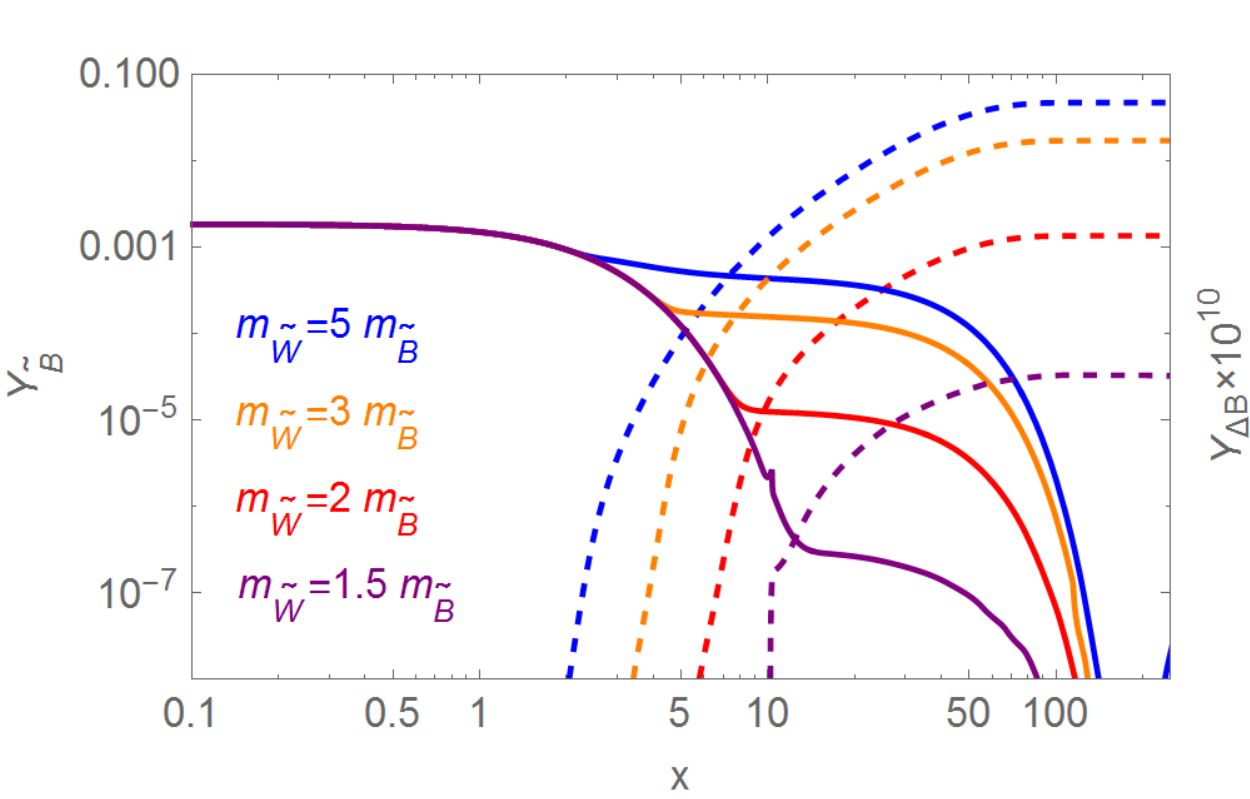}
\caption{\footnotesize{Evolution of the abundance of $Y_{\tilde{B}}$ (solid lines) and of the baryon abundance (dashed lines) for a definite assignment of $(\lambda, m_{\tilde{B}}, m_{\tilde{G}}, m_0, \mu$ and varying $m_{\tilde{W}}/m_{\tilde{B}}$), as reported in the plot.} } 
\label{fig:boltzman_wino}
\end{center}
\end{figure}

\noindent
Fig.~\ref{fig:fullcoannihilation} shows the evolution of the abundances of the Bino and of the baryon density, compared with the ones of the other two gauginos. As evident these two species tend to remain in thermal equilibrium (up to their decay) during the whole phase of generation of the baryon asymmetry. The four panels of fig.~(\ref{fig:fullcoannihilation}) differ in the assignments of the mass of the Wino, considered to be both below and above the mass of the Bino. The Wino has a profound impact in the generation of the baryon asymmetry. The case of a light Wino is, in particular, disfavored. Indeed in such a case, coannihilation effects turn to be very strong, keeping the Bino very close to the equilibrium distribution up to late times, with consequent suppression of the baryon abundance. Contrary to conventional WIMP coannihilation scenarios, the Bino abundance results altered even for sizable mass splitting with the Wino. This is consequence of the strong suppression of the Bino annihilation rates. This last effect is better evidenced in fig.~(\ref{fig:boltzman_wino}) where even higher values of the ratio $\frac{m_{\tilde{W}}}{m_{\tilde{B}}}$ have been considered.
\noindent
In order to maximize the production of the baryon asymmetry we need to invoke a strong hierarchy between the mass of the Bino and the one of the Wino, at least $\frac{m_{\tilde{W}}}{m_{\tilde{B}}} > 5$.
\noindent
As argued in the previous subsection, such a heavy Wino can copiously produce DM through freeze-in. This issue can be possibly avoided by requiring a very heavy Wino
with $m_{\tilde{W}} > T_{\rm R}$ or by checking that it is light enough to avoid overclosure as given by imposing the condition $ \Omega^{FI}_{DM} h^2 < 0.1$, i.e.
from eq.~(\ref{eq:FIMP})
\begin{equation}
m_{\tilde W} < 362\;\mbox{TeV} \left( \frac{m_{3/2}}{1\; \mbox{TeV}} \right)^{1/3} \; .
\end{equation}
In the next subsection we will focus on the case in which the decays of the Bino are the primary source of DM production and we will thus assume, for simplicity that the mass of the Wino is above the reheating temperature.

\subsection{Results}

\noindent
We will illustrate below the regions of the parameter space accounting for the experimentally favored values for the baryon and DM abundances.
\noindent
In our setup the baryon asymmetry depends on five parameters, namely the mass of the Bino $m_{\tilde{B}}$, the mass of the gluino $m_{\tilde{G}}$, the heavy scales $m_0$ and $\mu$, and the RPV coupling $\lambda$. The DM relic density depends on two additional parameters, the mass of the gravitino $m_{3/2}$ and, possibly, the mass of the remaining gaugino $m_{\tilde{W}}$. Regarding this latter parameter, as already discussed, a value close to the masses of the other gauginos is disfavoured by the correct baryon asymmetry. We will, from now on, implicitly assume, for simplicity, that the mass of the Wino is decoupled from the relevant phenomenology, i.e. $m_{\tilde{W}} > T_{\rm R}$. 

\noindent
As discussed above, the baryon density is the most difficult quantity to accommodate. We will thus determine it in the bidimensional plane $(m_{\tilde{B}},m_0)$ after having identified an optimal assignation for the remaining parameters. The correct DM abundance can be determined accordingly by a suitable choice of the mass of the gravitino.

\begin{figure}[htb]
\begin{center}
\subfloat{\includegraphics[width=6.8 cm]{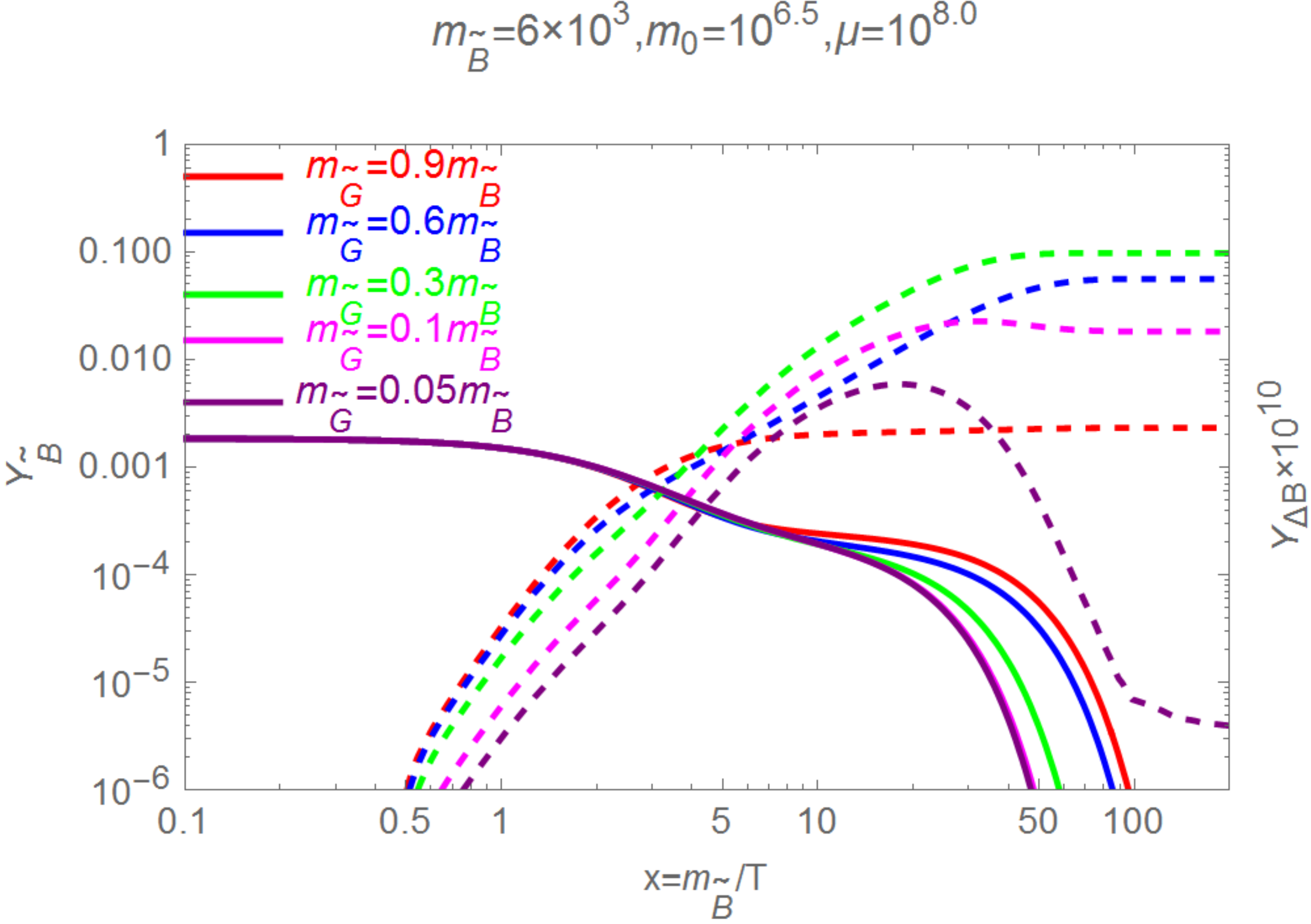}}
\subfloat{\includegraphics[width=6.8 cm]{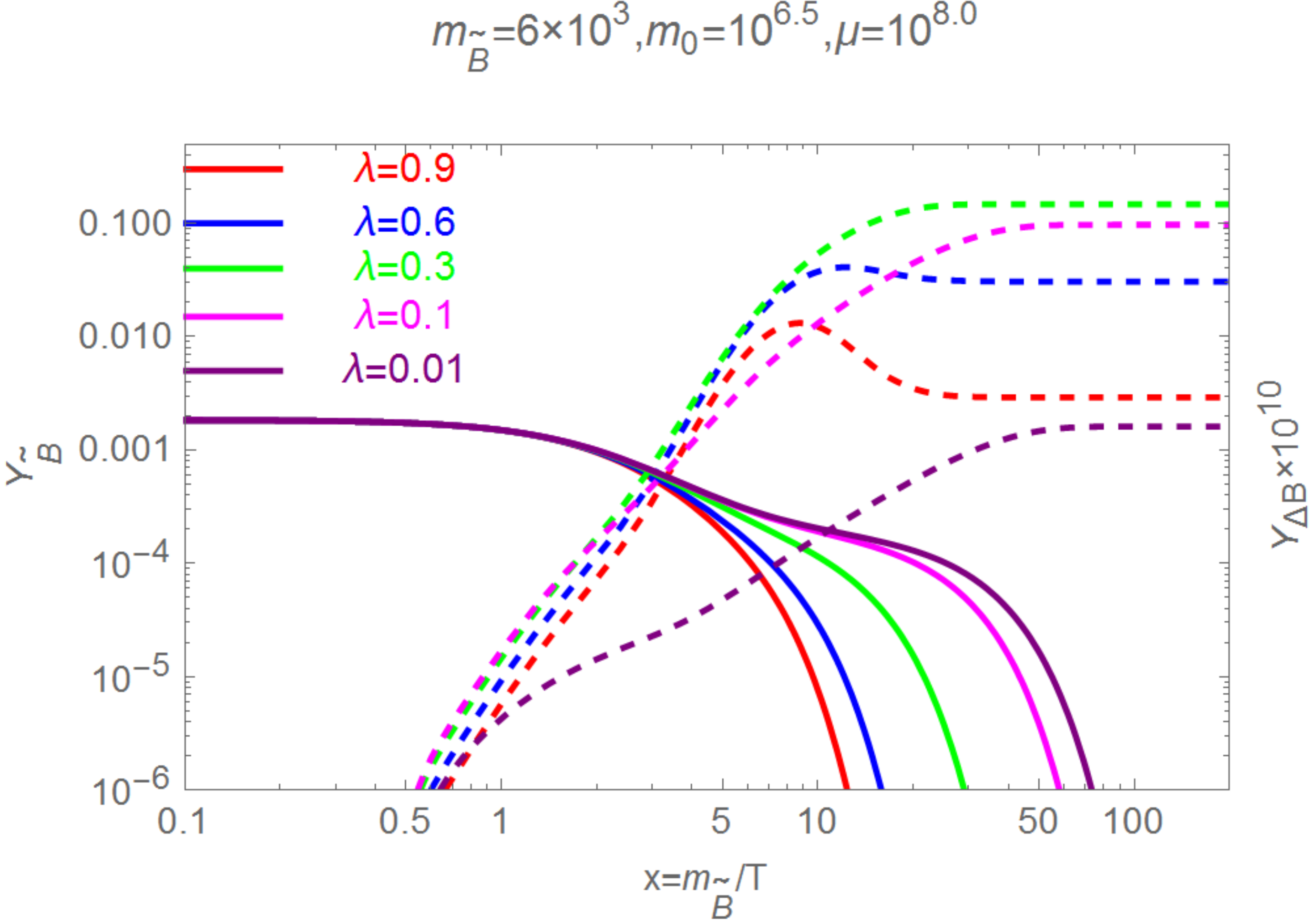}}
\caption{\footnotesize{Left panel: Bino (solid lines) and baryon yields (dashed lines) for a fixed assignment of $m_{\tilde{B}}$, $m_0$ and $\mu$, reported on the plot, for $\lambda=0.3$ and for four values of $m_{\tilde{G}}$ ranging from, $0.05 m_{\tilde{B}}$ and $0.9m_{\tilde{B}}$. Right panel: The same as left panel but with $m_{\tilde{G}}=0.35 m_{\tilde{B}}$ and $\lambda$ varying between 0.05 and 0.9.}}
\label{fig:lambda_and_glu}
\end{center}
\end{figure}
           
\noindent
Fig.~(\ref{fig:lambda_and_glu}) shows the evolution of the yields of the baryons and of the DM as the parameters $m_{\tilde{G}}$ (left panel) and $\lambda$ (right panel) are varied, while keeping fixed the others. As evident, in the case of both quantities there is a non-trivial interplay in the determination of the baryon abundance. A mass of the gluino very close to the one of the Bino determines a huge suppression of $\epsilon_{\rm CP}$ (in eq.~\ref{eq:epsilon_CP} $f_1\left(m_{\tilde{G}}^2/m_{\tilde{B}}^2\right) \ll 1$) while in the opposite scenario, i.e. $m_{\tilde{G}}/m_{\tilde{B}} \ll 1$ the baryon abundance is analogously suppressed by the factor $m_{\tilde{G}}/m_{\tilde{B}}$ in $\epsilon_{\rm CP}$ and, more important, the wash-out processes are efficient up to very late time scales, with respect to the one of decay of the Bino, substantially depleting the created asymmetry. This is then maximal for $\frac{m_{\tilde{G}}}{m_{\tilde{B}}}\sim 0.3-0.6$. For such values there is still a sizable kinematic suppression of the $B$-violating decay of the Bino as 
well as its abundance due to the effect of the coannihilations with the gluino as well as the single annihilations into a Gluino final state. 
\noindent
A similar situation occurs also for the $\lambda$ coupling, with a suppression of the baryon abundance both for $\lambda \sim 1$ and for $\lambda \ll 1$. The behaviour at high values of $\lambda$ is motivated by the fact that $\epsilon_{\rm CP}$ is independent of such coupling in this regime (see eq.~\ref{eq:epsilon_CP}). As a consequence the main effect is the increase of the rate of the single Bino annihilations, influencing both the Bino abundance and, directly, also the one of the baryons, through an enhancement of 
wash-out effects. In the regime of very low $\lambda$ the dominant effect is the suppression of the branching fraction of $B$-violating decays since the abundance of the Bino is controlled by the annihilations involving the Gluino as well as the pair annihilation processes. The optimal range for the $\lambda$ parameter is, again, the intermediate range $\lambda \sim 0.3-0.6$.

\begin{figure}[htb]
\begin{center}
\subfloat{\includegraphics[width=6.8 cm]{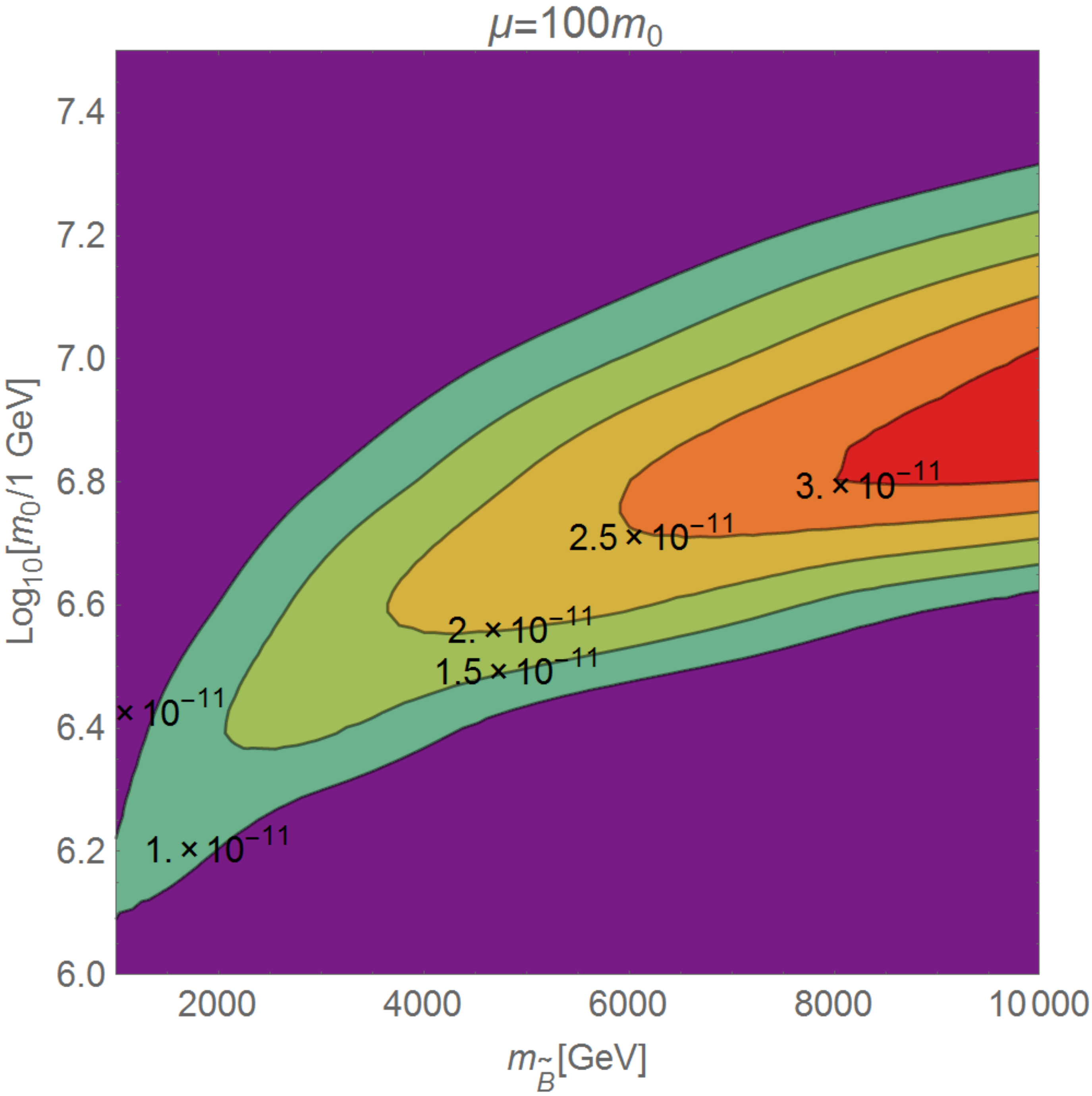}}
\subfloat{\includegraphics[width=6.8 cm]{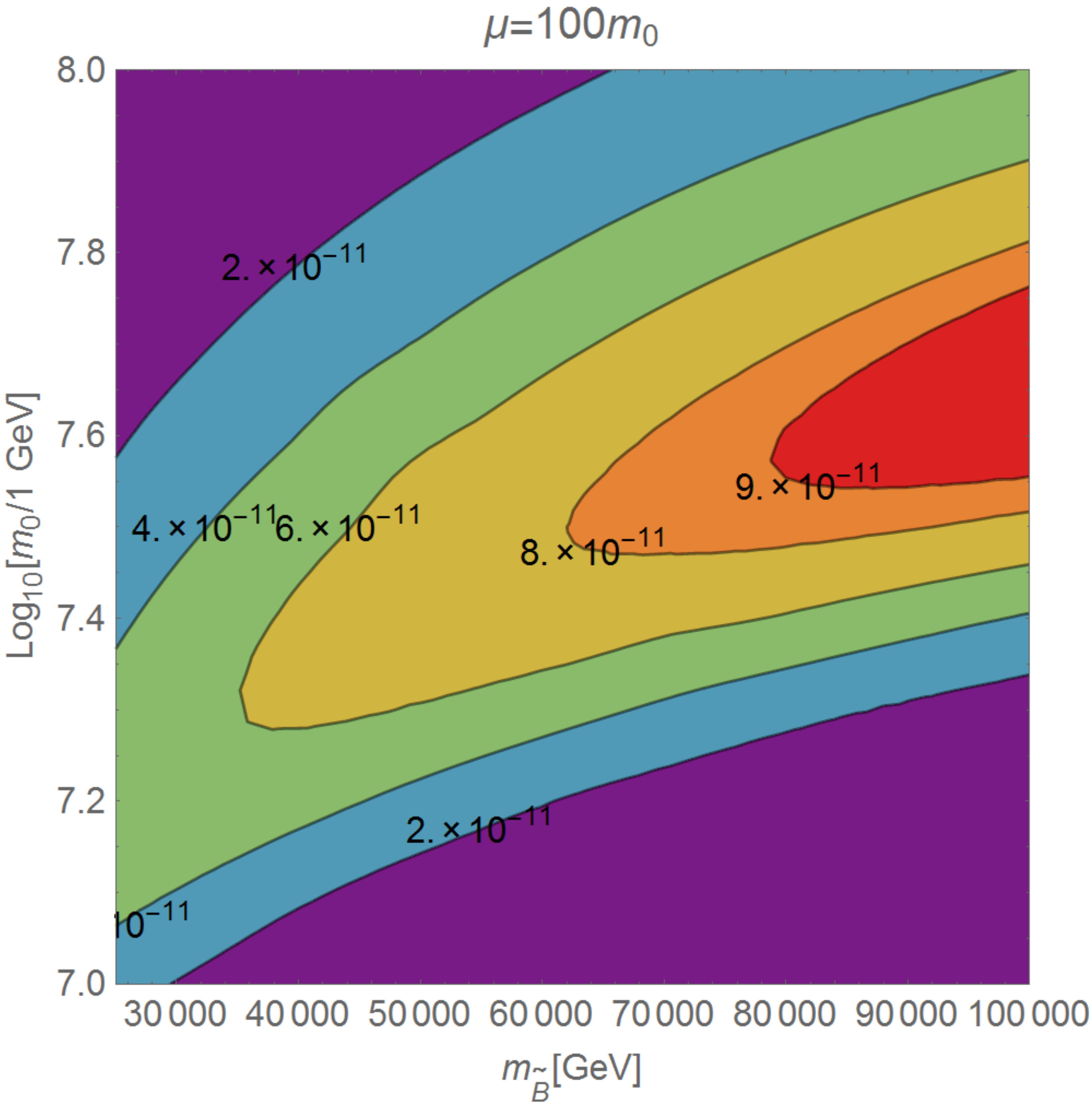}}
\end{center}
\caption{\footnotesize{Contours of values of the baryon abundance $Y_{\Delta B}$ in the plane $\left(m_{\tilde{B}},m_0\right)$. The $\mu$ parameter has been set to $100\,m_0$. In the left panel the Bino mass have been varied in the range $1-10$ TeV while in the right panel higher masses of the Bino, namely 20-100 TeV, have been considered. For both plots we have considered $m_{\tilde{G}}/m_{\tilde{B}}=0.4$ and $\lambda=0.4$.}}  
\label{fig:rainbows} 
\end{figure}
\noindent
Fig.~(\ref{fig:rainbows}) reports the isocontours of the baryon abundance $Y_{\Delta B}$ in the bidimensional plane $(m_{\tilde{B}},m_0)$ with $m_{\tilde{G}}$ and $\lambda$ fixed, according the discussion above, to, respectively, $0.4\; m_{\tilde{B}}$ and $0.4$. As already argued in our analytical study the correct order of magnitude is achieved only for a rather restricted range of values of $m_0$. Above this region there is an excessive suppression of $\epsilon_{\rm CP}$ while below the baryon abundance is erased by wash-out processes. The correct value of the baryon abundance is achieved for a rather heavy Bino, with mass $\sim 70\,\mbox{TeV}$, and 
$m_0 \sim 10^{7.5}\,\mbox{GeV}$ (by varying the ratio $m_{\tilde{G}}/m_{\tilde{B}}$ and $\lambda$ within the range indicated above it is possible to lower to 
approximately 50 TeV the minimal viable Bino mass). This is due to the suppression, direct and indirect, of the Bino density coming from the presence of a rather close-in-mass gluino, which requires high values of the scales $m_{\tilde{B}}$ and $m_0$ to be compensated.
\noindent
As can be seen from the second panel of fig.~(\ref{fig:rainbows}), the correct baryon abundance can be achieved also for $m_{\tilde{B}} > 100\,\mbox{TeV}$. However the consequent increase of the scale $m_0$ would create tension with the determination of the Higgs 
mass~\cite{Arvanitaki:2012ps,ArkaniHamed:2012gw,Bagnaschi:2014rsa,Vega:2015fna}. 

\noindent
As already mentioned we have assumed throughout this work the $\tan\beta \rightarrow 1$. As shown in~\cite{Vega:2015fna} it is possible to have the correct value of the Higgs mass also for $\tan\beta >50$ and $m_0 = |\mu|$. We have solved the Boltzmann system and found analogous contours as the ones shown in fig.~(\ref{fig:rainbows})
also for $\mu=m_0$. In this case $\tilde{B}\tilde{B} \rightarrow HH^{*}$ annihilations play no relevant role and we can thus reduce the number of free parameters, although the general results remain substantially unchanged.

\begin{figure}[htb]
\begin{center}
\includegraphics[width=8.5 cm]{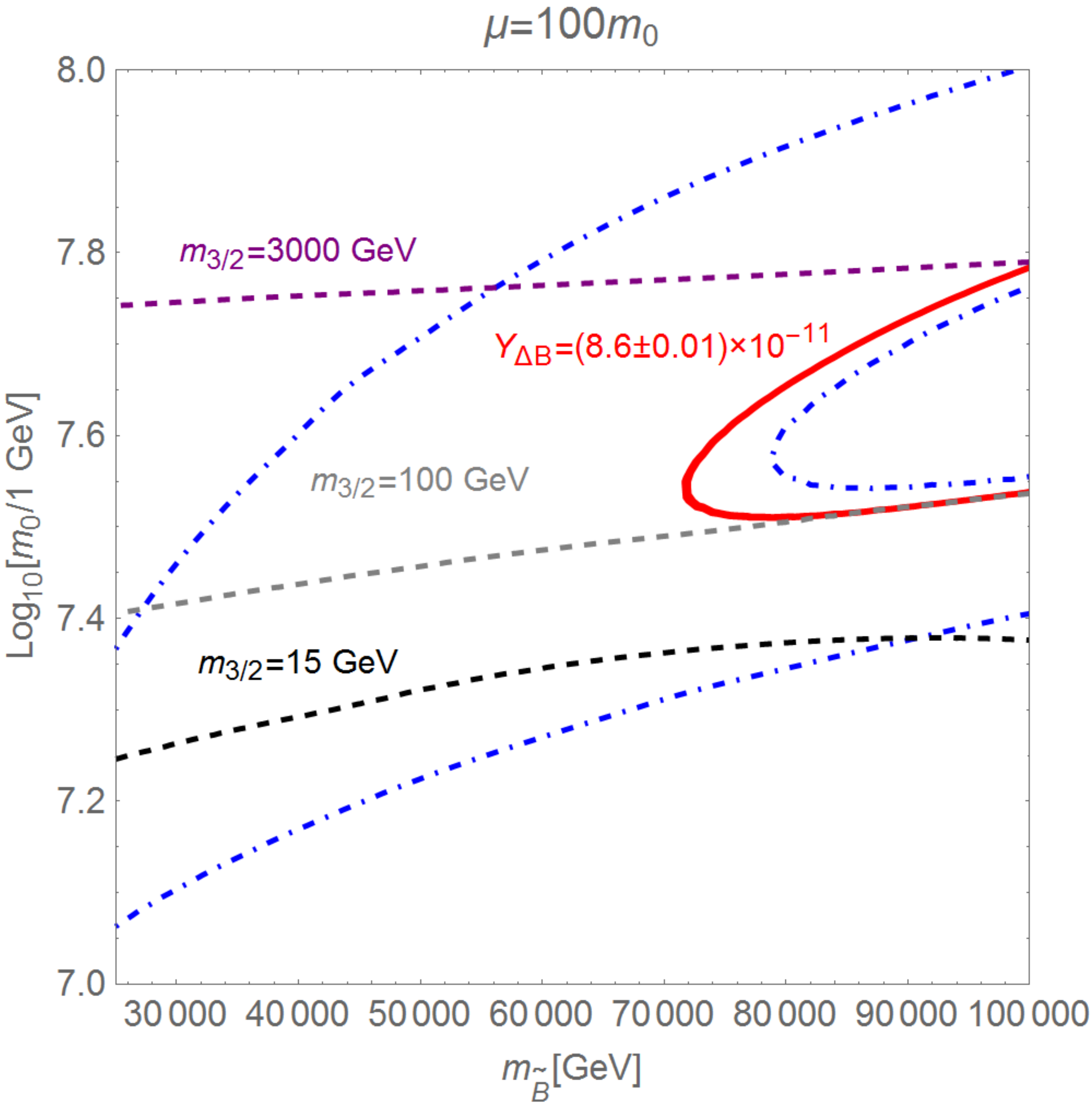}
\caption{\footnotesize{Isocontours of the baryon and DM Yields. The red band represents the value $Y_{\Delta B}=\left(0.86 \pm 0.01\right)\times 10^{-10}$ determined by CMB measurments~\cite{Ade:2013zuv}. The blue dashed lines represent the extrema $0.4 \times 10^{-10}$ and $0.9 \times 10^{-10}$ determined by BBN~\cite{Copi:1994ev}. The black, gray and purple dot-dashed lines represent the isocontours of the correct DM relic density for the reported values of the mass of the gravitino.}}
\label{fig:exp}
\end{center}
\end{figure}

\noindent
The baryon abundance is finally compared with the one of DM in fig.~(\ref{fig:exp}). Here we have reported the experimentally favored value $Y_{\Delta B}=\left(0.86 \pm 0.01\right)\times 10^{-11}$ and confronted it with isocontours of the correct DM relic density for some values of the gravitino mass. As we see the correct match between the two abundances occurs for a mass of the gravitino between, approximatively, 100 GeV and 3 TeV. A lower mass of the gravitino is achieved if a wider range of variation, like the one shown in the figure based on BBN measurements~\cite{Copi:1994ev}, is allowed.

\begin{figure}[htb]
\begin{center}
\subfloat{\includegraphics[width=7.5 cm]{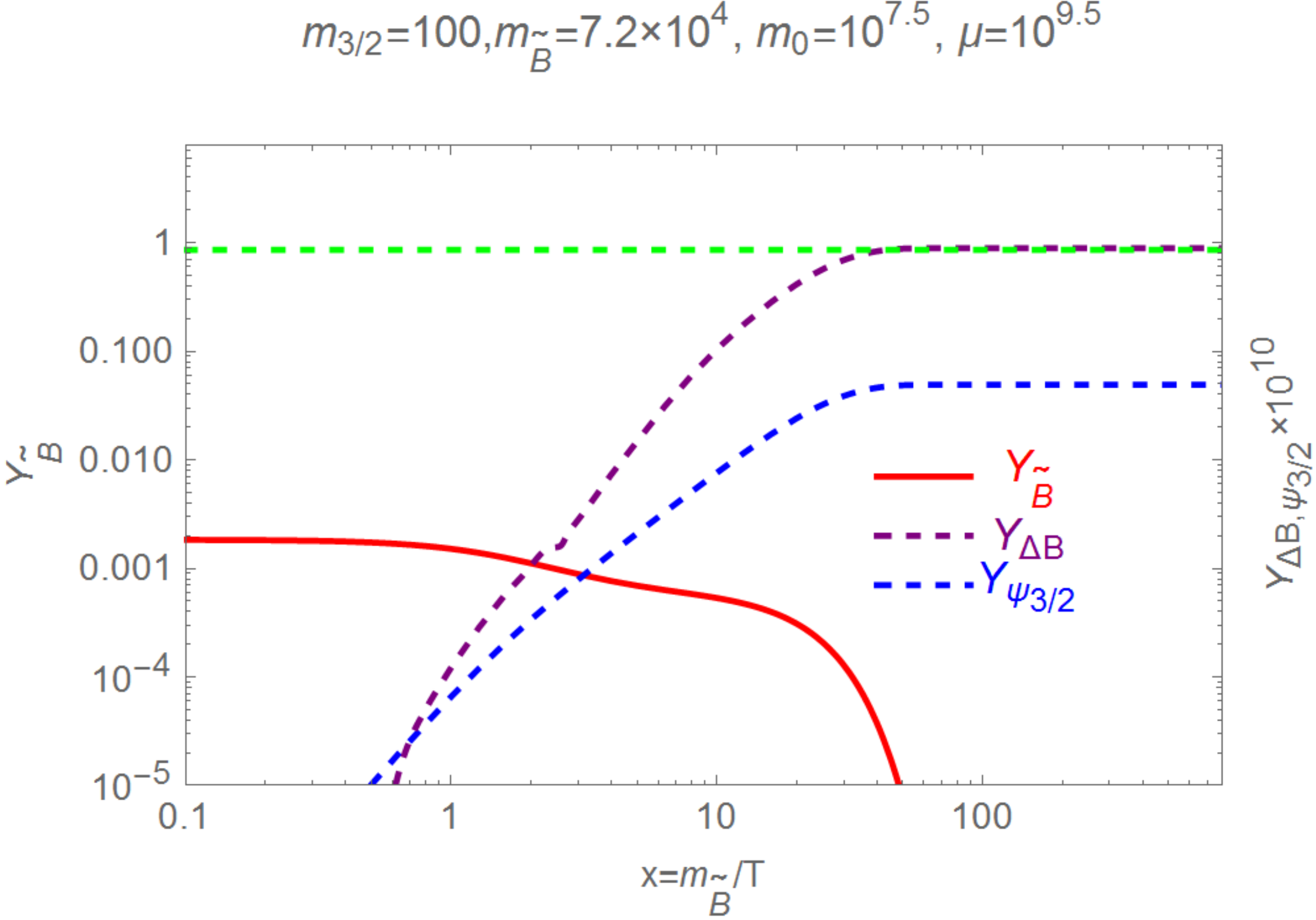}}
\subfloat{\includegraphics[width=7.5 cm]{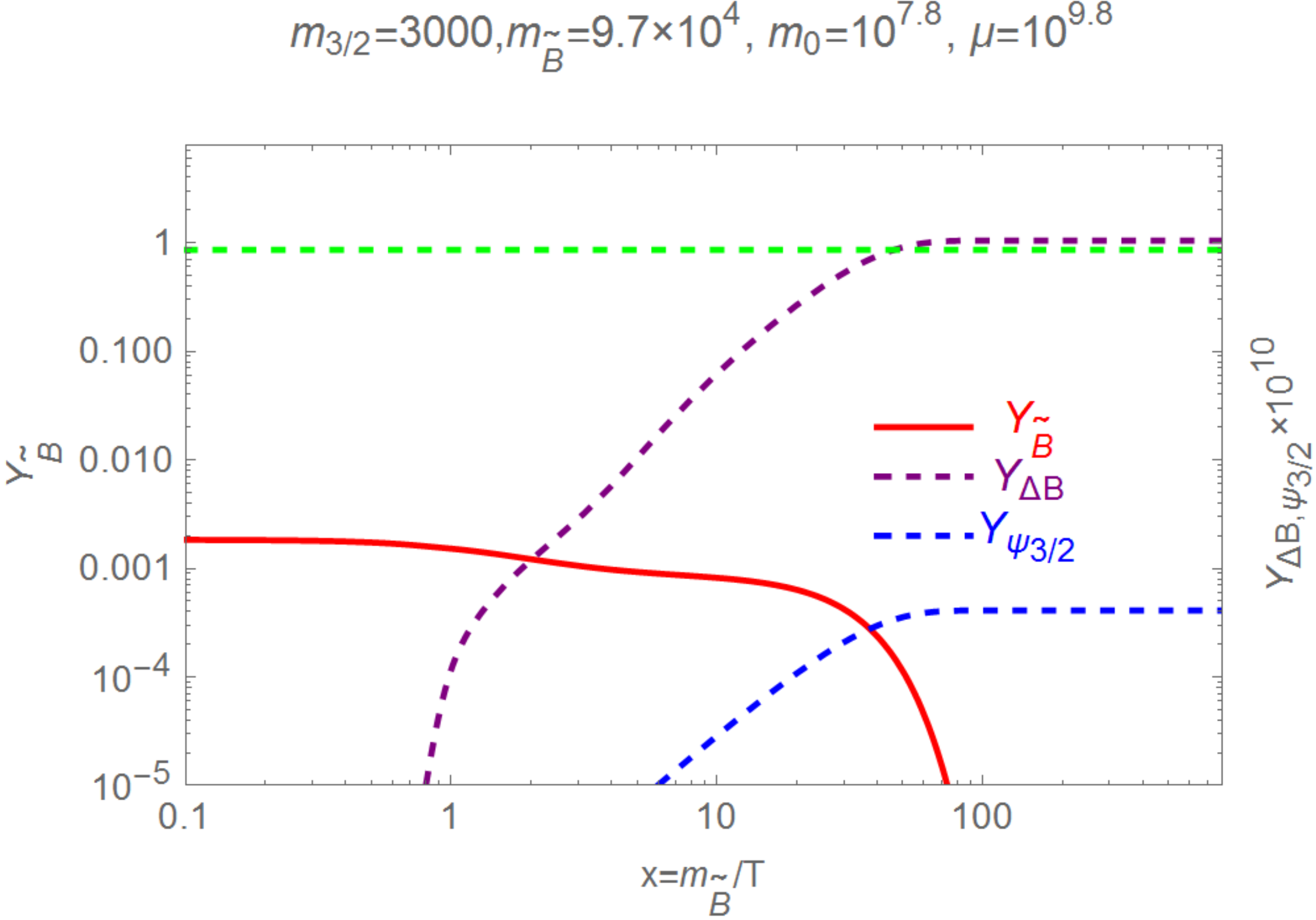}}
\end{center}
\caption{\footnotesize{Two benchmarks featuring the correct ratio between the DM and baryon abundances, as well as the correct agreement of the individual quantities with the experimental determination. In both cases the DM and the baryon asymmetry are produced by the out-of-equilibrium decay of a semi-relativistically decoupling Bino.}}
\label{fig:bplusg}
\end{figure}

\noindent
As shown in fig.~(\ref{fig:bplusg}), the optimal benchmarks highlighted in fig.~(\ref{fig:exp}) (namely the gray and purple lines), correspond to a contemporary production of the DM and of the baryon asymmetry from the out-of-equilibrium decay of the Bino with the latter featuring a substantially semi-relativistic decoupling. We also notice that the yield $Y_{3/2}$ of the DM is sensitively lower than the one of the baryons but it is compensated by the much higher mass, with respect to the one of the proton, such that the relic density results bigger, as expected. 

\noindent
The result obtained is sensitively different with respect to the scenario proposed in~\cite{Arcadi:2013jza}, consisting in accommodating the correct value of $\Omega_{\Delta B}/\Omega_{\rm DM}$ through similar values of $\epsilon_{\rm CP}$ and of the branching fraction of the mother particle into DM and, accordingly, similar values of the gravitino and proton masses. The reason of this resides in the more accurate determination of the abundance of the Bino as well as of $\epsilon_{\rm CP}$, in particular the inclusion of the kinematic functions $f_{1,2}$, and especially in the impact of wash-out processes. The combination of these effects leads to the need of a heavier supersymmetric spectrum
to achieve successful baryogenesis and therefore a substantially heavier gravitino is required to compensate the suppression of the Bino branching fraction into DM.
Indeed, the quantities $\epsilon_{\rm CP}$ and $BR\left(\tilde{B} \rightarrow \tilde{\psi}_{3/2}+X\right)$ differ in the allowed window by a few orders of magnitude 
and the similarity between DM and baryon densities cannot be directly related to their near-equality.
At the same time we remark that we could achieve a viable scenario in which the correct amounts of baryon asymmetry and of DM are contemporary produced by the 
decay of the Bino at relatively low value of the reheating temperature. 
Our predictions depend, apart a single assumption on the cosmological history of the Universe, on the masses of the superpartners, and we were able to identify rather 
definite ranges for the supersymmetric particles masses, in particular the gauginos. 

\noindent
Before concluding this section we just mention that an extensive exploration of the space of the parameters (in particular the flavor-mixing matrices $\Gamma_D$) involved in the generation of the baryon and DM densities is rather complex. Although we have found, in our analysis, that, as discussed above, there is no particular loss of generality in assuming contribution from only degenerate d-squarks to the relevant processes, we cannot completely exclude the presence of configurations, possibly involving also contributions from u-squarks, leading to $O(1)$ variations, with respect to the results presented here, might occur. We notice in particular from fig.~(\ref{fig:rainbows}) that an enhancement of a factor $2-3$ of the CP asymmetry, not compensated by annihilation or wash-out effects, would allow for smaller masses of the Bino and the Gluino, 
below 10 TeV, with the latter possible even lying in the LHC production range.

\section{Detection prospects}

\noindent
In this section we will briefly investigate possible experimental signatures  of this scenario and bounds associated to them. The main experimental signature of our scenario 
is the Indirect Detection of the decays of the gravitino DM.  Indeed due to the RPV coupling $\lambda$ , the gravitino has a three-body decay into SM quarks, possibly 
leading to signatures in the antiproton spectrum, with a rate~\cite{Moreau:2001sr}:
\begin{equation}
\Gamma\left(\tilde{\psi}_{3/2} \rightarrow u_k d_i d_j\right)= N_c\frac{\lambda^2}{6144 \pi^3}\frac{m_{3/2}^7}{m_0^4 M_{\rm Pl}^2}
\end{equation}
with $N_c$ being the number of channels giving a lifetime:
\begin{equation}
\tau_{3/2} \approx \frac{4.6}{N_c} \times 10^{28}\mbox{s} {\left(\frac{\lambda}{0.4}\right)}^{-2} {\left(\frac{m_0}{10^{7.5} \mbox{GeV}}\right)}^4 {\left(\frac{m_{3/2}}{1\mbox{TeV}}\right)}^{-7}
\end{equation}
Interestingly, for values of $m_{3/2}$ and $m_0$ of, respectively, $1\,\mbox{TeV}$ and $10^{7.5}\,\mbox{GeV}$, which provide the correct fit of the DM and baryon abundances, 
a DM lifetime of approximately $10^{28}\,\mbox{s}$ is achieved, which is exactly of the order of the current AMS-02 
sensitivity in the antiproton channel~\cite{Giesen:2015ufa,Hamaguchi:2015wga} and thus allows to test in the very next future our scenario. The decay into quarks of the gravitino can give rise as well to a sizable signal in $\gamma$-rays. Similar sensitivities, 
to the one discussed for AMS-02, are expected for $\gamma$-ray detectors like H.E.S.S. and CTA~\cite{Cirelli:2012ut}.
\noindent
The heavy supersymmetric spectrum does not offer, instead, very promising prospects for collider detection. The scenario proposed requires possibly a 
supersymmetric spectrum beyond the kinematical reach of LHC while the Gluino NLSP could be within the expected reach of a 100 TeV collider~\cite{Cohen:2013xda,Acharya:2014pua,Gori:2014oua}. As mentioned above, within the factor one uncertainty of our computations, we cannot exclude the possibility of having a slight
enhancement of the CP violating parameter $\epsilon_{\rm CP} $ allowing for viable baryogenesis and DM production in regions of parameter space with a 
lighter supersymmetric spectrum. In case of a mass of $m_{\tilde{B}} \lesssim 10\,\mbox{TeV}$ it would be possible to observe the Gluino NLSP at the LHC. 
In our scenario its main decay processes would be mediated by the RPV coupling $\lambda$ with typical decay length:
\begin{equation}
c \tau_{\tilde{g}} \approx \frac{2.75}{N_c} \mbox{m}  {\left(\frac{\lambda}{0.4}\right)}^{-2} {\left(\frac{m_0}{10^7 \mbox{GeV}}\right)}^4 {\left(\frac{m_{\tilde{G}}}{2\mbox{TeV}}\right)}^{-5}
\end{equation}
corresponding to displaced vertices or, most probably, a detector stable state. The prospects of detection can be inferred using the techniques discussed, 
for example, in~\cite{Covi:2014fba,Arcadi:2014tsa,Cui:2014twa}. The detection of EW gauginos requires anyway next future, higher center of mass energy 
facilities~\cite{Cohen:2013xda,Acharya:2014pua,Gori:2014oua}.

\noindent
We have not discussed here any particular flavour structure of the RPV couplings $\lambda$, but in principle those couplings can also contribute to flavour-violating
neutral current processes, as well as $B$-violating processes (other then proton decay) like neutron-antineutron oscillations and $\Delta B=2$ transitions (see e.g.~\cite{Barbier:2004ez} for an extensive discussion.). Since in our case we need a very large scale for the scalar quark partners, even if we need a large coupling, those rates remain well below the present limits and will be difficult to reach also in the future.

\section{Conclusions}

\noindent
We have presented a systematic approach for determining the contemporary production of the DM and the baryon asymmetry from the out-of-equilibrium decay of the 
same mother particle in a MSSM framework. These two quantities have been numerically determined through the solution of a system of coupled Boltzmann equations, 
accurately computing the abundance of the decaying state and taking, in particular, in account of the impact of wash-out processes. We have supported, whenever possible, 
our numerical results with analytical estimates.
\noindent
We have determined the ranges of the values of the relevant supersymmetric parameters which allow for an efficient production of the DM and baryon abundances. 
In the most simple (but rather general) limit of only right-handed d-squarks involved in the generation of the baryon and DM densities, the observed ranges for
those quantities are met for a value of the mass of the decaying Bino of 50-100 TeV, a mass of the gluino NLSP of 20-50 TeV and a mass for the DM gravitino 
LSP between 100 GeV and a few TeV, and all the other supersymmetric particles above the scale of $10^{7}\,\mbox{GeV}$ and not present in the primordial 
Universe because of the assumption of a lower reheating temperature, in order to avoid the overproduction of the DM. 
But note that a slight increase of the $ \epsilon_{\rm CP} $ parameter by a factor 2 or so, due to the presence e.g. of intermediate up-squarks,
if not completely compensate by an increase in the wash-out processes or the Bino annihilation rate, could allow to reduce the supersymmetric masses by 
a factor of a few.

\noindent 
We find moreover that the similarity of the DM and baryon densities cannot be explained by the relation 
$ \epsilon_{\rm CP} \sim BR\left(\tilde B \rightarrow DM+\mbox{anything} \right) $ and the gravitino mass has to be tuned to give the correct DM abundance. 
Nevertheless the common generation of baryon and DM density from the Bino neutralino after freeze-out can work and provide the right abundances for 
large values of the RPV coupling and in cosmologies with low reheating temperature. 

\noindent
The very heavy supersymmetric spectrum does not offer promising detection prospects at the LHC, but the Gluino LSP could be within the reach of a 100 TeV 
collider.  On the other hand a very promising signal in the near future might come from the decay of the gravitino whose lifetime can be within the present 
sensitivity of AMS-02 and gamma-ray detectors.

\acknowledgements

\noindent
We are grateful to Federico Mescia for the useful discussions.
\noindent
G.A. acknowledges support from the ERC advanced grant Higgs@LHC.
G.A. and L.C. acknowledge partial support from the European Union FP7 ITN-INVISIBLES (Marie Curie Actions, PITN-GA-2011-289442).

\appendix

\section{General expressions for the CP asymmetries}

\noindent
In this appendix we will provide some general expressions, including the dependence on the flavor matrices $\Gamma^{U,D}_{R,L}$ of the CP-asymmetry possibly 
originating the baryon abundance in our setup.
\noindent
As shown by our analysis the baryon asymmetry is mainly generated by the Bino decay. For simplicity we will just focus on the CP asymmetry in the decay processes. The computation can be actually straightforwardly extended to the $2 \rightarrow 2$ scatterings since the corresponding rates are cross-symmetric to the decay ones.

\noindent
As already mentioned in the main text the CP asymmetry is defined as:
\begin{equation}
\epsilon_{\rm CP} \equiv \frac{\Gamma\left(\tilde B \rightarrow u_k d_i d_k\right)-\Gamma\left(\tilde B \rightarrow \overline{u}_k \overline{d}_i \overline{d}_k\right)}{\Gamma\left(\tilde B \rightarrow u_k d_i d_k\right)+\Gamma\left(\tilde B \rightarrow \overline{u}_k \overline{d}_i \overline{d}_k\right)}
\end{equation}
The total CP asymmetry is given by the sum of the single CP-asymmetries in the different channels, weighted by the correspondent branching rations.
As well known a non-zero CP asymmetry requires the interference of tree and loop level contributions. In the scenario under considerations the relevant decays of the Bino are three-body processes in three SM quarks (un up-type quark and two d-type quarks) or a gluino and a SM quark pair (the two body decay into DM is irrelevant for the generation of the baryon asymmetry). These processes are mediated by down and up type squarks.

\noindent
For the case of the decay into only SM states we have that:
\begin{align}
& \Gamma\left(\tilde{B}\rightarrow u_k d_i d_k+\tilde{B}\rightarrow \overline{u}_k \overline{d}_i \overline{d}_k\right)= \frac{1}{128 \pi^3} g_1^2 m_{\tilde{B}}^5\sum_{\alpha,\beta}\left\{\frac{1}{m_{\tilde{q}_\alpha}^4}|\lambda_{lij}|^2\left(Q_u^2|\Gamma^U_{R\alpha i}\Gamma^{U\,*}_{R\alpha l}|^2\right. \right.\nonumber\\
& \left. \left.+(Q_u-T_3)^2 |\Gamma^U_{L\alpha i}\Gamma^{U\,*}_{R\alpha l}|^2\right)\right. \nonumber\\
&\left. +\frac{1}{m_{\tilde{q}_\alpha}^4}|\lambda_{klj}|^2\left(Q_d^2|\Gamma^D_{R\alpha i}\Gamma^{D\,*}_{R\alpha l}|^2+(Q_d-T_3)^2 |\Gamma^D_{L\alpha i}\Gamma^{D\,*}_{R\alpha l}|^2\right)\right.\nonumber\\
& \left. +\frac{1}{m_{\tilde{q}_\beta}^4}|\lambda_{kil}|^2\left(Q_d^2|\Gamma^D_{R\beta j}\Gamma^{D\,*}_{R\beta l}|^2+(Q_d-T_3)^2 |\Gamma^D_{L\beta j}\Gamma^{D\,*}_{R\beta l}|^2\right)   \right. \nonumber \\
& \left. - \frac{1}{m_{\tilde{q}_\alpha}^2m_{\tilde{q}_\beta}^2}\lambda_{lij} \lambda_{kpj}^{*} Q_u Q_d\left(\Gamma^U_{R\alpha l}\Gamma^{U\,*}_{R\alpha k}\right) \left(\Gamma^D_{R\beta p}\Gamma^{D\,*}_{R\beta i}\right) \right.\nonumber\\
& \left. - \frac{1}{m_{\tilde{q}_\alpha}^2m_{\tilde{q}_\beta}^2}\lambda_{lij} \lambda_{kip}^{*} Q_u Q_d\left(\Gamma^U_{R\alpha l}\Gamma^{U\,*}_{R\alpha k}\right) \left(\Gamma^D_{R\beta p}\Gamma^{D\,*}_{R\beta j}\right) \right.\nonumber\\
& \left. - \frac{1}{m_{\tilde{q}_\alpha}^2m_{\tilde{q}_\beta}^2}\lambda_{kip} \lambda_{klj}^{*} Q_d^2\left(\Gamma^D_{R\alpha l}\Gamma^{D\,*}_{R\alpha i}\right) \left(\Gamma^D_{R\beta p}\Gamma^{D\,*}_{R\beta j}\right)\right] 
\end{align} 
In the case that the squark matrices are diagonal and the down-type quark are sensitively lighter than up-type quarks we have that:
\begin{equation}
\Gamma\left(\tilde{B}\rightarrow u_k d_i d_k+\tilde{B}\rightarrow \overline{u}_k \overline{d}_i \overline{d}_k\right)=\frac{|\lambda_{kij}|^2 g_1^2 Q_d^2}{128 \pi^3}\frac{m_{\tilde{B}}^5}{m_0^4}
\end{equation}
The other relevant rate at the tree level is the one of decay into Gluino.
It is given by:
\begin{equation}
\Gamma\left(\tilde{B} \rightarrow \tilde{G} q \overline{q} \right)= \frac{m_{\tilde{B}}^5}{1024 \pi} \left[\left(\left(\mathcal{C}_{1,u}+\mathcal{C}_{1,d}\right)-1/2 \left(\mathcal{C}_{2,u}+\mathcal{C}_{2,d}\right)\right) f_2\left(\frac{m_{\tilde{G}}^2}{m_{\tilde{B}}^2}\right)+2 \frac{m_{\tilde{G}}}{m_{\tilde{B}}}\left(\mathcal{C}_{3,u}+\mathcal{C}_{3,d}\right)f_3\left(\frac{m_{\tilde{G}}^2}{m_{\tilde{B}}^2}\right)\right]
\end{equation}
where:
\begin{align}
& \mathcal{C}_{1,q}=\sum_{l} \left \{ |g_{\tilde{B}}^{\rm LL}|^2 |g_{\tilde{G}}^{\rm LL}|^2 |\frac{\Gamma^{q\,*}_{Lli}\Gamma^{q}_{Llj}}{m^2_{\tilde{q}_l}}|^2+|g_{\tilde{B}}^{\rm LL }|^2 |g_{\tilde{G}}^{\rm RR}|^2 |\frac{\Gamma^{q\,*}_{Lli}\Gamma^{q}_{Rlj}}{m^2_{\tilde{q}_l}}|^2+|g_{\tilde{B}}^{\rm RR}|^2 |g_{\tilde{G}}^{\rm LL}|^2 |\frac{\Gamma^{q\,*}_{Rli}\Gamma^{q}_{Llj}}{m^2_{\tilde{q}_l}}|^2\right.\nonumber\\
& \left. +|g_{\tilde{B}}^{\rm RR}|^2 |g_{\tilde{G}}^{\rm RR}|^2 |\frac{\Gamma^{q\,*}_{Rli}\Gamma^{q}_{Rlj}}{m^2_{\tilde{q}_l}}|^2\right \}\nonumber\\
& \mathcal{C}_{2,q}=\sum_{l,p}\frac{1}{m^2_{\tilde{q}_l}}\frac{1}{m^2_{\tilde{q}_p}}  Re\left \{ \left(g^{\rm LL\,*}_{\tilde{B}}g^{\rm RR\,*}_{\tilde{B}}g^{\rm RR}_{\tilde{G}}g^{\rm LL}_{\tilde{G}}\Gamma^{q\,*}_{\rm Llj}\Gamma^{\,* q}_{\rm Rpi}\Gamma^{q}_{\rm Rli}\Gamma^{q}_{\rm Lpj}+g^{\rm RR\,*}_{\tilde{B}}g^{\rm LL\,*}_{\tilde{B}}g^{\rm LL}_{\tilde{G}}g^{\rm RR}_{\tilde{G}}\Gamma^{q\,*}_{\rm Rlj}\Gamma^{q\,*}_{\rm Lpi}\Gamma^{q}_{\rm Lli}\Gamma^{q}_{\rm Rpj}\right)\right \}\nonumber\\
& \mathcal{C}_{3,q}= \sum_{l,p}\frac{1}{m^2_{\tilde{q}_l}}\frac{1}{m^2_{\tilde{q}_p}}  Re\left \{ \left(g^{\rm LL\,*}_{\tilde{B}}g^{\rm LL\,*}_{\tilde{B}}g^{\rm LL}_{\tilde{G}}g^{\rm LL}_{\tilde{G}}\Gamma^{q\,*}_{\rm Llj}\Gamma^{q\,*}_{\rm Lpi}\Gamma^{q}_{\rm Lli}\Gamma^{q}_{\rm Lpj}+g^{\rm RR\,*}_{\tilde{B}}g^{\rm RR\,*}_{\tilde{B}}g^{\rm RR}_{\tilde{G}}g^{\rm RR}_{\tilde{G}}\Gamma^{q\,*}_{\rm Rlj}\Gamma^{q\,*}_{\rm Rpi}\Gamma^{q}_{\rm Rli}\Gamma^{q}_{\rm Rpj}\right)\right \}
\end{align}

\noindent
We can now move to compute the CP asymmetry. Parametrizing the loop amplitude as $A_{\rm loop} F_{\rm loop}$ where $A$ is a numerical coefficient depending on the coupling and the effective CP-phase while $F$ is a suitable loop integral we have that:
\begin{equation}
\Delta \Gamma \equiv \Gamma\left(\tilde{B}\rightarrow u_k d_i d_k\right)-\Gamma\left(\tilde{B}\rightarrow \overline{u}_k \overline{d}_i \overline{d}_k\right)= 4 Im(A^{*}_{\rm tree} A_{\rm loop})Im(F_{\rm loop})
\end{equation}

\noindent
In order to properly identify the different contributions to the CP-asymmetry, in particular in relation to the flavor structure, we have performed our computation using the fully Supersymmetric Lagrangian, rather than~(\ref{eq:asymmetry_lagrangian}), and performed at the end of the computations the limit $m_{\tilde{q}} \gg m_{\tilde{B}},m_{\tilde{G}}$.

\noindent
In the most general case the CP asymmetry originates from a combination of several tree level and loop diagrams. We have first of all the diagrams in which only d-type squarks are exchanged. These have been reported in fig.~(\ref{fig:treediagrams}) and~\ref{fig:loopdiagrams} and consist in two tree-level diagrams, labeled as $T1$ and $T2$ and 4 loop diagrams, labeled L1, L2, L3 and L4, in which two d-squarks are exchanged. We have then a tree level diagram and loop diagram with the same topology as, respectively, T1 and L3, but with up-type squarks exchanged. We have finally diagrams with, again the same topology, as T1-T4, but with exchange of one up-squark and one d-squark. As already argued all the possible topologies of diagrams are already accounted by the case of exchange of only down-type quarks. For this reason we will focus on this case since the remaining contribution can be straightforwardly obtained from the expressions presented.
\noindent
We show below the values of the decay asymmetry originating from the combinations T1L1, T2L1, T3L1 and T3L2. The other combinations are obtained from this by exchanging the flavor indices of the two final state d-quarks.

\subsection{T1L1}

\begin{align}
& \mathcal{M}_{\rm T1L1} = - \frac{1}{8}c_f \sum_{\alpha \beta \gamma}\sum_{l p n} \frac{1}{m_{\tilde{q}_\alpha}^2}\frac{1}{m_{\tilde{q}_\beta}^2}Im\left[C_2 \lambda_{knj}^{*}\lambda_{kpl}\left(g^{RR\,*}_{\tilde{B}}g^{LL}_{\tilde{B}}g^{LL,\*}_{\tilde{G}}g^{RR}_{\tilde{G}}\Gamma^{D\,*}_{R\alpha i} \Gamma^{D}_{L\alpha n}\Gamma^{D\,*}_{R\gamma p}\Gamma^{D}_{R\gamma j}\Gamma^{D}_{L\beta i}\Gamma^{D\,*}_{R\beta l} \right. \right. \nonumber \\
& \left. \left. +g^{RR\,*}_{\tilde{B}}g^{RR}_{\tilde{B}}g^{RR\,*}_{\tilde{G}}g^{RR}_{\tilde{G}}\Gamma^{D\,*}_{R\alpha i} \Gamma^{D}_{R\alpha n}\Gamma^{D\,*}_{R\gamma p}\Gamma^{D}_{R\gamma j}\Gamma^{D}_{R\beta i}\Gamma^{D\,*}_{R\beta l}  \right)\right] Im\left[I_1\left(m_{\tilde{G}},m_{\tilde{q}_\gamma},m_{\tilde{B}},x_i,x_k\right)\right] 
\end{align}
where:
\begin{equation}
I_1\left(m_{\tilde{G}},m_{\tilde{q}_\gamma},m_{\tilde{B}},x_i,x_k\right)=\int \frac{d^4 l}{{\left(2 \pi\right)}^4} \frac{Tr\left[\slashed{p}_i\slashed{p}_{\tilde{B}}\right]Tr\left[\slashed{p}_j \slashed{p}_k \left(\slashed{l}-\slashed{p}_k\right) \left(\slashed{l}+\slashed{p}_j\right) \right]}{\left[l^2-m_{\tilde{q}_\gamma}^2\right]{\left(l-p_k\right)}^2 \left[{\left(p_j+l\right)}^2-m_{\tilde{G}}^2\right]}
\end{equation}
The imaginary part of this integral, as well as the others appearing in the expressions below, can be computed with the Cutkosky formalism putting the internal quark (with 
four-momentum $l-p_k$) and gluino (with four-momentum $l+p_j$) on-shell. 
The integration over the phase space leads to:
\begin{align}
& \Delta \Gamma_{\rm T1L1} = -\frac{1}{128 \pi^4}c_f \sum_{\alpha \beta \gamma}\sum_{l p n} \frac{1}{m_{\tilde{q}_\alpha}^2}\frac{1}{m_{\tilde{q}_\beta}^2} Im\left[C_2 \lambda_{knj}^{*}\lambda_{klp}\left(g^{RR\,*}_{\tilde{B}}g^{LL}_{\tilde{B}}g^{LL}_{\tilde{G}}g^{RR\,*}_{\tilde{G}}\Gamma^{D\,*}_{R\alpha i} \Gamma^{D}_{L\alpha n}\Gamma^{D\,*}_{R\gamma p}\Gamma^{D}_{R\gamma j}\Gamma^{D}_{L\beta i}\Gamma^{D\,*}_{R\beta l} \right. \right. \nonumber \\
& \left. \left. +g^{RR\,*}_{\tilde{B}}g^{RR}_{\tilde{B}}g^{RR}_{\tilde{G}}g^{RR\,*}_{\tilde{G}}\Gamma^{D\,*}_{R\alpha i} \Gamma^{D}_{R\alpha n}\Gamma^{D\,*}_{R\gamma p}\Gamma^{D}_{R\gamma j}\Gamma^{D}_{R\beta i}\Gamma^{D\,*}_{R\beta l}  \right)\right]\frac{1}{120} \frac{m_{\tilde{B}}^6}{m_{\tilde{q}_\gamma}^2}f_1\left(\frac{m_{\tilde{G}}^2}{m_{\tilde{B}}^2}\right)
\end{align}

\subsection{T2L1}
\begin{align}
& \mathcal{M}_{\rm T2L1} = \frac{1}{4}c_f \sum_{\alpha \beta \gamma} \sum_{lpn} \frac{1}{m_{\tilde{q}_\alpha}^2}\frac{1}{m_{\tilde{q}_\beta}^2} Im\left[\lambda_{kni}^{*}\lambda_{kpl} C_2 g^{RR}_{\tilde{B}} g^{RR\,*}_{\tilde{B}}g^{RR\,*}_{\tilde{G}}g^{RR}_{\tilde{G}}\Gamma^{D\,*}_{R\alpha j}\Gamma^{D}_{R\alpha n}\Gamma^{D\,*}_{R\gamma p}\Gamma^{D}_{R\gamma j}\Gamma^{D\,*}_{R\beta l}\Gamma^{D}_{R\beta i}\right] \nonumber\\
& Im\left[I_2\left(m_{\tilde{G}},m_{\tilde{B}},m_{\tilde{q}_\gamma},x_i,x_k\right)\right]  \nonumber \\
& + m_{\tilde{G}} m_{\tilde{B}} Im\left[\lambda_{kni}^{*}\lambda_{kpl} C_2 g^{RR}_{\tilde{B}} g^{LL\,*}_{\tilde{B}}g^{LL}_{\tilde{G}}g^{RR\,*}_{\tilde{G}}\Gamma^{D\,*}_{R\alpha j}\Gamma^{D}_{L\alpha n}\Gamma^{D\,*}_{L\gamma p}\Gamma^{D}_{L\gamma j}\Gamma^{D\,*}_{R\beta l}\Gamma^{D}_{R\beta i}\right]\nonumber\\
& Im\left[I_3\left(m_{\tilde{G}},m_{\tilde{B}},m_{\tilde{q}_\gamma},x_i,x_k\right)\right]
\end{align} 
where:
\begin{align}
& I_2\left(m_{\tilde{G}},m_{\tilde{q}_\gamma},m_{\tilde{B}},x_i,x_k\right)=\int \frac{d^4 l}{{\left(2 \pi\right)}^4} \frac{Tr\left[\slashed{p}_j \slashed{p}_i \slashed{p}_k \left(\slashed{l}-\slashed{p}_k\right) \right]}{\left[l^2-m_{\tilde{q}_\gamma}^2\right]{\left(l-p_k\right)}^2 \left[{\left(p_j+l\right)}^2-m_{\tilde{G}}^2\right]} \nonumber\\
& I_3\left(m_{\tilde{G}},m_{\tilde{q}_\gamma},m_{\tilde{B}},x_i,x_k\right)=\int \frac{d^4 l}{{\left(2 \pi\right)}^4} \frac{Tr\left[\slashed{p}_j \slashed{p}_{\tilde{B}} \slashed{p}_{i} \slashed{p}_k \left(\slashed{l}-\slashed{p}_k\right) \left(\slashed{l}+\slashed{p}_j\right) \right]}{\left[l^2-m_{\tilde{q}_\gamma}^2\right]{\left(l-p_k\right)}^2 \left[{\left(p_j+l\right)}^2-m_{\tilde{G}}^2\right]}
\end{align}    
The integration over the phase space leads to:
\begin{align}
& \Delta \Gamma_{T2L1}=-\frac{1}{128\pi^4}c_f \sum_{\alpha \beta \gamma} \sum_{l p n} \frac{1}{m_{\tilde{q}_\alpha}^2}\frac{1}{m_{\tilde{q}_\beta}^2} Im\left[\lambda_{kni}^{*}\lambda_{kpl} C_2 g^{RR}_{\tilde{B}} g^{RR\,*}_{\tilde{B}}g^{RR\,*}_{\tilde{G}}g^{RR}_{\tilde{G}}\Gamma^{D\,*}_{R\alpha j}\Gamma^{D}_{R\alpha n}\Gamma^{D\,*}_{R\gamma p}\Gamma^{D}_{R\gamma j}\Gamma^{D\,*}_{R\beta l}\Gamma^{D}_{R\beta i}\right]\nonumber\\
& \frac{1}{480} \frac{m_{\tilde{B}}^7}{m_{\tilde{q}_\gamma}^2}f_1\left(\frac{m_{\tilde{G}}^2}{m_{\tilde{B}}^2}\right)
\end{align}
This and the previous expression depend on the modulus square of the effective gauge couplings $g_{\tilde{G}}$ and $g_{\tilde{B}}$. A non null CP asymmetry from the corresponding diagrams might arise only in presence of flavor violation and CP violating phases in the $\Gamma^D_{R,L}$ matrices.

\subsection{T1L3}

\noindent
The contribution to the CP-asymmetry associated to this topology is:
\begin{align}
& \mathcal{M}_{\rm T1L3}=-\frac{1}{8}c_f \sum_{\alpha\,\beta\,\gamma}\sum_{l,p,n} \frac{1}{m_{\tilde{q}_\alpha}^2}\frac{1}{m_{\tilde{q}_\gamma}^2}Im\left[ \lambda_{kpj}\lambda^{*}_{knj}C_2\left(g^{LL\,*}_{\tilde{B}}g^{RR\,*}_{\tilde{B}}g^{LL}_{\tilde{G}}g^{RR}_{\tilde{G}}\Gamma^D_{R\alpha n}\Gamma^{D\,*}_{L\alpha i}\Gamma^{D\,*}_{R\beta l}\Gamma^D_{L\beta i}\Gamma^{D\,*}_{R\gamma p}\Gamma^D_{R\gamma l} \right. \right. \nonumber \\ 
& \left. \left. +g^{RR\,*}_{\tilde{B}}g^{LL\,*}_{\tilde{B}}g^{LL}_{\tilde{G}}g^{RR}_{\tilde{G}}\Gamma^D_{R\alpha n}\Gamma^{D\,*}_{R\alpha i}\Gamma^{D\,*}_{L\beta l}\Gamma^D_{R\beta i}\Gamma^{D\,*}_{R\gamma p}\Gamma^D_{L\gamma l}\right)\right] Im\left[I_4 \left(m_{\tilde{G}}^2,m_{\tilde{q}_\beta}^2,m_{\tilde{B}}^2,x_i,x_k\right)\right] \nonumber \\
& +m_{\tilde{B}}m_{\tilde{G}} \frac{1}{m_{\tilde{q}_\alpha}^2}\frac{1}{m_{\tilde{q}_\gamma}^2}Im\left[\lambda_{kpj}\lambda^{*}_{knj}C_2\left(g^{LL\,*}_{\tilde{B}}g^{LL\,*}_{\tilde{B}}g^{LL}_{\tilde{G}}g^{LL}_{\tilde{G}}\Gamma^D_{L\alpha n}\Gamma^{D\,*}_{R\alpha i}\Gamma^{D\,*}_{L\beta l}\Gamma^D_{L\beta i}\Gamma^{D\,*}_{R\gamma p}\Gamma^D_{L\gamma l} \right. \right. \nonumber\\ 
& \left. \left. +g^{RR\,*}_{\tilde{B}}g^{RR\,*}_{\tilde{B}}g^{LL}_{\tilde{G}}g^{RR}_{\tilde{G}}\Gamma^D_{R\alpha n}\Gamma^{D\,*}_{R\alpha i}\Gamma^{D\,*}_{R\beta l}\Gamma^D_{R\beta i}\Gamma^{D\,*}_{R\gamma p}\Gamma^D_{R\gamma l}\right)\right] Im\left[I_5 \left(m_{\tilde{G}}^2,m_{\tilde{q}_\beta}^2,m_{\tilde{B}}^2,x_i,x_k\right)\right]
\end{align}
where $C_2=4/3$ is a color factor arising from the coupling of the Gluino and $x_i=2 E_i/m_{\tilde{B}}$. As we notice the effective couplings of the Bino appear with the same conjugation opposite to the one of the coupling of the Gluino. The Maiorana phases are thus enough to originate a non null CP-asymmetry with effective phase $\phi=2\left(\phi_{g}-\phi_B\right)$. In this case however a suppression factor $m_{\tilde{G}}/m_{\tilde{B}}$ is however present. $I_4$ and $I_5$ are the loop integrals defined as:
\begin{align}
& I_4=\int \frac{d^4 l}{{\left(2 \pi\right)}^4} \frac{Tr\left[\slashed{p}_i \left(\slashed{l}-\slashed{q}\right)\right]Tr\left[\slashed{p}_j \slashed{p}_k\right]}{\left[l^2-m_{\tilde{G}}^2\right]{\left(l-q\right)}^2 \left[{\left(p_i+l\right)}^2-m_{\tilde{q}_\beta}^2\right]}\nonumber \\
& I_5=\int \frac{d^4 l}{{\left(2 \pi\right)}^4} \frac{Tr\left[\slashed{p}_i \slashed{p}_{\tilde{B}}\left(\slashed{l}-\slashed{q}\right)\slashed{l}\right]Tr\left[\slashed{p}_j \slashed{p}_k\right]}{\left[l^2-m_{\tilde{G}}^2\right]{\left(l-q\right)}^2 \left[{\left(p_i+l\right)}^2-m_{\tilde{q}_\beta}^2\right]}
\end{align}
Performing as well the integrations over the phase space it is possible to obtain, in the limit $m_{\tilde{q}_\beta} \gg m_{\tilde{B}}$:
\begin{align}
& \Delta \Gamma_{\rm T1L3}=\frac{1}{128 \pi^4} c_f \sum_{\alpha \beta \gamma} \sum_{lpn} \frac{1}{m_{\tilde{q}_\alpha}^2}\frac{1}{m_{\tilde{q}_\gamma}^2}Im\left[\lambda_{kpj}\lambda^{*}_{knj}C_2\left(g^{LL\,*}_{\tilde{B}}g^{RR\,*}_{\tilde{B}}g^{LL}_{\tilde{G}}g^{RR}_{\tilde{G}}\Gamma^D_{R\alpha n}\Gamma^{D\,*}_{L\alpha i}\Gamma^{D\,*}_{R\beta l}\Gamma^D_{L\beta i}\Gamma^{D\,*}_{R\gamma p}\Gamma^D_{R\gamma l} \right. \right. \nonumber \\ 
& \left. \left. +g^{RR\,*}_{\tilde{B}}g^{LL\,*}_{\tilde{B}}g^{LL}_{\tilde{G}}g^{RR}_{\tilde{G}}\Gamma^D_{R\alpha n}\Gamma^{D\,*}_{R\alpha i}\Gamma^{D\,*}_{L\beta l}\Gamma^D_{R\beta i}\Gamma^{D\,*}_{R\gamma p}\Gamma^D_{L\gamma l}\right)\right] \frac{1}{480}\frac{m_{\tilde{B}}^7}{m_{\tilde{q}_\beta}^2}f_1\left(\frac{m_{\tilde{G}}^2}{m_{\tilde{B}}^2}\right) \nonumber \\
& + \frac{1}{m_{\tilde{q}_\alpha}^2}\frac{1}{m_{\tilde{q}_\gamma}^2}Im\left[\lambda_{kpj}\lambda^{*}_{knj}C_2\left(g^{LL\,*}_{\tilde{B}}g^{LL\,*}_{\tilde{B}}g^{LL}_{\tilde{G}}g^{LL}_{\tilde{G}}\Gamma^D_{L\alpha i}\Gamma^{D\,*}_{R\alpha n}\Gamma^{D\,*}_{L\beta l}\Gamma^D_{L\beta i}\Gamma^{D\,*}_{R\gamma p}\Gamma^D_{L\gamma l} \right. \right. \nonumber\\ 
& \left. \left. +g^{RR\,*}_{\tilde{B}}g^{RR\,*}_{\tilde{B}}g^{RR}_{\tilde{G}}g^{RR}_{\tilde{G}}\Gamma^D_{R\alpha n}\Gamma^{D\,*}_{R\alpha i}\Gamma^{D\,*}_{R\beta l}\Gamma^D_{R\beta i}\Gamma^{D\,*}_{R\gamma p}\Gamma^D_{R\gamma l}\right)\right] \frac{1}{192} \frac{m_{\tilde{B}}^6 m_{\tilde{G}}}{m_{\tilde{q}_\beta}^2}f_2\left(\frac{m_{\tilde{G}}^2}{m_{\tilde{B}}^2}\right)
\end{align}

\subsection{T2L3}

\noindent
This contribution is computed analogously to the previous one and results to be:
\begin{align}
& \mathcal{M}_{\rm T2L3}=\frac{1}{4}c_f \sum_{\alpha \beta \gamma}\sum_{lpn} \frac{1}{m_{\tilde{q}_\alpha}^2}\frac{1}{m_{\tilde{q}_\gamma}^2}Im\left[\lambda_{kpj}\lambda^{*}_{kni}C_2\left(
  g^{RR\,*}_{\tilde{B}}g^{LL\,*}_{\tilde{B}}g^{LL}_{\tilde{G}}g^{RR}_{\tilde{G}}\Gamma^D_{R\alpha n}\Gamma^{D\,*}_{R\alpha j}\Gamma^{D\,*}_{L\beta l}\Gamma^D_{R\beta i}\Gamma^{D\,*}_{R\gamma p}\Gamma^D_{L\gamma l}\right)\right]\nonumber\\
&	Im\left[I_6 \left(m_{\tilde{G}}^2,m_{\tilde{q}_\beta}^2,m_{\tilde{B}}^2,x_i,x_k\right)\right] \nonumber \\
& +m_{\tilde{B}}m_{\tilde{G}} \frac{1}{m_{\tilde{q}_\alpha}^2}\frac{1}{m_{\tilde{q}_\gamma}^2}Im\left[\lambda_{kpj}\lambda^{*}_{kni}C_2\left(g^{RR\,*}_{\tilde{B}}g^{RR\,*}_{\tilde{B}}g^{RR}_{\tilde{G}}g^{RR}_{\tilde{G}}\Gamma^D_{R\alpha n}\Gamma^{D\,*}_{R\alpha j}\Gamma^{D\,*}_{R\beta l}\Gamma^D_{R\beta i}\Gamma^{D\,*}_{R\gamma p}\Gamma^D_{R\gamma l}\right)\right] \nonumber\\
& Im\left[I_7 \left(m_{\tilde{G}}^2,m_{\tilde{q}_\beta}^2,m_{\tilde{B}}^2,x_i,x_k\right)\right]
\end{align}
\noindent
%This contribution has opposite sign with respect to $T1L3$ since we have exchanged one fermion line.
The loop integrals are now:
\begin{align}
& I_6=\int \frac{d^4 l}{{\left(2 \pi\right)}^4} \frac{Tr\left[\slashed{p}_j \left(\slashed{l}-\slashed{q}\right)\slashed{p}_i \slashed{p}_k\right]}{\left[l^2-m_{\tilde{G}}^2\right]{\left(l-q\right)}^2 \left[{\left(p_i+l\right)}^2-m_{\tilde{q}_\beta}^2\right]}\nonumber \\
& I_7=\int \frac{d^4 l}{{\left(2 \pi\right)}^4} \frac{Tr\left[\slashed{p}_j \slashed{p}_{\tilde{B}}\left(\slashed{l}-\slashed{q}\right)\slashed{l}\slashed{p}_i \slashed{p}_k\right]}{\left[l^2-m_{\tilde{G}}^2\right]{\left(l-q\right)}^2 \left[{\left(p_i+l\right)}^2-m_{\tilde{q}_\beta}^2\right]}
\end{align}
\noindent
Performing the suitable integrations one obtains:
\begin{align}
& \Delta \Gamma_{\rm T2L3}=-\frac{1}{128 \pi^4} c_f \sum_{\alpha \beta \gamma} \sum_{l p n}\frac{1}{m_{\tilde{q}_\alpha}^2}\frac{1}{m_{\tilde{q}_\gamma}^2}Im\left[ \lambda_{kpj}\lambda^{*}_{kni}C_2\left(
  g^{RR\,*}_{\tilde{B}}g^{LL\,*}_{\tilde{B}}g^{LL}_{\tilde{G}}g^{RR}_{\tilde{G}}\Gamma^D_{R\alpha n}\Gamma^{D\,*}_{R\alpha j}\Gamma^{D\,*}_{L\beta l}\Gamma^D_{R\beta i}\Gamma^{D\,*}_{R\gamma p}\Gamma^D_{L\gamma l}\right)\right]\nonumber\\
	&\frac{1}{960}\frac{m_{\tilde{B}}^7}{m_{\tilde{q}_\beta}^2}f_1\left(\frac{m_{\tilde{G}}^2}{m_{\tilde{B}}^2}\right) \nonumber \\
& +m_{\tilde{B}}m_{\tilde{G}} \frac{1}{m_{\tilde{q}_\alpha}^2}\frac{1}{m_{\tilde{q}_\gamma}^2}Im\left[\lambda_{kpj}\lambda^{*}_{kni}C_2\left(g^{RR\,*}_{\tilde{B}}g^{RR\,*}_{\tilde{B}}g^{RR}_{\tilde{G}}g^{RR}_{\tilde{G}}\Gamma^D_{R\alpha n}\Gamma^{D\,*}_{R\alpha j}\Gamma^{D\,*}_{R\beta l}\Gamma^D_{R\beta j}\Gamma^{D\,*}_{R\gamma p}\Gamma^D_{R\gamma l}\right)\right]\nonumber\\
&  \frac{1}{384}\frac{m_{\tilde{B}}^6 m_{\tilde{G}}}{m_{\tilde{q}_\beta}^2}f_2\left(\frac{m_{\tilde{G}}^2}{m_{\tilde{B}}^2}\right)
\end{align}

\section{General expressions annihilation cross-sections}
\label{sec:crs_computation}

\noindent
We provide in this appendix the complete expressions of the scattering cross-section of the Bino, including the flavor structure as well as the exchange of $u$-type squarks, involving SM fermions.
\noindent
These are:
\begin{align}
&\langle \sigma v \rangle\left(\tilde{B}u_k\rightarrow \overline{d}_i \overline{d}_j+\tilde{B}d_i \rightarrow \overline{u}_k d_j\right)= \frac{m_{\tilde{B}}^2}{64 \pi} \frac{1}{x^4 K_2(x)} \int_x^\infty dz (z^2-x^2) (5 z^2+x^2) \mathcal{A}_1 = \frac{m_{\tilde{B}}^2}{8 \pi} \mathcal{A}_1 \left(5 \frac{K_4(x)}{K_2(x)}+1\right)\nonumber\\
&\mathcal{A}_1=\sum_{\alpha,\beta}\left\{\frac{1}{m_{\tilde{q}_\alpha}^4}|\lambda_{lij}|^2\left(Q_u^2|\Gamma^U_{R\alpha i}\Gamma^{U\,*}_{R\alpha l}|^2+(Q_u-T_3)^2 |\Gamma^U_{L\alpha i}\Gamma^{U\,*}_{R\alpha l}|^2\right) \right. \nonumber\\
&\left. \frac{1}{m_{\tilde{q}_\alpha}^4}|\lambda_{klj}|^2\left(Q_d^2|\Gamma^D_{R\alpha i}\Gamma^{D\,*}_{R\alpha l}|^2+(Q_d-T_3)^2 |\Gamma^D_{L\alpha i}\Gamma^{D\,*}_{R\alpha l}|^2\right)\right.\nonumber\\
& \left. +\frac{1}{m_{\tilde{q}_\beta}^4}|\lambda_{kil}|^2\left(Q_d^2|\Gamma^D_{R\beta j}\Gamma^{D\,*}_{R\beta l}|^2+(Q_d-T_3)^2 |\Gamma^D_{L\beta j}\Gamma^{D\,*}_{R\beta l}|^2\right)   \right. \nonumber \\
& \left. - \frac{1}{m_{\tilde{q}_\alpha}^2m_{\tilde{q}_\beta}^2}\lambda_{lij} \lambda_{kpj}^{*} Q_u Q_d\left(\Gamma^U_{R\alpha l}\Gamma^{U\,*}_{R\alpha k}\right) \left(\Gamma^D_{R\beta p}\Gamma^{D\,*}_{R\beta i}\right) \right.\nonumber\\
& \left. - \frac{1}{m_{\tilde{q}_\alpha}^2m_{\tilde{q}_\beta}^2}\lambda_{lij} \lambda_{kip}^{*} Q_u Q_d\left(\Gamma^U_{R\alpha l}\Gamma^{U\,*}_{R\alpha k}\right) \left(\Gamma^D_{R\beta p}\Gamma^{D\,*}_{R\beta j}\right) \right.\nonumber\\
& \left. - \frac{1}{m_{\tilde{q}_\alpha}^2m_{\tilde{q}_\beta}^2}\lambda_{kip} \lambda_{klj}^{*} Q_d^2\left(\Gamma^D_{R\alpha l}\Gamma^{D\,*}_{R\alpha i}\right) \left(\Gamma^D_{R\beta p}\Gamma^{D\,*}_{R\beta j}\right)\right], 
\end{align}

\begin{align}
&\langle \sigma v \rangle \left(\tilde{B}\tilde{G} \rightarrow u \overline{u}+\tilde{B}\tilde{G} \rightarrow d \overline{d}\right) =\frac{m_{\tilde{B}}^4}{256 \pi m_{\tilde{G}}^2} \frac{1}{x^6 K_2(x) K_2\left(x \frac{m_{\tilde{G}}}{m_{\tilde{B}}}\right)}\nonumber\\
& \int_{(x+y)}^\infty dz \sqrt{z^2-(x-y)^2}\sqrt{z^2-(x+y)^2}K_1(z) \nonumber\\ 
& \{\left[\left(2 z^4-(x^2+y^2) z^2 -(x^2-y^2)^2\right] \left(\mathcal{A}_{2,u}+\mathcal{A}_{2,d}\right)-\left[y^4+(z^2-x^2)^2-2 y^2 (x^2+z^2)\right] \left(\mathcal{A}_{3,u}+\mathcal{A}_{3,d}\right)\right. \nonumber\\
& \left. - 6 xyz^2 \left(\mathcal{A}_{4,u}+\mathcal{A}_{4,d}\right)\right \}\nonumber\\
& \mathcal{A}_{2,q}=\sum_{l} Re\left \{ |g_{\tilde{B}}^{\rm LL}|^2 |g_{\tilde{G}}^{\rm LL}|^2 |\frac{\Gamma^{q\,*}_{Lli}\Gamma^{q}_{Llj}}{m^2_{\tilde{q}_l}}|^2+|g_{\tilde{B}}^{\rm LL}|^2 |g_{\tilde{G}}^{\rm LL}|^2 |\frac{\Gamma^{q\,*}_{Lli}\Gamma^{q}_{Rlj}}{m^2_{\tilde{q}_l}}|^2+|g_{\tilde{B}}^{\rm LL}|^2 |g_{\tilde{G}}^{\rm RR}|^2 |\frac{\Gamma^{q\,*}_{Rli}\Gamma^{q}_{Llj}}{m^2_{\tilde{q}_l}}|^2\right. \nonumber\\
& \left.+|g_{\tilde{B}}^{\rm RR}|^2 |g_{\tilde{G}}^{\rm RR}|^2 |\frac{\Gamma^{q\,*}_{Rli}\Gamma^{q}_{Rlj}}{m^2_{\tilde{q}_l}}|^2\right \}\nonumber\\
& \mathcal{A}_{3,q}=\sum_{l,p}\frac{1}{m^2_{\tilde{q}_l}}\frac{1}{m^2_{\tilde{q}_p}}  Re\left \{ \left(g^{\rm LL\,*}_{\tilde{B}}g^{\rm RR\,*}_{\tilde{B}}g^{\rm RR}_{\tilde{G}}g^{\rm LL}_{\tilde{G}}\Gamma^{q\,*}_{\rm Lli}\Gamma^{q}_{\rm Rpj}\Gamma^{q\,*}_{\rm Rlj}\Gamma^{q}_{\rm Lpi}+g^{\rm RR\,*}_{\tilde{B}}g^{\rm LL\,*}_{\tilde{B}}g^{\rm LL}_{\tilde{G}}g^{\rm RR}_{\tilde{G}}\Gamma^{q\,*}_{\rm Rli}\Gamma^{q}_{\rm Lpj}\Gamma^{q\,*}_{\rm Llj}\Gamma^{q}_{\rm Rpi}\right)\right \} \nonumber\\
& \mathcal{A}_{4,q}=\sum_{l,p}\frac{1}{m^2_{\tilde{q}_l}}\frac{1}{m^2_{\tilde{q}_p}}  Re\left \{ \left(g^{\rm LL\,*}_{\tilde{B}}g^{\rm LL\,*}_{\tilde{B}}g^{\rm LL}_{\tilde{G}}g^{\rm LL}_{\tilde{G}}\Gamma^{q\,*}_{\rm Lli}\Gamma^{q}_{\rm Lpj}\Gamma^{q\,*}_{\rm Llj}\Gamma^{q}_{\rm Lpi}+g^{\rm RR\,*}_{\tilde{B}}g^{\rm RR\,*}_{\tilde{B}}g^{\rm RR}_{\tilde{G}}g^{\rm RR}_{\tilde{G}}\Gamma^{q\,*}_{\rm Rli}\Gamma^{q}_{\rm Rpj}\Gamma^{q\,*}_{\rm Rlj}\Gamma^{q}_{\rm Rpi}\right)\right \},
\end{align} 
where $y = x \frac{m_{\tilde{G}}}{m_{\tilde{B}}}$,

\begin{align}
& \langle \sigma v \rangle \left(\tilde{B}u \rightarrow \tilde{G} \overline{u}+\tilde{B}d \rightarrow \tilde{G} \overline{d}\right)=\frac{m_{\tilde{B}}^2}{1024\pi K_2(x) x^4}=\int_x^\infty dz \frac{z^2-y^2}{z^2}(z^2-x^2)\nonumber\\
&\left[\frac{(z^2-x^2)(z^2-y^2)\left(y^2 (2x^2+z^2)+z^2(8z^2+x^2)\right)}{z^4}\left(\mathcal{A}_{2,u}+\mathcal{A}_{2,d}\right)\right.\nonumber\\
& \left.+6 (x^2-z^2) (y^2-z^2) \left(\mathcal{A}_{5,u}+\mathcal{A}_{5,d}\right)-\frac{6xy(z^2-x^2)(z^2-y^2)}{z^2}\left(\mathcal{A}_{6,u}+\mathcal{A}_{6,d}\right) \right] \nonumber\\  
& \mathcal{A}_{5,q}=\sum_{l,p}\frac{1}{m^2_{\tilde{q}_l}}\frac{1}{m^2_{\tilde{q}_p}}  Re\left \{ \left(g^{\rm LL\,*}_{\tilde{B}}g^{\rm RR\,*}_{\tilde{B}}g^{\rm RR}_{\tilde{G}}g^{\rm LL}_{\tilde{G}}\Gamma^{q\,*}_{\rm Llj}\Gamma^{\,* q}_{\rm Rpi}\Gamma^{q}_{\rm Rli}\Gamma^{q}_{\rm Lpj}+g^{\rm RR}_{\tilde{B}}g^{\rm LL}_{\tilde{B}}g^{\rm LL}_{\tilde{B}}g^{\rm RR}_{\tilde{G}}\Gamma^{q\,*}_{\rm Rlj}\Gamma^{q\,*}_{\rm Lpi}\Gamma^{q}_{\rm Lli}\Gamma^{q}_{\rm Rpj}\right)\right \}\nonumber\\
& \mathcal{A}_{6,q}= \sum_{l,p}\frac{1}{m^2_{\tilde{q}_l}}\frac{1}{m^2_{\tilde{q}_p}}  Re\left \{ \left(g^{\rm LL\,*}_{\tilde{B}}g^{\rm LL\,*}_{\tilde{B}}g^{\rm LL}_{\tilde{G}}g^{\rm LL}_{\tilde{G}}\Gamma^{q\,*}_{\rm Llj}\Gamma^{q\,*}_{\rm Lpi}\Gamma^{q}_{\rm Lli}\Gamma^{q}_{\rm Lpj}+g^{\rm RR\,*}_{\tilde{B}}g^{\rm RR\,*}_{\tilde{B}}g^{\rm RR}_{\tilde{G}}g^{\rm RR}_{\tilde{G}}\Gamma^{q\,*}_{\rm Rlj}\Gamma^{q\,*}_{\rm Rpi}\Gamma^{q}_{\rm Rli}\Gamma^{q}_{\rm Rpj}\right)\right \},
\end{align}

and finally:
\begin{align}
& \langle \sigma v \rangle \left(\tilde{B}\tilde{B} \rightarrow q \overline{q}\right) =\frac{3 m_{\tilde{B}}^2}{32 \pi} \frac{1}{x^6 {\left[K_2(x)\right]}^2} \int_{2 x}^\infty dz z^3 (z^2-4 x^2)^{3/2}\nonumber\\ 
&\sum_l \left[ \left(|g^{LL}_{\tilde{B}}|^2 |g^{LL}_{\tilde{B}}|^2 |\frac{\Gamma^{\,*q}_{Lli}\Gamma^{\,*q}_{Llj}}{m_{\tilde{q}_l}^2}|^2+|g^{\rm RR}_{\tilde{B}}|^2 |g^{\rm RR}_{\tilde{B}}|^2 |\frac{\Gamma^{\,*q}_{Rli}\Gamma^{\,*q}_{Rlj}}{m_{\tilde{q}_l}^2}|^2\right)-|g^{\rm RR}_{\tilde{B}}|^2 g^{\rm LL}_{\tilde{B}}|^2 |\frac{\Gamma^{\*q}_{Rli}\Gamma^{\,*q}_{Llj}}{m_{\tilde{q}_l}^2}|^2\right].
\end{align}

\noindent
As evident there is a tight relation between the flavor structure of these expressions and the one of the decay rates shown in the previous appendix.
\noindent
In the limit $m_{\tilde{d}} \ll m_{\tilde{u}}$ and absence of flavor violation and left-right mixing, taking also for simplicity $m_{\tilde{G}}=0$, we obtain:
\begin{align}
& \langle \sigma v \rangle \left(\tilde{B}u_k\rightarrow \overline{d}_i \overline{d}_j+\tilde{B}d_i \rightarrow \overline{u}_k d_j\right)=\frac{\alpha_1 |\lambda_{kij}|^2}{54} \frac{m_{\tilde{B}}^2}{m_0^4}\left(5 \frac{K_4(x)}{K_2(x)}+1\right) \nonumber\\
& \langle \sigma v \rangle \left(\tilde{B}\tilde{G} \rightarrow u \overline{u}+\tilde{B}\tilde{G} \rightarrow d \overline{d}\right) =\frac{16 \pi \alpha_1 \alpha_s}{27} \frac{m_{\tilde{B}}^2}{m_0^4} \left(2 \frac{K_4(x)}{K_2(x)}+1\right) \nonumber\\
& \langle \sigma v \rangle \left(\tilde{B}u \rightarrow \tilde{G} \overline{u}+\tilde{B}d \rightarrow \tilde{G} \overline{d}\right) =\frac{4 \pi \alpha_1 \alpha_s}{27}\frac{m_{\tilde{B}}^2}{m_0^4}\left(8 \frac{K_4(x)}{K_2(x)}+1\right)\nonumber\\
& \langle \sigma v \rangle \left(\tilde{B}\tilde{B} \rightarrow q \overline{q}\right) =\frac{16 \pi}{27} \alpha_1^2 \frac{m_{\tilde{B}}^2}{m_0^4}\left[{\left(\frac{K_3(x)}{K_2(x)}\right)}^2-{\left(\frac{K_1(x)}{K_2(x)}\right)}^2\right].
\end{align}

\bibliography{bibfile}{}
\end{document}